\documentclass{pasa}%

\usepackage{graphicx}
\usepackage{newtxtext,amsmath}
\usepackage{graphicx}	
\usepackage{amssymb}	

\def\nyu{\mbox{NYU-VAGC}}

\usepackage{color} 
\definecolor{purple}{RGB}{160,32,240}
\usepackage {comment}
\usepackage{url}
\usepackage{IEEEtrantools}

\def\mha{\mbox{$M_{\rm HI}$}}
\def\mhm{\mbox{$M_{\rm H_{2}}$}}
\def\mg{\mbox{$M_{\rm gas}$}}
\def\mbar{\mbox{$M_{\rm bar}$}}
\def\msun{\mbox{M$_{\odot}$}}
\def\vmax{\mbox{$V_{\rm max}$}}
\def\RH2{\mbox{$R_{\rm H_{2}}$}}
\def\MH2MHIMs{\mbox{$M_{\rm H_{2}}/M_{\rm HI}$}}
\def\cotoh2{\mbox{$\rm CO\mbox{-to-}H_{2}$}}
\def\a_co{\mbox{$\alpha_{\rm CO}$}}
\def\HI{\mbox{$\rm H_{I}$}}
\def\H2{\mbox{$\rm H_{2}$}}

\def\lcdm{\mbox{$\Lambda$CDM}}
\def\ms{\mbox{$M_{*}$}}
\def\mhm{\mbox{$M_{H_{2}}$}}
\def\mg{\mbox{$M_{\rm gas}$}}
\def\mbar{\mbox{$M_{\rm bar}$}}
\def\msun{\mbox{M$_{\odot}$}}
\def\vmax{\mbox{V$_{\rm max}$}}
\def\h2toco{\mbox{$\rm H_{2}-\mbox{to}-CO$}}
\def\RH2{\mbox{$R_{\rm H_{2}}$}}
\def\HI{\mbox{$\rm HI$}}
\def\H2{\mbox{$\rm H_{2}$}}
\def\lMH2{\mbox{$\log_{10}(M_{\rm H_{2}}/M_{\odot})$}}

\def\gsmf{\mbox{GSMF}}

\def\lcdm{\mbox{$\Lambda$CDM}}

\def\ltsima{$\; \buildrel < \over \sim \;$}    
\def\lesssim{\lower.5ex\hbox{\ltsima}}           
\def\gtsima{$\; \buildrel > \over \sim \;$}    
\def\grtsim{\lower.5ex\hbox{\gtsima}}           

\defcitealias{Calette+2018}{Paper I}

\newcounter{storeeqcounter}
\newcounter{tempeqcounter}

\title[Bivariate gas--stellar mass functions at $z\sim0$]{The bivariate gas--stellar mass distributions and the mass functions of early- and late-type galaxies at $z\sim0$}

\author[Rodr\'iguez-Puebla et al.]{Aldo Rodr\'iguez-Puebla$^1$\thanks{apuebla@astro.unam.mx}, A. R. Calette$^1$, Vladimir Avila-Reese$^1$,Vicente Rodriguez-Gomez$^2$ and Marc Huertas-Company$^{3,4,5,6}$
\affil{$^1$ Instituto de Astronom\'ia, Universidad Nacional Aut\'onoma de M\'exico, A. P. 70-264, 04510, M\'exico, D.F., M\'exico}%
\affil{$^2$ Instituto de Radioastronom\'ia y Astrof\'isica, Universidad Nacional Aut\'onoma de M\'exico, A. P. 72-3, 58089 Morelia, M\'exico}
\affil{$^{3}$LERMA, Observatoire de Paris, PSL Research University, CNRS, Sorbonne Universit\'es, UPMC Univ. Paris 06,F-75014 Paris, France}
\affil{$^{4}$Univerist\'e de Paris, 5 Rue Thomas Mann - 75013, Paris, France}
\affil{$^{5}$Departamento de Astrof\'isica, Universidad de La Laguna, E-38206 La Laguna, Tenerife, Spain}
\affil{$^{6}$Instituto de Astrof\'isica de Canarias, E-38200 La Laguna, Tenerife, Spain}
}%

\jid{PASA}
\doi{10.1017/pas.\the\year.xxx}
\jyear{\the\year}

\usepackage{aas_macros}
\usepackage{hyperref} 
\hypersetup{colorlinks,citecolor=blue,linkcolor=blue,urlcolor=blue}

\hypersetup{draft}

\begin{document}
\label{firstpage}

\begin{frontmatter}
\maketitle

\begin{abstract}
We report the bivariate \HI- and \H2-stellar mass distributions of local galaxies in addition of an inventory of galaxy mass 
functions, MFs, for \HI, \H2, cold gas, and baryonic mass, separately into early- and late-type galaxies. 
The MFs are determined using the \HI\ and \H2\ conditional distributions
and the galaxy stellar mass function, GSMF. For the conditional distributions we use the results from the compilation 
presented in \citet{Calette+2018}. For determining the GSMF from $\ms\sim3\times10^{7}$ to $3\times10^{12}$ \msun,
we combine two spectroscopic samples from the SDSS at the redshift range 
$0.0033<z<0.2$. We find that the low-mass end  
slope of the \gsmf, after correcting from surface brightness incompleteness, is $\alpha\approx-1.4$, consistent with
previous determinations. The obtained  \HI\ MFs 
agree with radio {\it blind} surveys. Similarly, the \H2\ MFs are consistent with CO follow-up optically-selected 
samples. 
We estimate the impact of systematics due to mass-to-light ratios and find that our MFs are robust against systematic errors.
 We deconvolve our MFs from random errors to obtain the intrinsic MFs. 
 Using the MFs, we calculate cosmic density parameters of all the baryonic components.
  Baryons locked inside galaxies represent 5.4\% of the universal baryon 
content, while $\sim96\%$ of the \HI\ and \H2\ mass inside galaxies reside in late-type morphologies. Our results
imply cosmic depletion times of \H2\ and total neutral H in late-type galaxies of 
$\sim 1.3$ and 7.2 Gyr, respectively, which shows that late type galaxies are on average inefficient in converting 
\H2\ into stars and in transforming \HI\ gas into \H2. 
Our results provide a fully self-consistent empirical description of galaxy demographics 
in terms of the bivariate gas--stellar mass distribution and their projections, the MFs. 
This description is ideal to compare and/or to constrain galaxy formation models.
\end{abstract}

\begin{keywords}
galaxies: evolution -   galaxies: luminosity function - galaxies: mass function 
\end{keywords}
\end{frontmatter}

\section{INTRODUCTION }
\label{sec:intro}

The determination of the matter-energy content of the Universe is
one of the most important achievements from the recent advances in observational cosmology
\citep[e.g.,][]{Arnaud+2016,Aghanim+2018}. Current determinations are fully consistent with the 
spatially-flat $\Lambda$ Cold Dark Matter (\lcdm) cosmology, with a present-day matter-energy content
dominated by the cosmological constant, $\Omega_{\Lambda} = 0.689$, and contributions of 
cold dark matter and baryon matter of $\Omega_{\rm cdm} \approx 0.262$ and $\Omega_{\rm bar} =  0.049$,
respectively \citep[for a value of the normalized Hubble constant of $h=0.674$,][]{Aghanim+2018,Aver+2015,Cooke+2018}. 
Therefore, the universal baryon mass density fraction is $f_{\rm bar,U} \equiv \Omega_{\rm bar} / \Omega_{\rm m} = 0.158$,
where $\Omega_{\rm m}= \Omega_{\rm cdm} + \Omega_{\rm bar}$. 
How much of these baryons, and their different components, are locked inside galaxies? 
This paper addresses this question
by quantifying the contribution from stars, atomic and molecular gas in galaxies of different masses and morphological types.

According to the current paradigm of structure formation, non-baryonic dark matter played a major
role in the evolution of the non-linear structures that we see today.  Particularly,  
galaxies are believed to form and evolve within extended dark matter haloes,
where multiple physical mechanisms are responsible for self-regulating star formation and thus setting up 
their observed properties \citep[for reviews see,][]{Mo+2010,Frenk+2012,Somerville+2015}. 
As dark matter structures and galaxies evolve, baryons are redistributed from an initial 
smooth distribution to a more complex variety of structures. 
Of primordial importance for galaxy evolution is the amount of neutral hydrogen 
available for the formation of stars. Gas radiative cooling within the haloes regulates the inflow of cold gas to galaxies. 
The subsequent formation of stars is regulated by a complex 
interaction between cold gas inflows and the gas heating/outflows produced by the stars, a process that depends on
halo mass. 
In low-mass halos, the stellar feedback, mostly form Supernova (SN) explosions, is able not only to heat the interstellar medium (ISM)
but also to expel large gas fractions from the galaxy.
In high-mass haloes, the long cooling time of shock-heated gas and the powerful feedback from rapidly accreting supermassive black holes that
heats and/or expels the gas tend to suppress the star formation. Thus, it is not surprising that the expected
fraction of baryons inside galaxies will differ from the universal baryon fraction,  $f_{\rm bar,U}$. 
Therefore, constraining the fraction of baryons and their different components in galaxies (mainly stars, atomic and molecular gas),
is essential to constrain the processes  that have taken place during the evolution of the galaxies. 

One of the main properties of galaxies are their stellar masses \ms. Indeed, 
the abundance of galaxies as a function of \ms\ provides important 
clues regarding the evolution of the galaxy population \citep[e.g.,][]{Peng+2010,Peng+2012,Yang+2012,Rodriguez-Puebla+2017}. Over the last two decades,
there has been a remarkable progress in assembling large galaxy samples from
multi-wavelength sky surveys that have led to robust determinations of the 
galaxy stellar mass function \citep[\gsmf; for recent discussions, and compilations of observations up 
to high redshifts see,][]{Conselice+2016,Rodriguez-Puebla+2017}. While there have been similar
efforts in assembling galaxy samples for atomic gas mass, \mha, based on radio blind observations 
\citep[e.g.,][]{Zwaan+2003, Meyer+2004, Koribalski+2004,Kovac+2005,Martin+2010,Haynes+2011,Hoppmann+2015,Haynes+2018} 
or from follow-up subsamples based on optical/infrared surveys \citep[e.g.,][]{Springob+2005,vanDriel+2016}, 
these are relatively shallow and/or in small volumes compared to the optical/infrared sky surveys, 
as well as strongly affected by selection effects. 
Therefore, the demographical analysis of \mha\ is challenging
especially when determining the low- and high-mass ends of the \HI\ mass function, \HI\ MF \citep[for a more detailed
discussion, see][]{Jones+2018}\footnote{As we will discuss in Section \ref{sec:results},
studying the very low-mass end of the \HI\ MF is beyond the scope
of this paper.}, as well as other statistics like the \HI\ two-point correlation functions\footnote{Two point correlation
functions will be discussed in Calatte et al. in prep.}. 
The situation is not that different and even more challenging for the molecular gas as
there are not blind galaxy samples in \H2. 
Nonetheless, there are some notable efforts to use optically-selected 
samples combined with small and shallow CO surveys to indirectly derive, from the (uncertain)
CO-to-\H2\ mass conversion factor, the  galaxy mass function in \H2, \H2\ MF \citep[e.g.,][]{Keres+2003,Lagos+2014,Saintonge+2017,Andreani+2018}.
Unfortunately, these CO surveys are also subject to incompleteness and selection effects or 
subject to a large fraction of galaxies with upper limits reported due to flux detection limits. 

As mentioned above, galaxy formation is a non-linear and complex process. 
Remarkably, well-defined correlations (usually power-laws) are, however, found from the observations. 
Among these are the correlations between the
\HI\ and stellar mass, \mha--\ms, and the \H2\ and stellar mass,
$M_{\rm H2}$--\ms. While both correlations present large scatters, when divided into early- and late-type galaxies 
they tend to show different and tighter correlations \citep[e.g.,][, and more references therein]{Calette+2018}. 
This is not surprising given that the formation histories of early- and late-type galaxies were different.
Thus, understanding the contribution of these two populations to the abundance of galaxies traced by \HI\ and \H2
provides further key constrains to galaxy formation theory models. 

In a recent work, \citet[][hereafter \citetalias{Calette+2018}]{Calette+2018} 
were able to determine empirically not only the mean  \mha--\ms\ and \mhm--\ms\ relations and their scatters
for early/late-type galaxies but also the full conditional probability distribution functions of \mha\ and \mhm\ given \ms, 
hereafter \HI-CPDF and \H2-CPDF, respectively. 
In this paper, we combine the empirical CPDFs with the  $z=0$ \gsmf\ to derive the bivariate gas-to-stellar mass
distributions and the MFs for the \HI, \H2, cold gas, and  baryon components, for all galaxies as well as for early and late types. 
Thus, the present paper represent a natural continuation of \citetalias{Calette+2018} with some updates.
These updates include new constraints on the best fitting parameters to the observed CPDFs from \citetalias{Calette+2018}. 

In this paper we compute the GSMF and its decomposition into early- and late-type galaxies. 
While there are many studies that have determined the \gsmf\ in the past, they do not typically report systematic errors or do not 
deconvolve it from random errors \citep[with a few exceptions, e.g.,][]{Bernardi+2010,Bernardi+2017,Obreschkow+2018}
or they are limited in the dynamical range of \ms\ due to 
the limited depth of the sample and/or the cosmic variance in the galaxy sample \citep[but see][]{Wright+2017}. 
We combine here two large galaxy samples, the low redshift sample,  \texttt{low-z},
from the NYU SDSS DR4 \citep{Blanton+2005,Blanton+2005a}, and the new photometry pipeline for the SDSS DR7
from \citet{Meert+2015,Meert+2016}. The \texttt{low-z}  sample 
suffers from surface brightness incompleteness, but here we estimate and correct for  
the fraction of missing galaxies due to this selection effect. 
As for the SDSS DR7, the new photometry from \citet{Meert+2015} shows that galaxy magnitudes
were previously underestimated due to sky subtraction problems \citep[see also,][]{Simard+2011}; the impact of these new determinations 
has been studied previously in \citet{Bernardi+2017}. We extend the \citet{Bernardi+2017}
analysis by using  not only different definitions of galaxy stellar masses but by dividing into two morphology groups,
early- and late-type galaxies. 

The results reported in this paper integrate the \HI- and \H2-CPDFs with new determinations for the GSMF to 
offer a full statistical description of the local galaxy demographics traced by the stellar, \HI, \H2, total cold gas, and baryon mass components. 
This statistical description of the local galaxy demographics is much more complete than the typically employed \gsmf\ for constraining 
models and simulations of galaxy formation. The new generation of semi-analytic models \citep[e.g.,][]{Croton+2016,Lagos+2018,Henriques+2019,Yung+2019} 
and cosmological Hydrodynamics simulations \citep[e.g.,][]{Hirschmann+2014,Vogelsberger+2014,Schaye+2015,Pillepich+2018,Dave+2019}, and their post-processing 
outcomes, are now able to predict stellar, \HI, and \H2\ masses for large galaxy populations in cosmological boxes 
\citep[see e.g.,][]{Lagos+2015,Diemer+2018,Diemer+2019,Popping+2019}. The empirically-based results 
presented here are optimal for comparing with these predictions as well as for calibrating theoretical models of galaxy evolution 
(see e.g. \citealp{Romeo2020}).  
The results to be presented in this paper are the basis for further studies as the inference of the galaxy-halo connection extended
to \HI, \H2, cold gas, and baryon masses. 

The present paper is organised as follows. In Section \ref{the_method} we describe
our method of using CPDFs in order to derive galaxy MFs traced by atomic, molecular, and cold gas masses 
as well as by the baryonic mass. In Section \ref{local-GSMF+correlations}
we describe the samples we use to derive our local \gsmf\ divided into 
early- and late-type galaxies. In Section \ref{sec:results} we 
present the results for our inventory of galaxy MFs, and compare them with direct observational results. 
We also present our estimates
for the cosmic density parameters related to the different baryonic components in galaxies. 
Section \ref{Discussion} discusses the impact of systematics and random errors. 
In Section \ref{conclusions}
we present a summary and our main conclusions. 

In this paper we adopt cosmological parameter values that are close to the 
Planck mission: $\Omega_{\Lambda}  = 0.693$,  $\Omega_{\rm m}  = 0.307$,
 $\Omega_{\rm bar}  = 0.048$ and $h = 0.678$. All stellar masses are normalised 
 to a \citet{Chabrier2003} Initial Mass Function, IMF.

\section{Modeling The Bivariate Distributions and Mass Functions from The Conditional Distribution Functions}
\label{the_method}

In this Section we describe the statistical method for deriving the \HI\ and \H2 
mass functions, MFs (as well as the total cold gas and baryon MFs), from the GSMF and the respective correlations of \mha\ and \mhm\ with \ms,
or more generally, the respective full mass conditional distribution functions, CPDFs. 
In general, our approach allows to calculate bivariate distribution functions of the \HI\ or \H2\ mass
and the stellar mass.  
One can imagine that our methodology is equivalent to an optically-selected 
volume-limited sample that it is complete in stellar mass, with $\HI$ and $\H2$ gas masses determined for every galaxy in the sample, 
and for which any MF can be determined. When information about morphology is available,
the CPDFs are useful for deriving the corresponding MFs into different 
morphological components. Here, we consider that the galaxy population is divided  into
two main morphological groups: early- and late-type galaxies. 
Following \citetalias{Calette+2018}, our definition of early-type galaxies includes morphological types that comprises E and S0 galaxies or equivalently $T\leq 0$ from the \citet{Nair_Abraham2010} morphology classification. Late-type galaxies 
are just the complement, from Sa to Irr.
Below we briefly describe the basic ingredients for calculating the MFs: 

\begin{itemize}
	
	\item {\bf Conditional Distribution Functions}:  For a fixed morphology, a galaxy of mass \ms\ has the chance of having either a \HI\ or \H2 
	mass described by their corresponding CPDFs. We denote the CPDFs of early- and late-types 
	by $P_{E} (M_j|\ms)$ and $P_{L} (M_j|\ms)$, respectively,  where $ j = $ \HI\ 
	or \H2. The  HI-CPDF and H2-CPDF contain information about all the moments of the $\HI$- and $\H2$-to-stellar mass correlations. 
	We use the observed HI-CPDF and H2-CPDF from \citetalias{Calette+2018}. 
	In Section \ref{Models_for_HI_H2}, we describe the 
	functional forms for the CPDFs proposed in \citetalias{Calette+2018}.

	\item {\bf Galaxy Stellar Mass Function}: The GSMF
	is an important input since it allows us to project the CPDFs 
	into their corresponding MFs. 
	We derive the GSMF for all galaxies, as well for the early- and late-type,  based on the SDSS.
	Section \ref{local-GSMF+correlations}
	describes our methodology to compute the observed GSMF over $\sim5$ decades in \ms, as well as
	its decomposition into early- and late-type galaxies.

\end{itemize}

The reader interested in our resulting MFs and bivariate distributions may skip to Section \ref{sec:results}.

\subsection{Generalities}
\label{secc:generalities}

As discussed above, a CPDF, $P_j(M_j|\ms)$, determines the 
chances that a galaxy of mass \ms\ possess a specific galaxy property $M_j$,
with $ j = $ \HI, \H2, cold gas or baryonic mass. Note that the units of $P_j$ is per \msun. The relation
between the distribution $P_j$ in bins per \msun\ to dex$^{-1}$, $\mathcal{P}_{j} $, is given by
\begin{equation}
\mathcal{P}_{j} (M_j|\ms) = P_{j}(M_j|\ms) \times \frac{M_j}{\log e}.
\end{equation}
The advantage of using $P_j(M_j|\ms)$ is that it contains information about {\it all the moments}
of the distribution, in particular the mean $M_j-\ms$ relation and its standard deviation.

The {\it joint} distribution function of \ms\ and $M_j$, hereafter referred as the {\it bivariate} distribution function, is defined as:
\begin{equation} 
\Phi(M_{j},\ms) = \frac{d^2 N(M_{j}|\ms)} {V d\log M_{j} d\log \ms}= 
\mathcal{P}_{j}(M_{j}|\ms) \phi_* (\ms),
\label{2D_density}
\end{equation}
where $d^2 N$ is the bivariate number of galaxies within the mass range $\log \ms\pm d\log\ms/2$ and $\log M_{j}\pm d\log M_{j}/2$ 
in a given volume $V$, and $\phi_* (\ms)$ is the GSMF in units of ${\rm Mpc}^{-3} {\rm dex}^{-1}$. 
The integration (marginalisation) of $\Phi(M_{j},\ms)$ over \ms\ results in the total 
MF for $M_j$, $\phi_j(M_j)$, that is,
\begin{eqnarray}
\phi_{j} (M_j) =\int \Phi(M_{j},\ms)d\log\ms =  \cr
\int \mathcal {P}_{j}(M_j|\ms) \phi_* (\ms) d\log\ms.   
\label{phi_j}
\end{eqnarray}
The above equation shows how the CPDFs 
are projected into a number density function via the GSMF. 
Note that integration of $\Phi(M_{j},\ms)$ over $M_j$ gives the total GSMF\footnote{In the literature there
are different methods to determine multivariate joint distributions, one example is the copula approach. 
A copula is function that joint multivariate cumulative distribution 
functions to their corresponding marginal distributions. They are useful to model the dependence 
between random variables based on uniform marginals. According to the Sklar's theorem, any 
multivariate joint distribution is totally defined given the marginal distributions and a copula describing the 
structure between random variables. More details on the copula approach and the application to the galaxy luminosity 
function the reader is referred to \citet{Takeuchi2010} and \citet{Takeuchi+2013}. 
Here we use the CPDFs formalism for 
two reasons: 1)  the input data that we use 
are characterised on that format, see \citetalias{Calette+2018} and below; and 2) Our goal is
to determine the mass functions using the CPDFs.}.

As discussed previously, when studying the properties of galaxies it is useful to separate them into, at least, two morphological components
such as early types, or spheroid-dominated galaxies, and  late types, or disk-dominated galaxies. 
Thus, the total GSMF can be formally represented as the contribution of these two types 
\begin{equation}
\phi_*(\ms) = \phi_{*,E}(\ms) + \phi_{*,L}(\ms),
\end{equation}
denoted respectively by $\phi_{*,E}$, and $\phi_{*,L}$.
In terms of the fraction of early- and late-type galaxies ($f_E$ and $f_L$), their
corresponding galaxy stellar MFs are given respectively by $\phi_{*,E} = f_E\times\phi_*$, and $\phi_{*,L} = f_L\times\phi_*$, with 
$f_E+f_L = 1$. 

Early- and late-type galaxies are different in their $\HI$- and $\H2$-to-stellar mass distributions. Thereby, Equation (\ref{phi_j})
can be generalised in terms of the distribution 
$\mathcal {P}_{i,j}(M_j|\ms)$, where the subscripts indicate $i = $ early or late type, and $ j = $ \HI, \H2, cold gas or baryonic mass. 
Then, the
generalisation of Equation (\ref{phi_j}) to galaxies with morphological type $i$ and mass component $j$ is:
\begin{equation}
\phi_{j, i} (M_j) = \int f_i(\ms) \mathcal {P}_{i,j}(M_j|\ms) \phi_* (\ms) d\log\ms.
\label{phi_ij}
\end{equation}

Finally, the {\it total} CPDFs 
are calculated from the respective conditional distributions of early- and late-type galaxies as:
\begin{equation}
\begin{split}
\mathcal {P}_{j}(M_{j}|\ms) = & f_{E}(\ms)\times \mathcal {P}_{E,j} (M_{j}|\ms)+\\
& f_{L}(\ms)\times \mathcal {P}_{L,j} (M_J|\ms),
\label{PHI_all}
\end{split}
\end{equation}
with $ j = \HI, \H2,$ cold gas or baryonic mass.  

\subsection{The \HI\ and \H2\ Conditional Distribution Functions}
\label{Models_for_HI_H2}

As shown in Equations  (\ref{2D_density}) and (\ref{phi_ij}), the conditional 
or bivariate distribution functions
are useful to statistically determine the MFs. 
Evidently, in the case of atomic and molecular gas, we are assuming that for every galaxy
that is optically selected, there must exist  \HI\ and \H2\ counter parts. The discussion on 
the possible existence of pure \HI\ or \H2\
galaxies, those that will not be observed in optically selected samples but rather in radio blind surveys, 
is out of the scope of this paper.  Note that if they exist, the chance
of observing those galaxies is very low over the mass ranges that we will derive the MFs. 
For example, in the case of pure \HI\ galaxies, the ALFALFA survey has found $\sim1.5\%$
of \HI\ sources that were not clearly associated to an optical counterpart. Of those, $\sim75\%$ 
are likely tidal in origin \citep{Haynes+2011}. Thus, $\sim0.4\%$ of HI source observed
in the ALFALFA survey are purely gaseous galaxies candidates, most of them at
the mass range $10^{7}<\mha/\msun< 10^{10}$ \citep{Cannon+2015}. As we will show, our completeness 
limit for the \HI\ MF is $\mha\sim10^{8}\msun$. The above fraction, could be considered as an upper limit
as some of these sources have already detected optical counterparts revealing unusual high HI mass-to-light ratios
 \citep{Cannon+2015}. Thus we conclude that our results are unlikely to be affected by excluding pure gas galaxies
 in our analysis.

\subsubsection{The  \citet{Calette+2018} \HI\ and \H2\ Conditional Distribution Functions}
\label{secc:summary_C18}
Here we use the results from \citetalias{Calette+2018} \citep{Calette+2018} who determined the 
\HI- and \H2-to-stellar mass 
ratio distributions (CPDFs) as a function of \ms\ from a large compilation of optically-selected samples with 
radio observations. Next, we briefly describe the steps taken in  \citetalias{Calette+2018} 
to derive the  \HI\ and \H2\ CPDFs. The reader is referred to that paper for details. 

The compiled data described in  \citetalias{Calette+2018}  consist of a set of incomplete and
inhomogeneous samples. We first homogenised all these samples to a common 
IMF, cosmology, radio telescope configuration and sensitivity, and CO-to-luminosity conversion factor. Then, we selected only those samples 
without obvious biases due to selection effects such as environment. Radio non detections, reported
in the literature as upper limits, are an important source of uncertainty when deriving distributions
or correlations. In \citetalias{Calette+2018} we {\it included} non detections to derive the \HI\ and \H2\ CPDFs.
Below we briefly describe the treatment that we employed for radio non detections. 

In our compiled samples most of radio non detections are early-type galaxies representing 
a non negligible fraction of intermediate and massive galaxies, which are (typically) gas poor. 
An important fraction of those galaxies are from the GASS \citep{Catinella+2013} and the
COLD-GASS \citep{Saintonge+2011} surveys at distance of $109<D/{\rm Mpc}<222$. Compared to other
more nearby samples of intermediate and massive early-type galaxies with measurements of \HI\ and \H2\ mass, 
such as the ATLAS 3D \citep{Serra+2012}
at $\bar{D}\sim25 {\rm Mpc}$, we noted that the upper limits of the GASS/COLD-GASS samples are 
$\sim1-2$ orders of magnitude larger than nearby samples  (\citetalias{Calette+2018}). 
The above lead us to first introduce a correction for the upper limits of the GASS/COLD-GASS
surveys by a distance effect. Recall that radio non detections or upper limits depend not only on the 
sensitivity of the radio telescope or integration time but also on the distance to the object. In \citetalias{Calette+2018}
we corrected the upper limits of the GASS/COLD-GASS samples by a distance effect by using  
nearby samples such as the ATLAS 3D survey. Briefly, our correction consists in using the distances
and upper limits from nearby samples to estimate the upper limits in the GASS and COLD-GASS as if
these two samples were at the same distance as the nearby ones. 
We validated our procedure by using a mock galaxy survey by applying similar distance-sensitivity effects
as GASS/COLD-GASS surveys, for details see \citetalias{Calette+2018}. For late-type galaxies,
notice that most of them are detected in radio due to their
large fractions of gas and it is not necessary to introduce the above corrections. 
Next, we describe the treatment of the upper-limits to
derive the \HI\ and \H2\ CPDFs. 

In our analysis from \citetalias{Calette+2018} we {\it included} upper limits, or left-censored data, by using 
the \citet{Kaplan_Meire1958} non-parametric estimator. This  estimator provides a reconstruction of information 
lost by censoring. \citet{Feigelson+1985} adapted this estimator for astronomical samples. We used the \textsc{ASURV}
package based on \citet{Feigelson+1985} to derive 
the \HI\ and \H2\ CPDFs from our compiled samples. 
We have also applied the censoring \citet{BJ79} regression method to derive the relationship and 
standard deviations between the \HI- and \H2-to-stellar mass ratio and \ms. We note that the regression results 
are consistent with the (logarithmic) mean and standard deviation values obtained from the CPDFs based 
on the \citet{Kaplan_Meire1958} estimator.

\subsubsection{The functional forms of the \HI\ and \H2\ Conditional Distribution Functions}
\label{secc:functional_forms_HI_H2}


For the \HI\ and \H2\ CPDFs of late-type galaxies, in \citetalias{Calette+2018} we found that they are 
described by a Schechter function. In the case of early-type galaxies, the CPDFs are better described by a (broken) Schechter function plus a uniform distribution at the low$-\mathcal{R}_j$ values. 
Following, we describe in more detail these functional forms. 

We begin by introducing the following Schechter-type probability distribution function for the \HI- or \H2-to-stellar mass ratios, 
$ \mathcal{R}_j=M_j/\ms$, in the 
range $\log \mathcal{R}_j \pm d \log \mathcal{R}_j / 2$:
\begin{equation}
\mathcal{S}_{i,j} (\mathcal{R}_j) = \frac{\ln(10)}{\mathcal{N}_{i,j}}\left(\frac{\mathcal{R}_{j}}{\mathcal{R}^*_{i,j}}\right)^{\alpha_{i,j} + 1} 
\exp\left(-\frac{\mathcal{R}_{j}}{\mathcal{R}^*_{i,j}}\right),
\label{Sij}
\end{equation}
where the morphology is represented with $i = $ early or late type, and the galaxy property is 
represented with $ j = $ \HI\ or \H2. The parameters are:
the characteristic gas-to-stellar mass ratio,
$\mathcal{R}^*_{i,j}$, the normalisation parameter, $\mathcal{N}_{i,j}$, which constrains the probability to
be between zero and one,\footnote{For $\alpha_{i,j}>-1$ then $\mathcal{N}_{i,j} = \Gamma(1+\alpha_{i,j})$, with $\Gamma(x)$
the complete gamma function. In general $\mathcal{N}\propto \int_{-\infty}^{\infty} x^{\alpha} \exp(-x) dx$.}
and the power-law slope $\alpha_{i,j}$ for the part of the distribution of galaxies with low gas-to-stellar mass ratio. 

\begin{description}
\item[$\bullet$] Late-type Galaxies:
\end{description}

For late-type galaxies, that is $i = L$, in \citetalias{Calette+2018} we found that the HI-CPDF and H2-CPDF is described 
by the Schechter-type distribution function given by Eq. (\ref{Sij}) 
with the parameters 
$\alpha_{L,j}$ and $\mathcal{R}^*_{L,j}$ functions of \ms\ as follows:
\begin{equation}
\alpha_{L,j} =  \alpha_{0;L,j} \log \ms + \alpha_{1;L,j} ,
\label{alpha_lj}
\end{equation}
and 
\begin{equation}
\mathcal{R}^*_{L,j} = \frac{\mathcal{R}^*_{0;L,j}}{\left(\frac{\ms}{\mathcal{M}^*_{L,j}} \right)^{\beta_{L,j}} + \left(\frac{\ms}{\mathcal{M}^*_{L,j}} \right)^{\gamma_{L,j}}}.
\label{Rchar}
\end{equation}
Consider that $\mathcal{S}_{L,j} (\log\mathcal{R}_j) d\log \mathcal{R}_j =  \mathcal{S}_{L,j} (\log M_j - \log\ms) d\log (M_j / \ms)$. 
By definition \ms\ is fixed, thus the \HI\ and \H2\ CPDFs of late-type galaxies are given by:
\begin{equation}
\mathcal {P}_{L,j}(M_j | \ms) d\log M_j = \mathcal{S}_{L,j} (\log M_j - \log\ms) d\log M_j.
\label{CDF-LTG}
\end{equation}
The above explicitly shows that the integration over conditional distribution functions can also be interpret as convolutions in Equation (\ref{phi_j}).

\begin{description}
\item[$\bullet$] Early-type Galaxies:
\end{description}

In the case of early-type galaxies, $i = E $, we showed in \citetalias{Calette+2018} that both for the HI-CPDF and H2-CPDF 
are described as the sum of two distribution functions; the Schechter-type distribution function, $\mathcal{S}_{E,j}$, and a
uniform function, $\mathcal {U}_{0,j}$,
\begin{equation}
\mathcal {E}_{j} (\mathcal{R}_j) = \left\{ 
\begin{array}{l l}
\mathcal {U}_{0,j} & \mathcal{R}_{0,j} \le\mathcal{R}_j < \mathcal{R}_{1,j} \\
A \times \mathcal{S}_{E,j} (\mathcal{R}_j) & \mathcal{R}_{1,j} \le \mathcal{R}_j 
\end{array},\right.
\label{Pej}
\end{equation}
where $\mathcal{R}_{0,j} = \mathcal{R}_{1,j} / 10,$\footnote{As discussed in \citetalias{Calette+2018}, 
	the observed data imply that the \HI- and \H2-to-\ms\ ratios will not be lower than $\sim10^{-4}-10^{-5}$. This seems plausible 
	since even for galaxies that transformed all their gas into stars, the gas mass recycled to the ISM by stellar evolution could provide the above 
	minimal floor for the gas mass ratios.}
and $ \log \mathcal{R}_{1,j} = r_{0,j} \log\ms + r_{1,j}$, while the uniform 
distribution is given by
\begin{equation}
\mathcal {U}_{0,j}  (\ms) = \frac{p_{0,j} \log\ms + p_{1,j}}{\Delta},
\label{Poj}
\end{equation}
and 
\begin{equation}
A = \left( 1 - \mathcal {U}_{0,j} \times \Delta\right) \times \frac{\mathcal{N}_{i,j}}{\eta_{i,j}(\mathcal{R}_{1,j})},
\label{normA}
\end{equation}
where in \citetalias{Calette+2018} we assumed that $\Delta = \log 10 = 1$ dex, 
the symbol $\eta_{i,j}(\mathcal{R}_{1,j})$ takes into account the fraction of galaxies in the Schechter-type 
mode for galaxies with gas ratio above 
$\mathcal{R}_{1,j}$.\footnote{Similarly to late-types, in the case that $\alpha_{i,j}>-1$  then $\eta_{i,j}(\mathcal{R}_{1,j}) = \gamma(1+\alpha_{i,j},\mathcal{R}_{1,j})$, with $\gamma(x,a)$
	as the incomplete gamma function. In general $\eta_{i,j}(a)\propto \int_{a}^{\infty} x^{\alpha} \exp(-x) dx$.} 
	The HI-CPDF and \H2-CPDF of early-type galaxies are:
\begin{equation}
\mathcal {P}_{E,j}(M_j | \ms) d\log M_j =  \mathcal{E}_{j}(\log M_j - \log\ms) d\log M_j.
\label{CDF-ETG}
\end{equation}

\subsubsection{Constraints on the best fitting parameters}
\label{secc:fitting_data}

In \citetalias{Calette+2018} the best fit parameters for late-type galaxies, Equations (\ref{alpha_lj})-(\ref{Rchar}),
and for early-type galaxies, Equations (\ref{Pej})-(\ref{normA}), were constrained using the observed
\HI- and \H2-CPDFs on various stellar mass bins. Computing CPDFs over \ms\ bins requires of
the GSMF in addition of the fraction of early/late-type galaxies (see Section \ref{CDF-constraints} for more details). Since we
are using slightly different inputs, namely the GSMF and the fractions of early/late-type galaxies, than in \citetalias{Calette+2018}, we 
prefer to perform our own fits to the same data, for consistency. The results are presented in Section \ref{CDF-constraints}. 
The differences with the parameters reported in \citetalias{Calette+2018} are actually small.

\subsection{The Cold Gas and Baryonic Conditional Distribution Functions}

Once we have constructed the HI-CPDF and \H2-CPDF we can now define the conditional distributions
for the cold gas and baryon masses, \mg\ and \mbar.

The total cold gas content in a galaxy is 
composed of \HI, \H2, helium, and metals; helium and metals account for roughly 30\% of the cold gas, $M_{\rm He} + M_{\rm Z}\approx0.3$\mg.
Therefore, $\mg = \mha  + \mhm + M_{\rm He} + M_{\rm Z} = 1.4 \times (\mha + \mhm)$. 
For simplicity, let \mha\ and \mhm\ be two independent random 
variables. Section \ref{Discussion} discusses the validity of this assumption. 
Then, \mg\ is a random variable with the conditional distribution function:
\begin{equation}
\begin{split}
P_{\rm gas}(\mg|\ms) =\frac{1}{1.4}  & \int P_{\rm HI}\left(0.71 \mg - \mhm|\ms\right) \times  \\ 
& P_{\rm H_2}(\mhm|\ms) d \mhm,  \\ 
=\frac{1}{1.4}  &\int P_{\rm HI}(\mha|\ms) \times   \\
&P_{\rm H_2}(0.71 \mg - \mha|\ms) d \mha,
\label{Pgas_lin}
\end{split}
\end{equation}
or after some algebra, the same distribution function but per bin in log space is:  
\begin{equation}
\begin{split}
\mathcal{P}_{\rm gas}(\mg|\ms) =  &\int  \frac{\mathcal{P}_{\rm HI}(0.71\mg - \mhm|\ms)}{1 - 1.4\ \mhm/\mg}\times \\
& \mathcal{P}_{\rm H_2}(\mhm|\ms) d \log \mhm, \\
=  &\int \mathcal{P}_{\rm HI}(\mha|\ms)\times \\
& \frac{\mathcal{P}_{\rm H_2}(0.71 \mg -\mha|\ms)}{1 - 1.4\ \mha/\mg}  d \log \mha.
\label{Pgas}
\end{split}
\end{equation}

For the baryonic conditional distribution functions, we again assume that $M_{\rm gas}$ and \ms\ are two
independent random variables. Thus $M_{\rm bar} = M_{\rm gas} + \ms$ is a random variable with a
distribution function given by
\begin{equation}
\begin{split}
P_{\rm bar}(M_{\rm bar}|\ms) = & \int P_{\rm gas} (M_{\rm bar} - \mathcal{M}_*|\ms) \times \\
& \delta(\mathcal{M}_* - \ms) d \mathcal{M}_*\\
= & P_{\rm gas} (M_{\rm bar} - \ms|\ms),
\end{split}
\end{equation}
where $P_{\rm gas}$ is the conditional distribution function for gas, Equation (\ref{Pgas_lin}), and the Dirac-$\delta$ function
appears explicitly for the \ms\ term. Similarly as above, we find that
\begin{equation}
\mathcal{P}_{\rm bar}(M_{\rm bar}|\ms) = \frac{\mathcal{P}_{\rm gas}(M_{\rm bar} - \ms|\ms)}{1 - \ms / M_{\rm bar}}.
\label{Pbar}
\end{equation}

Finally, we derive the gas and baryon MFs using Equations (\ref{phi_ij}), (\ref{Pgas}) and 
(\ref{Pbar}), the last two valid for early- and late-type galaxies.

\section{The GSMF of all, early- and late-type galaxies}
\label{local-GSMF+correlations}

The preceding Section described a methodology to use the GSMF as an interphase that transforms 
galaxy CPDFs into MFs, see Equation (\ref{phi_ij}). 
In this Section we briefly describe how we determine the local \gsmf\ for masses above 
$\ms\sim3\times10^{7} \msun$, as well as the \gsmf's for early- and late-type galaxies.
For a more detailed description of the galaxy samples utilised here and the different corrections we apply, 
the reader is referred to Appendices \ref{SDSS_DR7_GSMF}--\ref{GSMF-correction}.

\subsection{The Galaxy Samples and the GSMF}
\label{Sec:GSMF}

To estimate the \gsmf\ over a large dynamical range we use two galaxy samples. 
Next, we shortly describe the procedure and our determinations.

\begin{description}

\item[1)] For masses above $\ms = 10^9$ \msun, we use the SDSS DR7 based on the photometric catalog from \citet{Meert+2015}
and \citet{Meert+2016}\footnote{Available at \url{http://www. physics.upenn.edu/?ameert/SDSS_PhotDec/}} at the 
redshift interval  $0.005<z<0.2$. 
Previous studies have concluded that the measurements of the apparent brightnesses based on the 
standard SDSS pipeline photometry are underestimated due to sky subtraction problems, particularly,
in crowded fields \citep{Bernardi+2010,Blanton+2011,Simard+2011,Bernardi+2013,He+2013,Mendel+2014,Kravtsov+2014,Meert+2015,DSouza+2015,Bernardi+2016,Meert+2016}. New determinations of the \gsmf\ based on the new algorithms 
for obtaining more precise measurements of the sky subtraction, and thus to improve the photometry, have concluded that 
the bright end of the luminosity/mass function has been systematically underestimated \citep{Bernardi+2017}.
While there are various groups working in improving the determination of galaxy apparent brightnesses, see 
references above, \citet{Bernardi+2017} showed that all those studies agreed up to 0.1 dex in the \gsmf.  
In this paper we use the apparent S\'ersic $r$, $g$, and $i$ band luminosities reported in \citet{Meert+2015} and \citet{Meert+2016} derived for 
the SDSS DR7 based on the \textsc{PyMorph} software pipeline \citep{Vikram+2010,Meert+2013}. This software has been extensively tested
in \citet{Meert+2013} and shows that it does not suffer from sky subtraction problems.  
All magnitudes and colours are K+E corrected at a redshift rest-frame $z=0$, see Appendixes \ref{SDSS_DR7_GSMF} and \ref{Apendix_Kcorr}.
As described in Appendix \ref{SDSS_DR7_GSMF}, for every galaxy we estimate \ms\ from five colour-dependent mass-to-light ratios
but we define as our fiducial \ms\ the geometric mean of all the determinations. Using the $1/\vmax$ method, we derive
six $\gsmf$s based on the mass definitions described above. Consistent with \citet{Bernardi+2017}, we find that 
the differences in mass-to-light ratios introduce large discrepancies in the \gsmf, especially at the high-mass end. In
Figure \ref{fig:GSMF_comparison} from Appendix \ref{SDSS_DR7_GSMF}, we find that a shift of $\sim\pm0.15$ dex in the \ms\ axis 
recovers systematic errors  in the \gsmf\ due to different mass-to-light ratios.

\item[2)]  For masses below $\ms =10^9$ \msun, we use the SDSS DR4 NUY-VAGC 
\texttt{low-z} sample,\footnote{Available at \url{http://sdss.physics.nyu.edu/vagc/lowz.html}} at the redshift interval $0.0033<z<0.005$, and ideal to study the 
low mass/luminosity galaxies \citep{Blanton+2005,Blanton+2005a}.  
As before, all absolute magnitudes and colours were K+E corrected at a redshift rest-frame $z=0$.
Also, we derive \ms\ from five colour-dependent mass-to-light ratios
and, again, we define our fiducial \ms\ as the geometric mean of all the determinations.  
We construct the \gsmf\
using the $1/V_{\rm max}$ method and include missing galaxies due to surface brightness incompleteness,
as described in Appendix \ref{GSMF-correction}. For surface brightness incompleteness we follow closely the methodology 
described in \citet{Blanton+2005a}. The latter correction is relevant for the low-mass end. 
Based on the conclusions from \citet{Baldry+2012}, we use a simple correction for the low-mass end  
in order to correct for the local flow model distances from \citet{Willick+1997} to the one by  \citet{Tonry+2000}. 

\end{description}

\begin{figure} 
	\vspace{-120pt}
	\hspace{-15pt}
	\includegraphics[height=8.in,width=6.in]{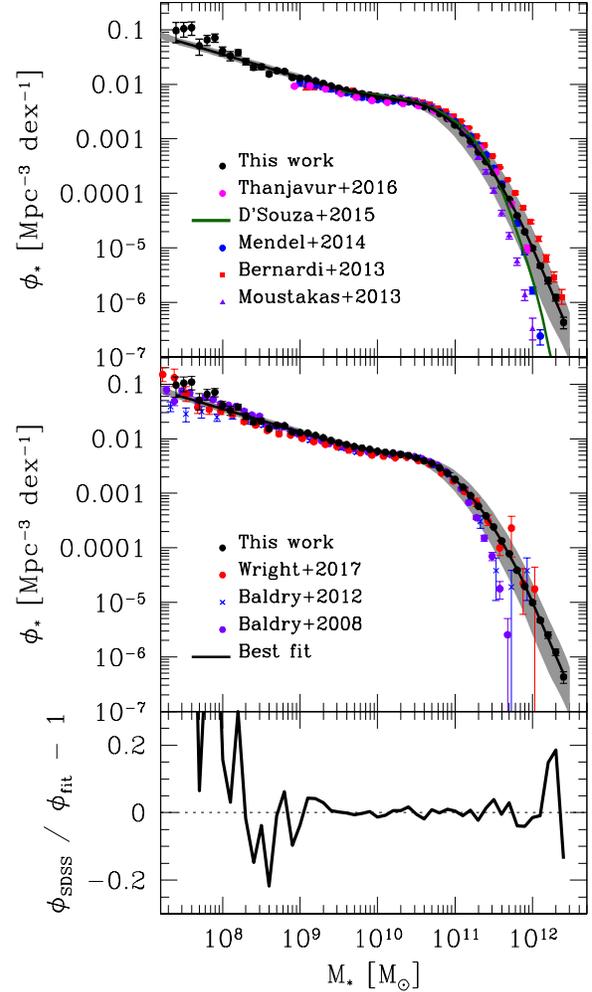}
	\vspace{-130pt}
	\caption{Observed GSMF when combining the SDSS NYU-VAGC low-redshift sample and
		the SDSS DR7 sample, black filled circles with error bars. We reproduce our results in the upper and
		the middle panels. The best fit model composed of a Schechter function with a sub-exponential slope 
		and a double power-law function is shown 
		as the black solid line. The shaded area shows an estimate of the
		systematic errors with respect to the best fitting model. 
		The bottom panel shows the residuals for our best fitting model as a function of \ms.
		We include comparisons to some previous observational determinations of the GSMF: in the upper panel we show
		determinations that are complete down to $\sim 10^9$ \msun, mostly based on the SDSS DR7, while in the 
		middle panel we show determinations based on the GAMA survey, which are complete down to 
		$\sim 3-5\times 10^7$ \msun, but suffer from cosmic variance at high masses due to the small volume. 
		} 
	\label{GSMF} 
\end{figure}

Our final \gsmf\ is the result of combining the SDSS NUY-VAGC \texttt{low-z} sample, for galaxies with masses 
$\ms\approx 3\times 10^{7}\msun$ to $\ms\approx10^{9}\msun$, and the SDSS DR7 sample for galaxies with $\ms\grtsim 10^{9}\msun$, based 
on our fiducial \ms\ determination. 
Figure \ref{GSMF} presents our final \gsmf\ with the black solid circles and error bars. 
The black solid line shows the best fit to the data (described below), and the grey shaded area shows a shift in the
\ms\ axis of $\pm 0.15$ dex. As discussed above, in Appendix \ref{SDSS_DR7_GSMF} we find that this is a 
good approximation to the systematic errors in the \gsmf\ due to differences in the mass-to-light ratios. In the same figure,
we include comparisons to previous works. In order to account for differences in cosmologies, we scale previous
studies to our cosmology using the following relations: 
\begin{equation}
\phi_{*, {\rm us}} = \phi_{*, {\rm lit}}\left(\frac{h_{\rm us}}{h_{\rm lit}}\right)^3,
\end{equation}
and 
\begin{equation}
M_{*, {\rm us}} = M_{*, {\rm lit}}\left(\frac{h_{\rm lit}}{h_{\rm us}}\right)^2,
\end{equation}
where $h_{\rm us} = 0.678$ and $h_{\rm lit}$ is the respective value reported in the literature. 
Nonetheless, the impact of accounting for different cosmologies is small. 

In the upper panel of Figure \ref{GSMF}, we reproduce the $\gsmf$s from previous determinations with stellar mass 
completeness above $\ms\sim 10^9$ \msun. The violet triangles with error bars are the determinations from \citet{Moustakas+2013},
who used a spectroscopic sample of SDSS DR7 galaxies from the NYU-VAGC with redshifts $0.01<z<0.2$ combined with observations 
from GALEX. The red squares with error bars are the estimation obtained in \citet{Bernardi+2013} from 
a sample of SDSS DR7 galaxies with photometry based on the \textsc{PyMorph} software pipeline at $z\sim0.1$. Here
we reproduce their result based on S\'ersic luminosities. Additionally, we compute the \gsmf\ using the stellar mass
estimates from S\'ersic photometry from \citet{Mendel+2014} who used the \citet{Simard+2011} SDSS DR7 sample of $g$ and $r$ band photometry
and extended to $u$, $i$ and $z$ bands, blue filled circles with error bars. We show the best fitting model from \citet{DSouza+2015},
who estimated the \gsmf\ by stacking images of galaxies with similar stellar masses and concentrations 
to correct \textsc{Model} magnitudes from the SDSS DR7, dark green solid line. Finally, we compare 
our result to \citet{Thanjavur+2016}, who derived the \gsmf\ using the analysis from \citet{Mendel+2014}.   

Our \gsmf\ agrees well with previous 
determinations at the $\sim 10^{9.3}-10^{11}$ \msun\ range. At the high mass end, 
it is shallower than previous determinations \citep[e.g.,][]{Moustakas+2013} except to \citet{Bernardi+2013}, who use 
S\'ersic photometry from the SDSS DR7. As extensively discussed in \citet{Bernardi+2017}, there are two systematic effects that could lead to
differences when comparing to previous determinations from the literature; assumptions on mass-to-light ratios and 
estimations of galaxy surface brightness. In the case of \citet{Moustakas+2013}  and \citet{DSouza+2015}, who used 
\textsc{cmodel} and \textsc{Model} magnitudes, the comparison is not obvious due to systematic effects in both mass-to-light ratios 
and photometry \citep{Bernardi+2017}. In the case of \citet{Mendel+2014} and \citet{Bernardi+2013}, effects
on photometry are not the dominant ones but mass-to-light ratios. Nonetheless, those differences are within the 
expected systematic effect, especially at the massive-end, \citep[][see also Figure \ref{fig:GSMF_comparison}]{Bernardi+2017}.
We therefore conclude that when comparing to other previous determinations, the differences that we observe are
consistent with the differences expected from systematic effects. 
Indeed, Figure \ref{GSMF} shows that most of the previous determinations
are within our region of systematic errors. Thus, hereafter we will assume that our shift of  $\pm0.15$ dex
in the \ms\ axis approximately captures systematics not only from stellar population models but also from photometry. 

The middle panel of Figure \ref{GSMF} presents comparisons to some previous determinations from deep but small-volume samples.
The purple dots with error bars are from \citet{Baldry+2008}, who used  
the SDSS NYU-VAGC \texttt{low-z} sample but did not include missing galaxies due to surface brightness 
incompleteness. In addition, we compare to \citet{Baldry+2012}, who used the GAMA survey for galaxies at $z<0.06$, and complete
down to $r=19.4$ mag for two thirds of the galaxy sample and to $r=19.8$ for one third of the sample.   
Finally, we reproduce the observed \gsmf\ from \citet{Wright+2017}, who also used the GAMA survey to estimate the \gsmf. 

At low masses our results are in excellent agreement with the GAMA $\gsmf$s. This is encouraging since the GAMA survey 
does not suffer from surface brightness incompleteness, at least within the stellar mass range that we are comparing our results.
This is an indication that the surface brightness corrections described in Appendix \ref{GSMF-correction} 
are able to recover the slope of the \gsmf\ at low masses. 
Consistent with the values reported in \citet{Baldry+2012} and \citet{Wright+2017}, we find that 
the faint-end slope of the GSMF is $\alpha \approx -1.4$, below we describe in more detail the fitting model for the
GSMF. The above is also in good agreement with \citet{Sedgwick+2019} who recently determined the low mass-end
of the GSMF by identifying low surface brightness galaxies based on data of core-collapse supernovae. The authors
used the IAC Stripe 82 legacy project \citep{Fliri_Trujillo2016} and the SDSS-II Supernovae Survey \citep{Frieman+2008}.

At the massive end we notice, however, some apparent tension between our and the GAMA results. 
Effects due to cosmic variance (due to the small redshift and angular coverage of the 
GAMA sample) could explain those differences as well as systematics in the mass-to-light ratios. 
Indeed, we see that some of the data are within 
the systematic errors. In addition, note that Figure \ref{fig:gsmf_final} from Appendix \ref{GSMF-correction} shows that using 
the mass-to-light ratios from \citet{Taylor+2011}, utilised in the \citet{Baldry+2012} \gsmf, 
tend to underestimate the high-mass end of the \gsmf.

\begin{table*}
	\centering
	\caption{Best fitting parameters for the GSMF (Eqs. \ref{eq:Fit_MF_comp_z0}-\ref{model_gsmf})}
	\resizebox{17.0cm}{!}{
		\begin{tabular}{c c c c c c c c c c}
			\hline
			$\log\phi^*_S$ $\left[{\rm Mpc}^{-3} {\rm dex}^{-1}\right]$ & $\alpha_S$ & $\beta$ & $\mathcal{M}_D=\mathcal{M}_S$ $\left[\msun\right]$ & $\log\phi^*_D$ $\left[{\rm Mpc}^{-3} {\rm dex}^{-1}\right]$ & $\alpha_D$ & $\delta$ & $\gamma$   \\
			\hline
			$-3.019  \pm 0.067$ & $-1.418  \pm 0.025 $ & $0.660 \pm 0.011$ & $10.897  \pm 0.036$ & $-2.267  \pm 0.120$ & $-0.207  \pm 0.169$ & $3.660  \pm 0.347$ & $1.236  \pm 0.080$   \\
			\hline
		\end{tabular}
	}
	\label{T1}
\end{table*}

\subsection{Best Fitting Model to the GSMF}
\label{best-fitGSMF}

To provide an analytic form to our \gsmf\  we choose to use a function composed of a Schechter function with a
sub-exponential decreasing slope and a double power-law function. Note that the resulting high-mass end of
our \gsmf\ is shallower than an exponential function, and, thus, better fitted
to a power-law \citep[see also][]{Tempel+2014}. 
The Schechter sub-exponential function is given by:
\begin{equation}
\phi_{*, {\rm S}}(M_{*})= 
\phi^{*}_S\ln 10 \left(\frac{M_{*}}{\mathcal {M}_S}\right)^{1+\alpha_S} \exp\left[{-\left(\frac{M_{*}}{\mathcal {M}_S}\right)^{\beta}}\right],
\label{eq:Fit_MF_comp_z0}
\end{equation}
where $\phi^{*}_S$ is the normalisation parameter in units of Mpc$^{-3}$ dex$^{-1}$, $\alpha$ is the slope at the low-mass end, $\mathcal {M}_S$ is the characteristic mass, 
and $\beta$ is the parameter that controls the slope at the massive end; note that $\beta=1$
corresponds to a Schechter function. The double power-law function is given by: 
\begin{equation}
\phi_{*, {\rm D}}(M_{*}) = 
\phi^{*}_D\ln 10\left(\frac{M_{*}}{\mathcal {M}_D}\right)^{1+\alpha_D}\left[1 + \left(\frac{M_{*}}{\mathcal {M}_D}\right)^{\gamma}\right]^{\frac{\delta-\alpha_D}{\gamma}},
\label{eq-DPL}
\end{equation}
where $\phi^{*}_D$ is the normalization parameter in units of Mpc$^{-3}$ dex$^{-1}$, 
$\alpha$ and $\delta$ control the slope at low and high masses, respectively, while $\gamma$ determines 
the speed of the transition between the low and high mass regimes; and $\mathcal {M}_D$ is the characteristic mass of the transition. Finally, the 
analytic form for fitting the observed GSMF is given by
\begin{equation}
\phi_{*, {\rm model}}(M_{*}) = \phi_{*, {\rm S}}(M_{*}) + \phi_{*, {\rm D}}(M_{*}),
\label{model_gsmf}
\end{equation}
where we assumed that $\mathcal {M}_S  =\mathcal {M}_D $.

We find the best fit parameters $\vec{p}_{\rm GSMF} = (\phi_{S}^{*}, \alpha_{S}, \mathcal{M}_{S}, \beta,\phi_{D}^{*}, \alpha_{D}, \delta,\gamma)$,  that maximize the likelihood function $\mathcal{L}\propto \exp({-\chi^2/2})$ by using the Markov chain 
Monte Carlo (MCMC) method algorithm described in \citet{Rodriguez-Puebla+2013}. Here
\begin{equation}
\chi^2 = \sum_{i=1}^{N_{\rm obs}}\left(\frac{\phi^{i}_{*,{\rm SDSS}} - \phi_{*, {\rm model}}^i }{\sigma^{i}_{\rm SDSS}}\right)^2,
\end{equation}
with $N_{\rm obs} $ as the number of observational data points of the \gsmf\ each with an $i$th value of $\phi^{i}_{*,{\rm SDSS}}$ and 
an error of $\sigma^{i}_{\rm SDSS}$. The $i$th value of our model is given by $\phi_{*, {\rm model}}^i$.

We sample the best-fit parameters by running a set of ten chains with $1\times 10^{5}$ MCMC models each. Table \ref{T1}
lists the best fit parameters. For our best fitting model we find that $\chi^2=85.42$ from a number of $N_{\rm obs} = 50$ observational data points. Our model
consist of $N_{\rm p} = 8$ free parameters, thus the reduced $\chi^2$ is $\chi^2 / {\rm d.o.f.} = 2.03$.
The upper and middle panels of Figure \ref{GSMF} show our best fitting model as the black solid line and
the bottom panel shows the residuals as a function of \ms. Our best fitting model has an error of $\sim2\%$
at the range $\ms\sim2\times10^{9}-5\times10^{11}\msun$ and an error lower than 
$\sim10\%$ at the mass range $\ms\sim7\times10^{8}-1\times10^{12}\msun$. For lower masses
errors can be up to $\sim20\%$. 

A valid question is how much we improve the analytic prescription when using a 
Schechter sub-exponential plus a double power-law function model confronted to a double Schechter function model,
commonly employed by previous authors \citep[see e.g.,][]{Baldry+2012,Wright+2017}. 
We have explored this possibility but assuming a Schechter function, $\beta=1$ in Equation (\ref{eq:Fit_MF_comp_z0}), and 
a Schechter sub-exponential function, that is, we are adding a extra degree of freedom due to the shallow decay at the high mass-end. 
Based on this alternative, we repeat our fitting procedure but this time finding that 
$\chi^2=662.817$ from a number of $N_{\rm obs} = 50$ observational data points. Now, our model
consist of $N_{\rm p} = 6$ free parameters resulting in a reduced $\chi^2$ of $\chi^2 / {\rm d.o.f.} = 15.06$. This is
considerably worse when combining Schechter sub-exponential and double power-law functions. Thus, hereafter we will consider only 
the latter model.

\subsection{The GSMFs of Early- and Late-Type Galaxies}
\label{fraction_ETG}

\begin{figure*}
	\vspace{-200pt}
	\hspace{-30pt}
	\includegraphics[height=10in,width=8in]{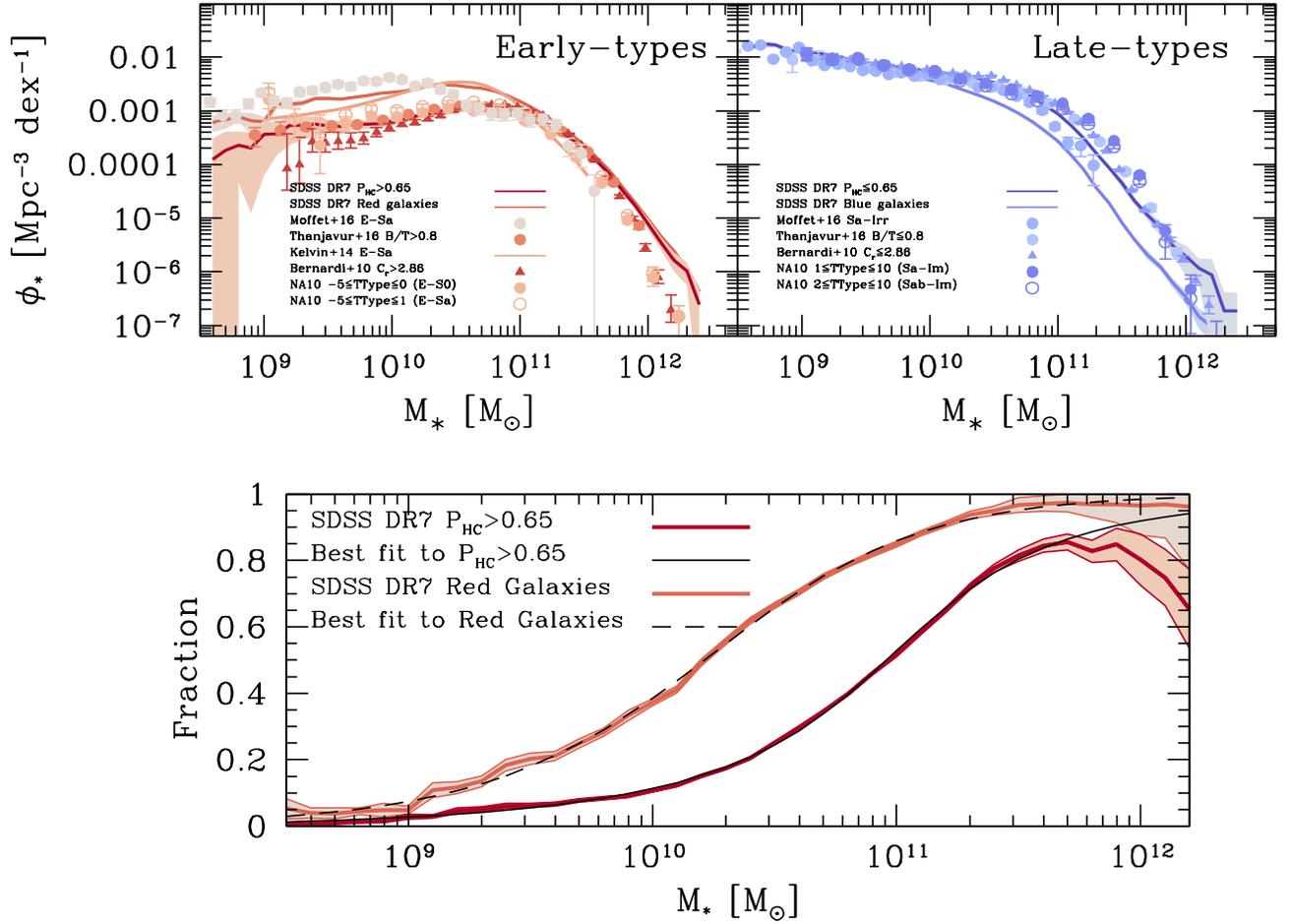}
	\vspace{-160pt}
	\caption{SDSS DR7 GSMFs for early- and late-type galaxies, left and right upper panels, respectively. 
		Early- (late-)type galaxies are defined as those with $P(E)>0.65$ ($P(E)\le0.65$) from the tabulated 
		probabilities of \citet{Huertas-Company+2011}. This is equivalent to morphological types that comprises 
		E and S0 galaxies or $T\le0$ (Sa to Irr galaxies or $T>0$). We compare
		to various previous determinations from the literature as indicated by the legends, see also the text for details. 
		Our determinations are in general in good agreement with previous determinations from SDSS
		spectroscopic samples, while a tension is evident with determinations from the GAMA survey. 
		We also present our resulting GSMFs for blue and red galaxies. These GSMFs follow closely those by morphology from the GAMA 
		survey. The bottom panel shows our number density-weighted fractions of early-type and
		red galaxies as a function of \ms. Their corresponding best fit models (Eq. \ref{eq:Frac_e}) are shown with solid and dashed lines,
		respectively.}
	\label{fig:fetgs}        
\end{figure*}

\begin{table*}
	\centering
	\caption{Best fit parameters to the fraction of early-type and red galaxies}
	\begin{tabular}{c c c c c c c c}
		\hline
		\hline
		Sample & $A$ & $\gamma_1$ & $ \log\mathcal{M}_{C,1} $ $[\msun]$ & $x_{0,1}$ & $\gamma_2$ & $ \log\mathcal{M}_{C,2} $ $[\msun]$  & $x_{0,2}$ \\
		\hline
		$P_{\rm HC}>0.65$ &  0.46 &  3.75 &  11.09 & 0.09  & 1.51  & 10.38 & 0.462 \\
		Red galaxies &  0.21 &  2.44 &  10.66 & 0.36  & 1.81  & 9.68 & 0.070 \\
		\hline
	\end{tabular}
	\label{T2}
\end{table*}

\begin{table*}
	\centering
	\caption{Best fit parameters of the \HI\ and \H2\ mass CPDFs for late- and early-type galaxies}
	\resizebox{\textwidth}{!}{\begin{tabular}{c c c c c c c c c c c}
			\hline
			\multicolumn{11}{c}{Late-Type Galaxies (Eqs. \ref{alpha_lj}--\ref{CDF-LTG})} \\  
			\hline
			\hline
			& & Component & $\alpha_{0;l,j}$ & $\alpha_{1;l,j}$ & $\mathcal{R}_{0;l,j}^*$ & $\mathcal{M}_{l,j}^*$ & $\beta_{l,j}$ & $\gamma_{l,j}$  & &  \\
			& &  \HI & -0.127$\pm$0.036 & 1.279$\pm$0.345 &  2.598$\pm$0.745 & 8.646$\pm$0.399 &  -0.018$\pm$0.108 & 0.577$\pm$0.053  & &  \\
			& &  \H2 & -0.085$\pm$0.120  & 0.830$\pm$1.213 & 0.122$\pm$0.037 & 10.595$\pm$0.301  & 0.841$\pm$0.195 & 0.063$\pm$0.089  & &  \\
			\hline
			\multicolumn{11}{c}{Early-Type Galaxies (Eqs. \ref{Pej}--\ref{CDF-ETG})} \\  
			\hline
			\hline
			Component & $\alpha_{0;e,j}$ & $\alpha_{1;e,j}$ & $\mathcal{R}_{0;e,j}^*$ & $\mathcal{M}_{e,j}^*$ & $\beta_{e,j}$ & $\gamma_{e,j}$ & $p_{0,j}$ & $p_{1,j}$ & $r_{0,j}$ & $r_{1,j}$ \\
			\HI & -0.052$\pm$0.067 & -0.074$\pm$0.6840& 1.573$\pm$0.533 & 8.354$\pm$0.258 & -0.820$\pm$0.272 & 0.468$\pm$0.077 & 0.060$\pm$0.032 & -0.113$\pm$0.338 & -0.259$\pm$0.015 & -0.310$\pm$0.160 \\
			\H2 & 0.059$\pm$0.069 & -1.491$\pm$0.725 & 0.674$\pm$0.229 & 8.182$\pm$0.317 & -0.686$\pm$0.412 & 0.375$\pm$0.156 & 0.017$\pm$0.074 & 0.515$\pm$0.785 & -1.084$\pm$0.074 & 7.980$\pm$0.724 \\
			\hline
		\end{tabular}}
		\label{T0}
	\end{table*}

Our main goal for this paper is to construct bivariate distributions as well as mass function based on the observed gas mass CPDFs and the $\gsmf$s of early- and late-type galaxies.  
In this section, we determine the \gsmf\ of early- and late-type galaxies from the SDSS DR7 spectroscopic sample with the public automated morphological galaxy 
classification by \citet{Huertas-Company+2011}.\footnote{\url{http://gepicom04.obspm.fr/sdss_morphology/ Morphology_2010.html}}
The morphological classification in \citet{Huertas-Company+2011} was determined based on support vector machine algorithms. Here we use their 
tabulated probabilities for each SDSS galaxy as being classified as an early type, $P(E)$. For masses below the completeness of the
SDSS DR7 sample, we use an extrapolation of the observed fraction of early type galaxies. We will come back to this point later in this section.  

From a catalog of galaxies with visual morphological classification 
\citep[UNAM-KIAS;][]{Hernandez-Toledo+2010} we find that galaxies with types $T\le0$ are mostly those with 
$P(E)>0.65$, and those with $P(E)\le0.65$ correspond mostly to $T>0$; here $T$ is the \citet{Fukugita+2007} 
notation.\footnote{\citet{Huertas-Company+2011} define as
	elliptical galaxies those objects with $T\le0$, S0s as $T=1$, Sabs as $2< T < 4$ and, Scd as $4\le T < 7$ 
	based on the \citet{Fukugita+2007} morphology classification.
	\citet{Huertas-Company+2011} included elliptical galaxies and S0s as early-type galaxies which corresponds to  galaxies with types $T\le1$
	in the   \citet{Fukugita+2007}  notation, and $T\le0$ when using the \citet{Nair_Abraham2010} notation, see below. In the de Vacouleours notation
	this is equivalently to $T=0$.} Based on the above, we consider as
early-type galaxies those with a probability $P(E)>0.65$ while late-type galaxies those with $P(E)\le0.65$. 
We checked that our morphology definition 
between early- and late-type galaxies is consistent with the morphological classification based on the concentration parameter $c=R_{90}/R_{50}$. That is,
the division between early- and late-types is approximately at $c=2.85$ \citep[see below and also,][]{Hyde_Bernardi2009,Bernardi+2010}. 

We calculate the SDSS DR7 \gsmf\ of early- and late-type galaxies using the $1/\vmax$ method 
described in Appendix \ref{SDSS_DR7_GSMF}. Figure \ref{fig:fetgs} shows the corresponding $\gsmf$s of
early and late types in the upper left and right panels, respectively. For comparison we show the $\gsmf$s for red and blue galaxies
based on a color cut limit in the $(g-r)^{0.0}-\ms$ diagram. In this diagram, we find that a rough division criteria from blue to red galaxies
is given by the color limit of $(g-r)^{0.0}=0.66$.\footnote{
	While this is just a rough division line, we used it as a practical method for decomposing the \gsmf\ into two main groups. Notice that 
	in Appendix \ref{GSMF-correction} we apply a more
	sophisticated method to derive the distribution of blue and red galaxies. Additionally, we checked
	that both methods give similar results.} 
In the same figure, we compare our results to different determinations from the literature as we describe below. 
All the data in this figure have been renormalised to our cosmology.

Recently, \citet{Moffett+2016} visually classified morphologies in the GAMA survey, and reported the \gsmf\
for different morphologies. Here we reproduce their \gsmf\ from E to Sa galaxies as
early types, and the complement as late types. Contrary to our definition,
Sa galaxies are included in the early-type group; this is because the authors report
S0 and Sa galaxies as one morphology group. As shown in Figure \ref{fig:fetgs}, the \gsmf\ of early-type galaxies 
from \citet{Moffett+2016} results in an overabundance of low-mass
galaxies compared to other studies. We reproduce the results from \citet{Thanjavur+2016} with bulge-to-total ratios 
of $B/T>0.8$ as early types, and $B/T\leq0.8$ as late types. \citet{Thanjavur+2016} used the bulge-to-disc decomposition from
\citet{Simard+2011} SDSS DR7 spectroscopic sample, and stellar masses derived from \citet{Mendel+2014}. We also include
results from \citet{Kelvin+2014}. Similarly to 
\citet{Moffett+2016}, \citet{Kelvin+2014} visually classified morphologies in the GAMA survey. 
We again use their \gsmf\ from E to Sa galaxies for early types since the authors 
combined S0-Sa galaxies as in \citet{Moffett+2016}. The filled triangles with error bars show the \gsmf\ from
\citet{Bernardi+2010} for galaxies with concentration parameter $c>2.86$ for early types, and $c\leq2.86$ for late 
types.\footnote{Figure 5 in \citet{Bernardi+2010} shows that using $c=2.85$ separates galaxies into earlier
	and later morphologies. While this selection criteria is not perfect, their Figure 18 shows that using the above concentration is very similar
	to the E+S0 \gsmf\ based on the \citet{Fukugita+2007} sample.}
Finally, using the \citet{Nair_Abraham2010} morphology catalog, who visually classified 14,034 
spectroscopic galaxies from the SDSS DR4, we derive the \gsmf\ for early-type galaxies.\footnote{We construct 
	volume-limited samples that are complete in \ms\ and compute the \gsmf\ as described in Appendix \ref{GSMF-correction}. 
	In this case we slightly modified Eq. (\ref{Mste_lim}) by shifting our stellar mass limit by 0.4 dex, that is, 
	$\log M_{*,{\rm lim, NA10}}(z) =\log M_{*,{\rm lim}}(z) + 0.4$.}
We utilise their morphological notation and define early-type galaxies as those objects with $-5\le T\le0$ (E-S0s), 
equivalent to $T\le1$ in the  \citet{Fukugita+2007} notation. We additionally
derive the \gsmf\ with morphologies between $-5\le T\le1$ in the \citet{Nair_Abraham2010} notation, which include Sa galaxies. 

In general our results agree with previous determinations, especially with those from the SDSS spectroscopic samples. 
In contrast, the \gsmf\ of early-type galaxies from the visual classification of the GAMA survey are systematically above
our results at the low-mass end, $\ms\lesssim2\times10^{10}\msun$, but closer to our classification based on galaxy colour. 
While it is not clear the reason of the differences outlined above (the inclusion of Sa galaxies as early-types, environment, etc.), 
in Appendix \ref{impact_of_color} we will discuss the impact of using 
galaxy colour instead of morphology when deriving the \HI, \H2, cold gas, and baryonic MFs separated into two main galaxy populations.

Finally, the bottom panel of Figure \ref{fig:fetgs} shows the resulting fraction of early-type galaxies 
as a function of stellar mass,
$f_E(\ms)$. In addition, we show the fraction of red galaxies 
when using our $g-r$ colour cut limit, $f_r(\ms)$. We find that 
the fraction at which early-type galaxies is $50\%$ is 
at $\ms\sim10^{11}\msun$, while at $\ms\sim1.6\times10^{10}\msun$ and $\ms\sim8\times10^{11}\msun$ the fractions are 
16\% and 84\%, respectively. For red galaxies, the fraction of $50\%$ is at 
$\ms\sim 10^{10}\msun$, while at $\ms\sim3\times10^{9}\msun$ and $\ms\sim10^{11}\msun$ the fractions are 16\% and 84\%,  
respectively. Note that the characteristic mass at which the fraction of early-type
galaxies is $50\%$ is a factor of $\sim10$ larger than for red galaxies. 
In general, $f_E(\ms)$ rises slower than the fraction $f_r(\ms)$. 
In the same figure we present the best fit model to the data. After exploring different functions, we find that
two {\it sigmoid} functions accurately describe the functionality of $f_E(\ms)$ or $f_r(\ms)$:
\begin{equation}
f_{k}(\ms) = \frac{1-A}{1+e^{-\gamma_1(x_{C,1}+x_{0,1})}}
+ \frac{A}{1+e^{-\gamma_2(x_{C,2}-x_{0,2})}},
\label{eq:Frac_e}
\end{equation}
where $k=E$ or $r$, $x_{C,i} = \ms / \mathcal{M}_{C,i}$, with $i = 1,2$. 
The best-fit parameters for the two fractions are listed in Table \ref{T2}.

To derive the analytic model for the \gsmf\ of early- and late-type galaxies 
we use the best fit model to our \gsmf, Section \ref{best-fitGSMF}, and the best fit model for $f_E(\ms)$.
For masses below $5\times 10^8$ \msun\ we extrapolate $f_E(\ms)$.
This is an acceptable approximation since as seen in Fig. \ref{fig:fetgs}, the fraction $f_E(\ms)$ 
tends to $\sim0$ below $\ms=10^9$ \msun.
Recall that our main goal in this paper is to derive the MFs for \HI, \H2, cold gas, and baryons
by combining the observed \HI\ and \H2\ CPDFs with the GSMF, both for early- and late-type galaxies, 
over a large mass range. 
Thus, at this point we are in a position to determine these MFs. 

\section{Results}
\label{sec:results}

In this Section we present our fits to the \HI- and \H2-CPDFs for early- and late-type galaxies from Paper I, the corresponding 
correlations (first and second moments), the bivariate \HI- and \H2-stellar mass distributions, and the \HI\ and \H2\ MFs. 
We also present the total cold gas and baryonic MFs.
We will show that our empirically-inferred \HI\ and \H2\ MFs agree with direct determinations from {\it blind} or optically/infrared (selected) radio
galaxy surveys. Previous works related to our approach are, e.g., \citet[][]{Obreschkow+2009,Lemonias+2013}; and \citet{Butcher+2018}.

For those interested in using our results, we provide a \textsc{Python} code containing all the necessary information 
to reproduce the results presented here, for details see Section \ref{conclusions}.

\subsection{The \HI\ and \H2\ Conditional Distribution Functions}
\label{CDF-constraints}

\begin{figure} 
	\vspace{-70pt}
	\hspace{-15pt}
	\includegraphics[height=8.in,width=6.in]{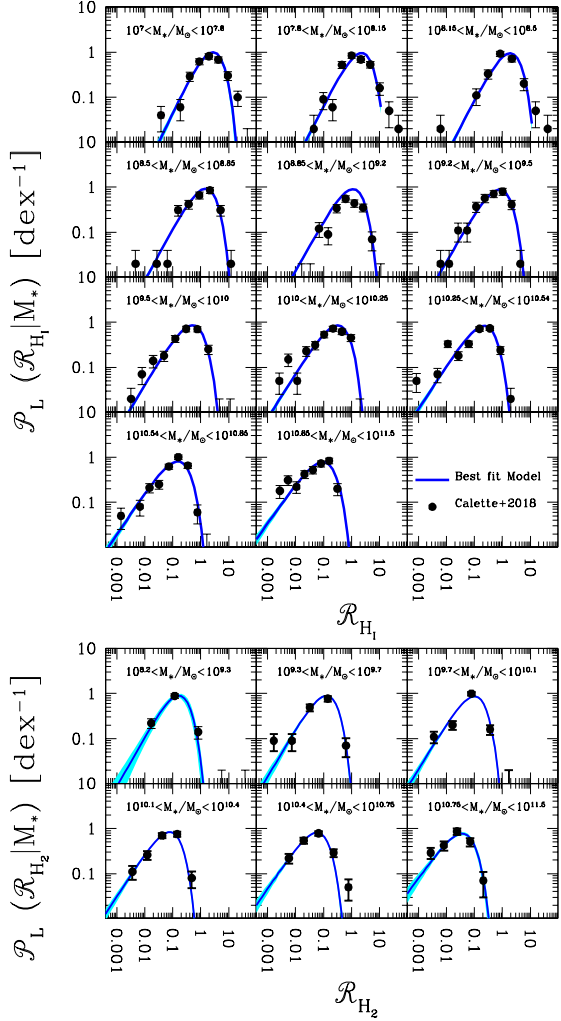}
	\vspace{-130pt}
	\caption{\HI\ and \H2\ mass CPDFs for late-type galaxies. The results for the compilation sample from \citetalias{Calette+2018} are 
		shown as filled circles with error bars. Note that the above results include non-detections  since the authors used the \citet{Kaplan_Meire1958} estimator 
		for uncensored data in their analysis. Our best fitting models are shown as the solid lines.} 
	\label{CMF_L} 
\end{figure}

\begin{figure} 
	\vspace{-170pt}
	\hspace{-15pt}
	\includegraphics[height=8.in,width=6.in]{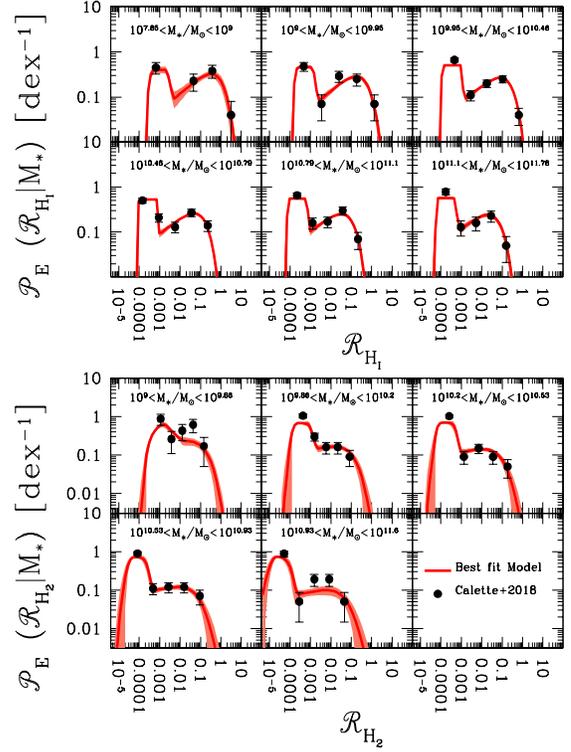}
	\vspace{-130pt}
	\caption{Same as Figure \ref{CMF_L} but for early-type galaxies. Note that the CPDFs 
	of early-type galaxies reported in \citetalias{Calette+2018} account for upper limits corrected 
	by distance selection effects when necessary and the treated with the \citet{Kaplan_Meire1958} estimator, 
	see Section \ref{secc:summary_C18}.} 
	\label{CMF_E} 
\end{figure}

\begin{figure} 
	\vspace{-80pt}
	\hspace{-15pt}
	\includegraphics[height=5in,width=3.7in]{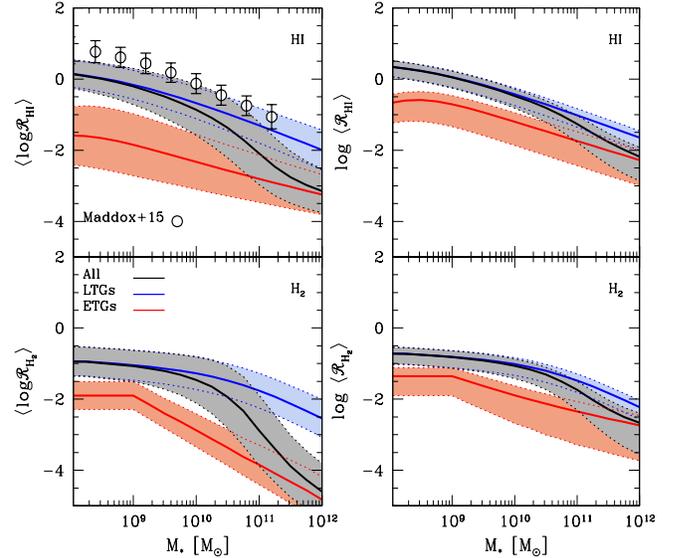}
		\vspace{-80pt}
	\caption{Logarithmic, left panels, and arithmetic, right panels, averaged mass ratios $\mathcal{R}_j$ as a function of \ms\ from our analysis, with $j=\HI, \H2$.  Blue and red lines are for early- and late-type galaxies, respectively, while the black lines correspond to all galaxies. The shaded areas show the respective standard deviations. Notice that $\log \langle \mathcal{R}_j (\ms)\rangle \geq \langle \log \mathcal{R}_j (\ms)\rangle$ and the dispersion reduces for the  arithmetic mean. The open circles with error bars in the upper left panel correspond to the data from
	ALFALFA galaxies with SDSS spectral and stellar mass counterparts according to \citet{Maddox+2015}.
	} 
	\label{fig:moments_from_pdfs} 
\end{figure}

Section \ref{Models_for_HI_H2} describes the functional forms for the \HI- and \H2-CPDFs of early- and late-type galaxies
proposed in \citetalias{Calette+2018}. 
Here, by using the determinations of the CPDFs for early- and late-type galaxies from \citetalias{Calette+2018}, 
we find the best fit parameters of the proposed functional forms: a Schechter-type function and
a Schechter-type + Uniform function, respectively (see Section \ref{Models_for_HI_H2}). 
While \citetalias{Calette+2018} reported their corresponding best fit parameters, 
here we opt for an update based on our own determinations of the GSMFs, for consistency. There are two reasons for doing this: {\it i)}
When fitting a CPDF that is determined over stellar mass bins, one should take into account contributions to this CPDF 
from the different masses. 
Weighting the conditional distributions by the \gsmf\ takes care
on that, see equations 6 and 8 of Section 5 from \citetalias{Calette+2018}; and {\it ii)} \citetalias{Calette+2018} used the fraction of 
bulge-dominated galaxies from  \citet{Moffett+2016a} as a proxy to the fraction of early-type galaxies.
As discussed in Section \ref{fraction_ETG}, the results from \citet{Moffett+2016}, and thus  \citet{Moffett+2016a}, 
overestimate the fraction of early-type galaxies compared to the
SDSS morphological catalogues. The above could be due to the inclusion of Sa galaxies into the group of early-types. 
We used the above to argue in favor of our derived fraction of early-type galaxies based on the automated 
morphological classification from \citet{Huertas-Company+2011}. 

Following \citetalias{Calette+2018}, we use the Bayesian approach described previously through a MCMC method 
applied jointly to all the data (the CPDFs in different \ms\ bins)
to find the best fit parameters of the proposed functions. These are listed in Table \ref{T0}. 
Figure \ref{CMF_L} shows our best fitting models for late-type galaxies compared 
to the CPDFs from \citetalias{Calette+2018}. 
Figure \ref{CMF_E} shows the same but for early-type galaxies.  We notice that our best fit parameters are very similar to those determined 
in \citetalias{Calette+2018}.

\subsection{The \HI- and \H2-to-stellar mass correlations} 

Next, we explore the resulting first and second moments from our best fitting models to the observed \HI- and \H2-CPDFs,
shown in Figure \ref{fig:moments_from_pdfs}. 
The left panels of the figure present the logarithmic mean $\langle \log \mathcal{R}_j\rangle$ and its corresponding standard 
deviation, $\sigma_{\log \mathcal{R}_j}$, $j=$ \HI\ or \H2,  as a function of \ms\ for early- and late-type galaxies as well as for all the galaxies. 
At low masses the correlation of all galaxies approaches the one of late-type galaxies while at 
high-mass end it approaches early types. The above trends are just the consequence of the observed fraction of early/late types.   
Figure \ref{fig:moments_from_pdfs} shows that early- and late-type galaxies follow different $\langle \log \mathcal{R}_j\rangle-\ms$ correlations. 
Therefore, due to the strong bimodality of these correlations conclusions based on some subset of galaxies as representative
of all galaxies would lead to incorrect results. 

In the literature,  sometimes  the gas-to-stellar mass relations are reported using the arithmetic mean (though the results are 
plotted in logarithmic diagrams). The right panel of Figure \ref{fig:moments_from_pdfs} shows $\log \langle \mathcal{R}_j\rangle$ 
vs. \ms\ from our empirical CPDFs. As is clearly seen, there are notable differences when computing different ways of averaging the distributions:
	 {\it i)} $\log \langle \mathcal{R}_j(\ms)\rangle > \langle \log\mathcal{R}_j(\ms)\rangle$, being larger the difference for the
	 early-type galaxies;\footnote{In the case of the arithmetic mean, the contribution of low values, even if they dominate in number, 
	 could be in some cases significantly lower than higher values. Then, for the arithmetic mean the contribution of low $\mathcal{R}_j$ values is
	minimised contrary to the  logarithmic mean of  $\mathcal{R}_j$.} 
	and {\it ii)} the standard deviations from the arithmetic mean is smaller than from the logarithmic mean.

\begin{figure*} 
	\vspace{-54pt}
	\hspace{40pt}
	\includegraphics[height=5.5in,width=5.9in]{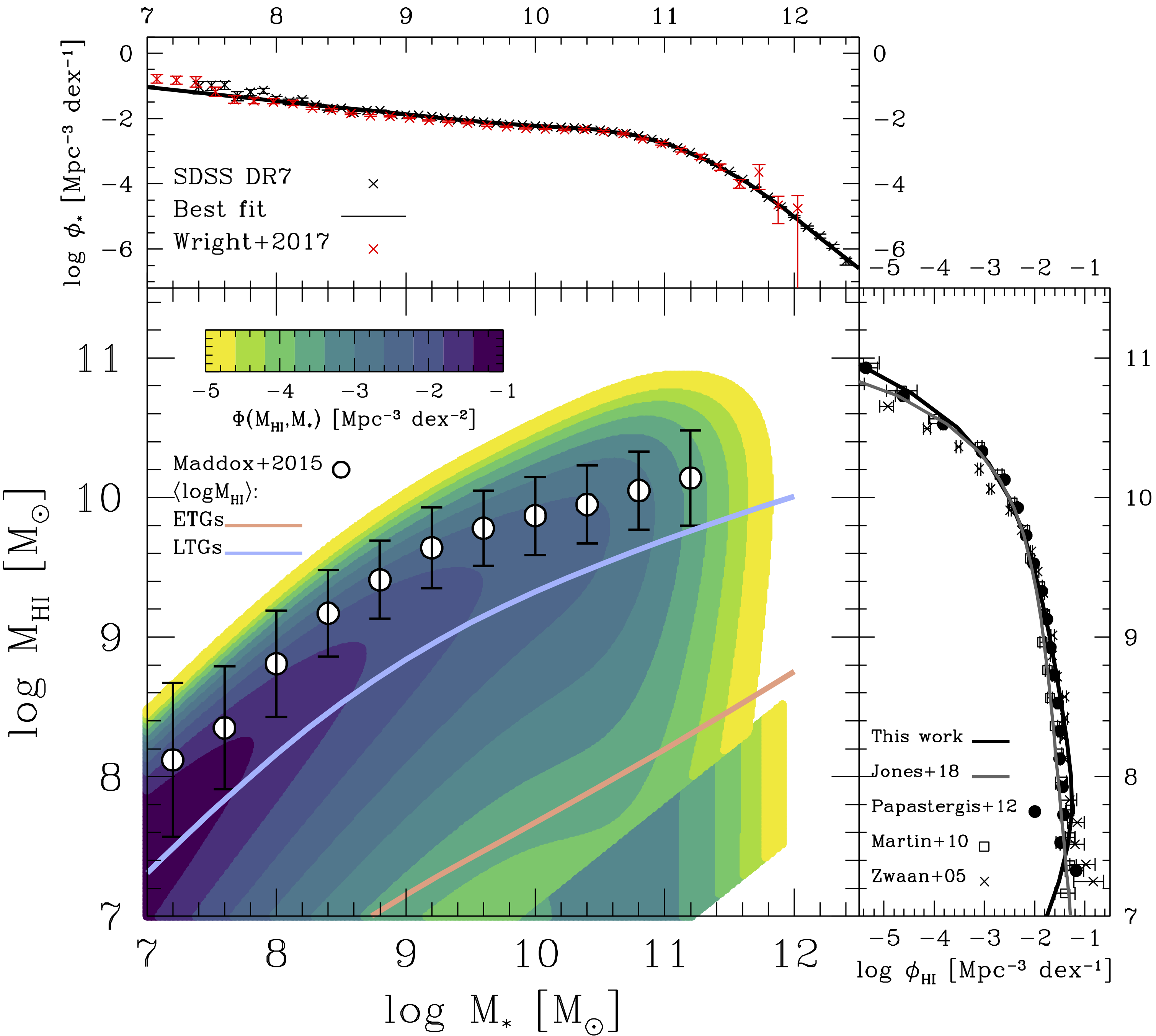}
	\caption{Atomic gas--stellar mass bivariate (joint) distribution function. The color code shows various number density levels as 
		indicated by the legends. Due to the rising slope of the MFs at low masses most of the galaxies are located at small \HI\ and stellar
		masses.Note that the discontinuity seen at the low-\HI\ and high-stellar masses is due to the assumption of an uniform function 
		for the lowest values of gas-to-stellar mass ratios of early-type galaxies where non-detections piled up.  
		Recall that in our analysis non detections (upper limits) are included by using the non-parametric 
		estimator \citet{Kaplan_Meire1958} for censored data in \citetalias{Calette+2018}. The solid lines show the mean 
		$\langle\log\mha\rangle$ as a function of \ms, both for early- and late-type galaxies. 
		The upper panel shows the GSMF which is the result of integrating the bivariate
		distribution function along the \mha\ axis, while the bottom right panel shows the same but for the \HI\ 
		MF which results from integrating along the \ms\ axis. We  compare to some previous observational determinations of the MFs
		and the relationship between \mha\ and \ms\ derived in \citet[][]{Maddox+2015} for the ALFALFA survey with SDSS spectral and stellar mass counterparts.
		} 
	\label{IsoHI} 
\end{figure*}

In the left upper panel of Fig.  \ref{fig:moments_from_pdfs} we reproduce the results from \citet[][]{Maddox+2015} for the ALFALFA 
galaxies with SDSS spectral and stellar mass counterparts. It is clear that the ALFALFA survey is biased towards galaxies 
with high \HI\--to--\ms\ ratios.  
In other words, the ALFALFA survey mainly detects galaxies in the upper envelope of the full distribution of  $\mathcal{R}_{\HI}$ 
(see also \citealp{Maddox+2015}) and is dominated mostly by late types \citep[see also e.g.,][]{Haynes+2011}.

\begin{figure*} 
	\vspace{-54pt}
	\hspace{40pt}
	\includegraphics[height=5.5in,width=5.9in]{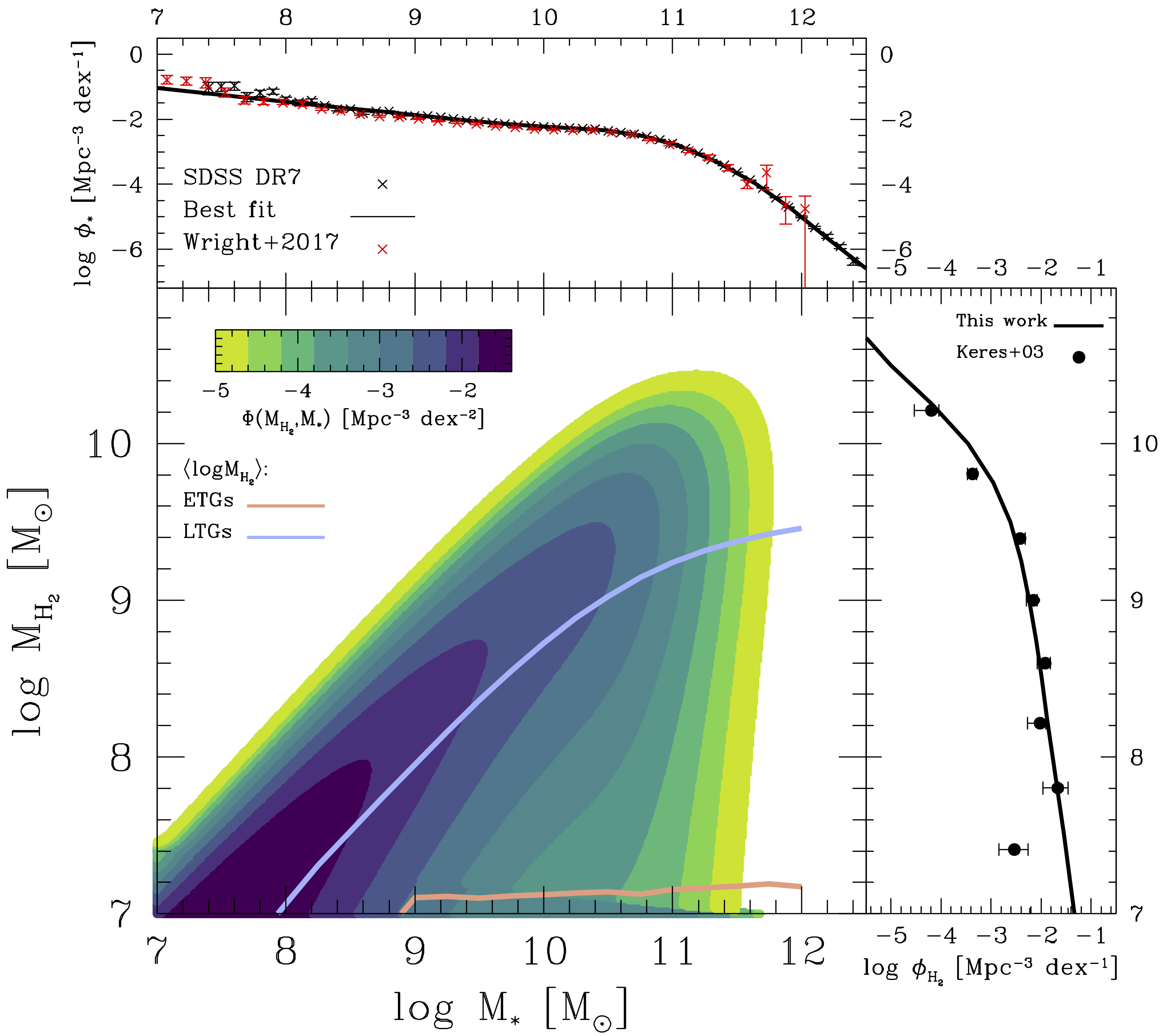}
	\caption{Same as Figure \ref{IsoHI} but for molecular gas. Note that while there are more 
		non-detections for \H2\ observations these are mostly from early-type galaxies that represent only a
		small fraction overall in the \H2\ mass bivariate distribution function. We also compared to previous determinations 
		from \citet{Keres+2003}. 
		} 
	\label{IsoH2} 
\end{figure*}

\subsection{The Bivariate Mass Distribution Functions}
\label{sec:bivariate-distribution}

Figure \ref{IsoHI} shows the resulting bivariate stellar-\HI\ mass distribution function, $\Phi(M_{\rm HI},\ms)$, see Equation (\ref{2D_density}). 
The color code shows various number density levels for $\Phi(M_{\rm HI},\ms)$. Notice that $\Phi(M_{\rm HI},\ms)$
is for all galaxies, that is, it includes the contribution from early- and late-type galaxies. The
discontinuity in the number density isocontours at the bottom right of the diagrams is related to contributions from the non-detections from early types.
Recall that for the CPDFs of early types we assumed an uniform function (or top-hat)
for the lowest values of the gas-to-stellar mass ratios $\mathcal{R}_{\rm HI}$, 
where the non-detection piled up, \footnote{Note that the top-hat is not the result 
of applying the \citet{Kaplan_Meire1958} estimator as we a posteriori redistributed
the lowest values of $\mathcal{R}_{\rm HI}$ (including upper limits) into a uniform function.}
see Section \ref{secc:functional_forms_HI_H2} and Figure \ref{CMF_E}. 
In the bottom right and upper panels of the same figure we present respectively our measurements of the \HI\ MF and GSMF 
with the solid black lines. We compare the \HI\ MF with 
 {\it blind} \HI\ galaxy surveys based on 
ALFALFA \citep{Jones+2018,Papastergis+2012,Martin+2010} and on \HI\ Parkes All Skye Survey HIPASS \citep{Zwaan+2005}. 
While in the next subsection we discuss in more detail the comparison with previous works, for the moment we note that our 
total \HI\ MF is in good agreement with the above direct observations. In the case of 
ALFALFA this is a revealing result given the strong selection bias of this survey towards \HI-rich and late-type galaxies as seen Figure \ref{IsoHI} \
(open circles reproduce the results from \citealp{Maddox+2015}, 
see also the discussion of the previous subsection and Figure \ref{fig:moments_from_pdfs}).
As we will discuss in the next section, the above reflects that the total \HI\ MF is dominated by late-type galaxies. 

Figure \ref{IsoHI} explicitly shows the contribution of galaxies of different stellar masses to the \HI\ MF. Particularly we observe that
the low mass-end of the \HI\ MF is composed mainly by low \ms\ galaxies but there is also a non-negligible contribution from a population
of high \ms\ galaxies. Most of these high \ms\ galaxies are early-type (quenched) galaxies 
for which there is a significant fraction of non-detections ($\sim 55\%$). 
In \citetalias{Calette+2018} we included non-detections for the determination  
of the HI-CPDF based on methods of censored data \citep{Kaplan_Meire1958}. 
Nonetheless, the contribution of non-detections is only marginal
because the fraction of early-type galaxies at those masses is low, see the bottom panel of Figure \ref{fig:fetgs}.  In addition, Figure \ref{IsoHI} 
shows that the completeness limit in the \HI\ MF,
due to our stellar mass limit of $\ms=10^{7}\msun$, is at $M_{\rm HI}\sim 10^{8}\msun$ (see below), 
which excludes a large region of galaxies with non-detections. 

Similarly to Figure \ref{IsoHI}, Figure \ref{IsoH2}  presents the resulting bivariate stellar--\H2\ mass distribution function for all galaxies 
and the mean $\langle\log M_{\rm H2}\rangle$ for early- and late-type galaxies.
The resulting total \H2\ MF is shown with the solid line in the bottom  right panel and 
compared to the \citet{Keres+2003} \H2\ MF based on the CO 
luminosity function. 
At the low-mass end there is a substantial population of galaxies with non-detections, 
roughly $\sim 78\%$, which are mostly early-type galaxies. As above, non-detections
have been included in the statistical analysis of the H2-CPDFs, for details see \citetalias{Calette+2018}. Nevertheless, from the 
contour density level their contribution is marginal. Finally, we can conclude that our \H2\ MF is complete for
$M_{\rm H2}\grtsim10^{7}\msun$.

\begin{figure*} 
	\vspace{-110pt}
	\hspace{20pt}
	\includegraphics[height=7.5in,width=6.5in]{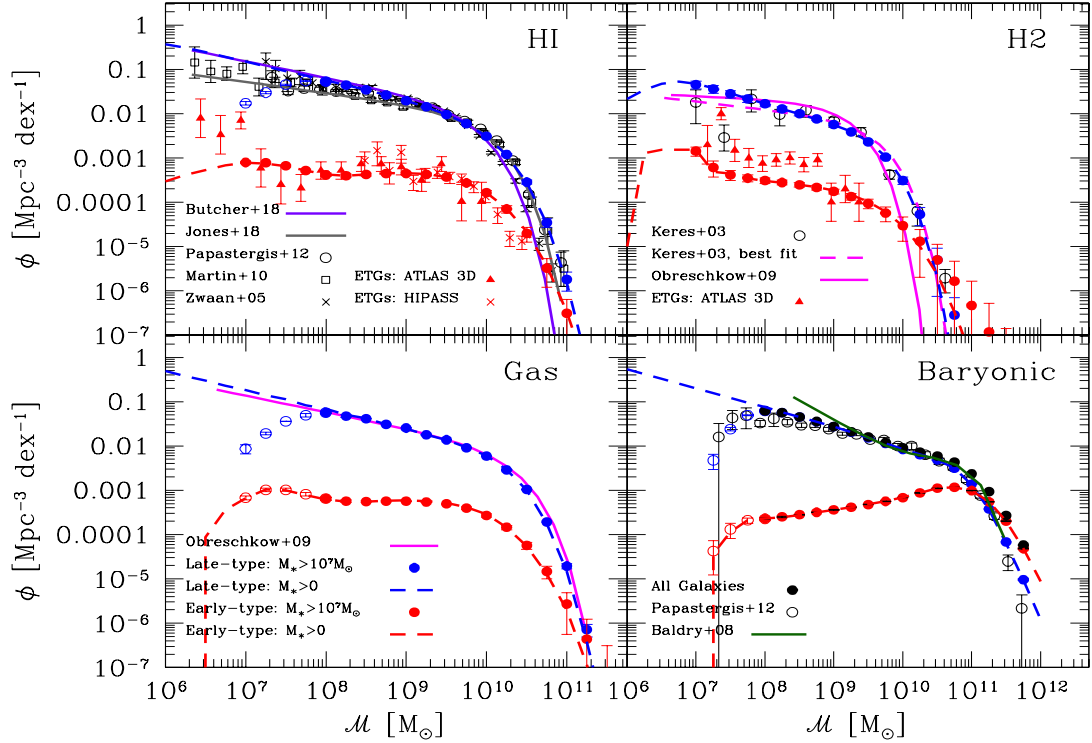}
	\vspace{-120pt}
	\caption{Results on the galaxy MFs of early- and late-type galaxies for atomic gas, left upper panel, molecular gas, right upper panel, 
	cold gas, bottom left, and baryons, bottom right panel. In all the panels,
		late-type galaxies are shown as the blue circles with error bars while early-type galaxies are
		shown as the red circles with error bars, when using a stellar mass limit of $\ms=10^{7}\msun$. 
		Filled blue/red circles indicate when the MFs are complete, 
		while open circles clearly show that the MFs became incomplete. The dashed lines are for MFs
		when using a stellar mass limit of $\ms=0$. The total MFs for \HI\ and cold gas are not shown because they are practically indistinguishable
		from the respective MFs of late-type galaxies.  Our results are in good agreement with observational determinations of the total MFs. 
		For only early-type galaxies, we compare our results with those from the ATLAS 3D sample (red triangles). While we observe some tension
		we suspect that selection effects are more likely to artificially increase the amplitude of their MF at low masses.} 
	\label{mfs_comp_to_obs} 
\end{figure*}

\subsection{The Mass Functions}
\label{MFs_results}

Next, we discuss in detail the MFs presented above. In particular, we focus on the determinations separately for early- and late-type galaxies
based on the morphology classification described in Section \ref{fraction_ETG}.  
The various panels in Figure \ref{mfs_comp_to_obs} present the MFs for atomic and molecular gas, total cold gas, and baryons, 
as indicated by the labels. In all the panels,
the MFs for late-type galaxies are shown as blue filled circles with error bars, while for early-type galaxies are
shown as red circles with error bars. These symbols are for a stellar mass limit of $\ms>10^{7}$ \msun. 
Instead, we use blue/red open circles when the MFs is incomplete. We also calculate the MFs in the 
hypothetical case of a mass limit of $\ms=0$ shown with the blue/red dashed lines. 
Note that the total MFs for \HI\ and cold gas are not plotted. This is because in these cases the MFs of late-type galaxies
are hardly distinguishable from the total one at all masses. 
We also omit the total \H2\ MF. This is because it is hardly distinguishable from the MF of late-type galaxies 
at $M_{\rm H2}\lesssim 2\times 10^{10}$ \msun\ while for larger masses it is indistinguishable from the MF of the early-type galaxies.
In the case of the baryon masses, the total MF is plotted with black  filled circles. 

We note that our determinations for the MFs are the result of the convolution between random errors and 
the intrinsic MFs, similarly as it happens with the direct observational determinations of MFs. 
In Section \ref{random_and_systematic_errors} we discuss the impact from random errors and present the
intrinsic MFs, after deconvolving by these errors.

\subsubsection{The \HI\ Mass Function}

We compare our results with previous direct observational determinations of the {\it total} \HI\ MFs. 
According to our results, late-type galaxies dominate the \HI\ MF, even at the highest masses, so that it is practically
equivalent to the total \HI\ MF. In Figure \ref{mfs_comp_to_obs}
we reproduce the best fit to the \HI\ MF from \citet{Jones+2018} based on the final catalogue of the blind \HI\  
ALFALFA Survey, dark grey solid line. The violet solid line shows the best fit model reported in \citet{Butcher+2018} to the \HI\ MF
from the Nan\c{c}ay Interstellar Baryons Legacy Extragalactic Survey (NIBLES), which is a project that complemented recent and/or 
ongoing large blind \HI\ surveys. Results from \citet{Papastergis+2012} and \citet{Martin+2010} based on the 40 per cent sample of the ALFALFA Survey
are shown respectively with black open circles and squares. The skeletal symbols with error bars show the results from \citet{Zwaan+2005}, who used
the \HI\ Parkes All Sky Survey HIPASS. Note that our \HI\ MF for late-type galaxies, which dominates the total MF, is in good agreement 
with direct inferences from \HI\ blind surveys, particularly those based on ALFALFA as discussed in Figure \ref{IsoHI}. 
As for the \HI\ MF of early-type galaxies, we plot the determinations from 
the ATLAS 3D \citep{Serra+2012} and HIPASS \citep{Lagos+2014} samples shown as the red solid triangles and skeletal symbols, respectively. 

Our resulting \HI\ MFs are in good agreement with direct determinations from radio observations. 
This is particularly true for \HI\ masses above the completeness limit corresponding to our \ms\ limit for the GSMF. These masses are 
$\mha\sim 10^{8}\msun$ for late-type galaxies, and  $\mha\sim 10^{7}\msun$ for early-type galaxies. Even when extrapolating to a limit mass
of $\ms=0$,  at the low-mass end we find a good agreement with direct determinations, though the early-type galaxies from the ATLAS 3D sample 
present a higher amplitude for masses below $10^{7}\msun$.
However, those extrapolations should be taken with caution as it is not clear whether
we can extrapolate our bivariate distribution functions to such low masses.  
In conclusion, we consider that the remarkable consistency between {\it our \HI\ MFs  and radio {\it blind}  surveys
above $\mha\sim 10^{8}\msun$ is reassuring} and 
validates the HI-CPDFs determined in \citetalias{Calette+2018}. 
Recall that the observational data used in that paper were derived from various heterogeneous samples, 
affected by many selection effects, including those related to the non radio detections. 
Therefore, the agreement between the \HI\ MF with that of the blind radio observations is far from trivial, 
unless adequate corrections are introduced and the data are adequately analysed from the statistical point of view. 

\subsubsection{The \H2\ Mass Function}

In the upper right panel of Figure  \ref{mfs_comp_to_obs} we present the results for the \H2\ MF. 
The \H2\ MF is largely dominated by late-type galaxies below $\mhm\sim 2\times10^{10}\msun$, but for larger masses, early-type galaxies are more abundant.
In the same panel we reproduce the total \H2\ MF from \citet{Keres+2003}, who used a CO luminosity function from FIR and $B-$band 
limited galaxy samples and adopting a constant CO-to-\H2\ conversion factor of $\alpha_{\rm CO} = 4.76$, 
open black circles with error bars. The dashed
line shows the best fit to a Schechter function derived in \citet{Obreschkow+2009}. Additionally, 
we show the results from the ATLAS 3D sample for early-type galaxies \citep{Lagos+2014} with the filled triangles. 
The magenta solid line shows the results from \citet{Obreschkow+2009},
who derived the \H2\ MF by introducing a phenomenological model for the \H2-to-\HI\ mass ratio that depends on the galaxy morphological
type and its total cold gas mass.  

When comparing to the \H2\ MF from \citet{Keres+2003} we observe a good agreement with our results. 
At the low mass end, though, the \citet{Keres+2003} MF seems to be slightly shallower than ours. It is not clear the origin
for this discrepancy. One possibility is
due the constant $\alpha_{\rm CO}$ factor employed by the authors. Based on previous empirical studies,  
\citetalias{Calette+2018} showed that ignoring the dependence of $\alpha_{\rm CO}$ with \ms\footnote{In Appendix
C of \citetalias{Calette+2018}, we have constrained the CO-to-\H2\ conversion factor to be mass dependent:
 $\log(\alpha_{\rm CO})=0.15 +0.35 [1 +0.1(3\times 10^{10}/\ms)^{0.64}]$ down to $\ms=10^8$ \msun\ and for lower masses 
 the value of $\alpha_{\rm CO}$ remains constant. Therefore, $\alpha_{\rm CO}$ increases as \ms\ decreases saturating to a 
 value of $\approx 250$ for $\ms<10^8$ \msun. This is due to the empirical dependences of  $\alpha_{\rm CO}$ on the gas-phase metallicity,  
and the dependence of the latter with \ms} flattens the  low-mass end of the \H2\ MF,
consistent with the results from \citet{Keres+2003}. Another possibility is an effect of the incompleteness of the 
CO luminosity function. 
As for \citet{Obreschkow+2009}, our results are consistent for masses below $\mhm\sim 3\times10^{9}\msun$. 
For larger masses, the \citet{Obreschkow+2009} MF falls much stepper than ours. 
Similar to our analysis, \citet{Obreschkow+2009} used the relationships between  
galaxy properties to derive their MF.  As mentioned above, their phenomenological 
model employed the dependence of the \H2-to-\HI\ mass ratio with
morphology and cold gas mass. While the above differences could be arguably referred to the use of different estimators for the
\H2\ gas masses, it could be also an indication that random errors are larger when using only one galaxy parameter. 
Recall that in this paper we are using \ms. In Section \ref{random_and_systematic_errors}, we will show that after 
deconvolving from random errors, our intrinsic \H2\ MF becomes steeper at the high-mass end, and more consistent 
with the derivation from \citet{Obreschkow+2009}. Nonetheless, it is difficult to conclude the origin of the above differences given the  
different nature of the models employed in both studies. 

As for early-type galaxies, our results are consistent with those from the ATLAS 3D \citep{Lagos+2014} at the high mass end
but they lie slightly below at the low mass end.  It is unclear the reason of the above discrepancy for low-$M_{\rm H_2}$ early-type galaxies, though
large-scale and environmental selection effects could boost the inferences of the MF when using the 1/V$_{\rm max}$ 
estimator, see for example, Appendix \ref{GSMF-correction} and \citet{Baldry+2008}. Recall that in the case of \HI, the ATLAS 3D 
also presents an slightly enhancement at the low-mass end of the \HI\ MF. Thus, selection effects are more likely to 
artificially increase the amplitude of the MF for low-mass galaxies in the ATLAS 3D sample. 

\subsubsection{The Gas and Baryonic Mass Functions} 

Similarly to the \HI\ MF, the cold gas MF is completely dominated by late-type galaxies, even at the high-mass end. 
In Figure \ref{mfs_comp_to_obs} we compare our results with the phenomenological determination from \citet[][pink solid line]{Obreschkow+2009}. 
These authors combined their inference of the \H2\ MF  with the \HI\ MF from \citet{Zwaan+2005} to derive the gas MF.
Despite the differences mentioned above for the \citet{Obreschkow+2009} \H2\ MF, their total cold gas MF is in excellent agreement
with our one. This is not surprising as it is just reflecting that \HI\ is much more abundant than \H2. 

Finally, we show our results for the baryonic MFs in the bottom right panel of Figure \ref{mfs_comp_to_obs}. The baryonic MF is very
similar to the GSMF at the high-mass end but the at low-mass end is steeper as the contribution of cold gas becomes more relevant. On the other
hand, late-type galaxies dominate the baryonic MF for $\mbar<10^{11}$ \msun, while at the high-mass end, early-type galaxies 
are more abundant than the late-type ones.  We reproduce with 
the green solid line the baryonic MF from \citet{Baldry+2008}. These authors combined the \gsmf\ from the \texttt{low-z} survey
of the SDSS DR4, the same galaxy survey used here for low masses, with a closed-box model and the mass-metallicity relation 
to derive cold gas masses for their baryonic MF.  The open black circles show the MF from \citet{Papastergis+2012}, who defined baryonic mass as
$\mbar=1.4 \times \mha+\ms$. 
We notice that these previous baryonic MF determinations are in good agreement with our results at the mass range $\sim 2\times 10^9-2\times 10^{11}$ \msun, 
while for lower an higher masses there are some differences, which are easy to understand. 

The MF from \citet{Baldry+2008} is steeper than our MF at low masses. This might be a consequence of the fact that the \citet{Baldry+2008} \gsmf\  itself 
is steeper compared to other determinations, e.g., \citet{Baldry+2012}. As these authors discuss, the disagreement between the \citet{Baldry+2008}
and  \citet{Baldry+2012} $\gsmf$s is just the result of different flow models for distances, which affect significantly to nearby low-mass galaxies. 
Recall that our GSMF has been corrected to be consistent with the flow model of \citet{Tonry+2000}. 
Additionally, the gas masses in \citet{Baldry+2008} were actually obtained from a close-box model constrained with the empirical mass-metallicity relation. 
The combination of these two assumptions are likely the result of a steep baryonic MF at low masses, which differs from our results and those of \citet{Papastergis+2012}. 

Regarding the high mass end, our baryonic MF falls shallower than those of  \citet{Papastergis+2012} and \citet{Baldry+2008}. This
is because our GSMF is shallower. As discussed in Section \ref{local-GSMF+correlations} there are two main
systematic effects that could lead to different GSMFs, mass-to-light ratios and the determination of galaxy surface brightness (especially
due to sky subtraction problems). Both effects are likely to affect the high-mass end of the baryonic MF. In addition, 
due to the small volumes of the surveys used in \citet{Baldry+2008} and \citet{Papastergis+2012}, cosmic variance 
enhances the differences. 

\subsection{Cosmic density parameters and relevant timescales}

\subsubsection{Cosmic density parameters}

\begin{table*}
	\centering
	\hspace{-15pt}
	\caption{Cosmic density of \HI, \H2, gas, stars and baryons for all, LTGs and ETGs.. 
	The fraction of each component is denoted as $f_j = \Omega_j / \Omega_{\rm bar,U}$ with  $\Omega_{\rm bar,U} = 0.048$.}
	\begin{tabular}{ccccccccccc}
		\hline
		\hline
		& $\Omega_{\rm H_2}/10^{-4}$ & $f_{\rm H_2}$ & $\Omega_{\rm HI}/10^{-4}$ & $f_{\rm HI}$  & $\Omega_{\rm gas}/10^{-4}$ & $f_{\rm gas}$  & $\Omega_{\ast}/10^{-4}$ & $f_{\ast}$  & $\Omega_{\rm bar}/10^{-4}$ & $f_{\rm bar}$  \\ \hline
		All & $0.86 \pm 0.05$ & $0.18\%$ & $4.24 \pm 0.10$ & $0.88\%$ &  $6.85 \pm 0.92$ &  $1.43\%$ & $20.40 \pm 0.08$ & $4.25\%$ & $26.01 \pm 0.13$ & $5.42\%$ \\
		LTG & $0.82 \pm 0.04$ & $0.17\%$ & $4.09 \pm 0.10$ & $0.85\%$ & $6.59 \pm 0.89$ & $1.37\%$ & $13.20 \pm 0.05$ & $2.75\%$ & $18.25 \pm 0.12$ & $3.80\%$ \\
		ETG & $0.04 \pm 0.01$ & $\sim0.01\%$ & $0.15 \pm 0.02$ & $0.03\%$ & $0.21 \pm 0.03$ & $0.04\%$ & $7.21 \pm 0.03$ & $1.50\%$  & $7.76 \pm 0.37$ & $1.62\%$ \\ \hline
	\end{tabular}
	\label{tab:densities}
\end{table*}

The cosmic density parameter measures the average mass density of some matter species 
with respect to the critical density,  
$\rho_c$.  
Here, we determine the mass density in stars, \HI, \H2, cold gas, and baryons
that are locked inside galaxies 
by using their MFs. 
The differential mass density function $d\rho_j(M_j)$ for some galaxy mass component 
$M_{j}$ in the mass range $\log M_j\pm d\log M_j /2$
is: $d\rho_j(M_j ) = M_{j} \times \phi_j (M_j ) d\log M_j$, where 
$\phi_{j}$ is in units of Mpc$^{-3}$ dex$^{-1}$. Thus the cosmic mass density is given by:
\begin{equation}
\rho_j = \int_{-\infty}^{\infty} d\rho_j(M_j ) 
\label{mass_density}
\end{equation}
with the cosmic density parameter
\begin{equation}
\Omega_j =  \frac{\rho_j}{\rho_c},
\end{equation}
where the critical density is 
$\rho_c = 2.775\times 10^{11} h^{-1}\msun/(h^{-1}{\rm Mpc})^3 = 1.2756 \times 10^{11} \msun/ {\rm Mpc}^3 h_{67.8}^{2}$.\footnote{We
use this symbol to emphasised that $H_0=67.8$ km s$^{-1}$ Mpc$^{-1}$ in our cosmology.}
The limits of integration in Equation (\ref{mass_density}) reflect that we are considering all the
spectrum of masses from galaxies. In reality, this is not possible, due to completeness limits in galaxy samples. Here, we 
consider the following mass limits for all our components: $M_{\rm inf} = 10^7$ \msun\  and  $M_{\rm upper} =  10^{12.6}$ \msun.
We notice that using values smaller than  $M_{\rm inf}$ and/or larger than $M_{\rm upper}$ do not substantially
change our results. This is because the multiplicity functions, $\propto M_{j} \times \phi_j (M_j )$, have a maximum
around the knee of the MFs. 

\begin{figure} 
	\vspace{-200pt}
	\hspace{-15pt}
	\includegraphics[height=6.4in,width=5.2in]{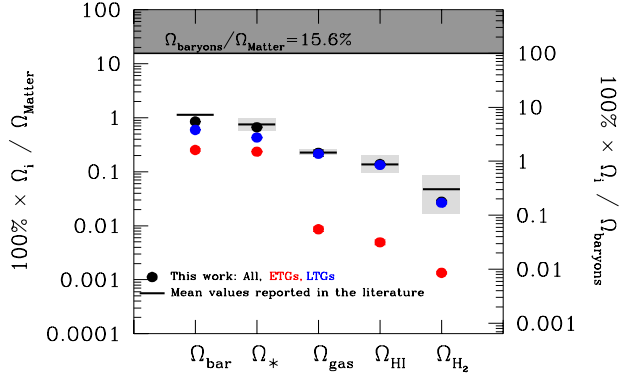}
	\vspace{-120pt}
	\caption{Density parameter $\Omega$ of \HI, \H2, cold gas, and baryonic mass locked in all galaxies as well as in early- and  late-type 
	galaxies (coloured filled circles; the errors are smaller than the circle size).  The $\Omega$ parameter values are reported as fractions
	in per cents of the universal matter (left axis) and baryonic (right axis) densities. The gray boxes show the range of values from previous determinations  
	and the horizontal lines correspond to the mean of these values.} 
	\label{fig:Omegas}        
\end{figure}

Figure \ref{fig:Omegas} shows the different values of each $\Omega_{j}$ corresponding to all galaxies and separately for
early- and late-type galaxies, listed in Table \ref{tab:densities}. 
The $\Omega_j$ values are presented as the fractions in per cents with respect to the universal matter density ($\Omega_{\rm m}=0.307$,
left axis) and the universal baryonic density ($\Omega_{\rm bar,U}=0.048$, right axis).
To estimate errors in our cosmic density parameters, we use all our MCMC models for the HI-CPDF and H2-CPDF in addition of all our MCMC 
models to the fit of the GSMFs. 
We found that the largest uncertainties arise from the uncertainties in the CPDFs. 

In the past, there have been some assessments of the cosmic density parameters at $z\sim0$.
Usually, these studies do not report cosmic density parameters for different populations and 
for different components at the same time. As mentioned in the Introduction, it is important to distinguish between different populations 
given that late- and early-type galaxies follow different formation histories. Studies close to ours 
are the ones by \citet{Fukugita+1998,Fukugita+2004}, and \citet{ReadTrentham2005}.
Below we 
present and compare our results with many previous determinations from the literature. 

$\bullet$ \textit{\HI\ cosmic density:}  The atomic hydrogen in late-type galaxies is 
$\sim 27$ times larger than in early-type galaxies, which means 
that $\sim96\%$ of \HI\ mass is in late-type galaxies. 
Previously, \citet{Zwaan+2005,ReadTrentham2005,Martin+2010,Braun+2012,Delhaize+2013,Hoppmann+2015,Butcher+2018}, and \citet{Jones+2018}
have derived the \HI\ cosmic density parameter by using either blind \HI\ galaxy surveys (HIPASS and ALFALFA) or indirect techniques. 
The mean value from these determinations 
is $\Omega_{\rm HI} = 4.2\times10^{-4}$ with a lower bound of 
$\Omega_{\rm HI}^- = 3\times10^{-4}$ and an upper bound of $\Omega_{\rm HI}^+ = 6.2\times10^{-4}$, shown
as the grey box in Figure \ref{fig:Omegas}\footnote{All the 
	values for the papers listed above have been renormalised to a units of $h^{-1}_{67.8}$.}. 
	Recently, using a spectral stacking technique and from WSRT observations of 1895 galaxies crossed with the SDSS,
	\citet{Hu+2019} found $\Omega_{\rm HI} = (4.15\pm 0.26)\times10^{-4}$. 
	Our determined value is in good agreement with these previous determinations, in particular with the latter one.

$\bullet$ \textit{\H2\ cosmic density:} The molecular hydrogen cosmic density in late-type galaxies is 
$\sim21$ times larger than in early-type galaxies. This implies that $\sim95\%$ of \H2\ mass is in 
late-type galaxies. Using the CO surveys from \cite{Young+1995}, \citet{Keres+2003} determined that 
$\Omega_{\rm H2}=(1.64\pm0.63)\times10^{-4}$, while from the observations in CO from \citet{Maeda+2017}
they report $\Omega_{\rm H2}=0.51\times10^{-4}$.
\cite{Obreschkow+2009} used a phenomenological model to derive 
$\Omega_{\rm H2}= (1.01\pm0.39)\times10^{-4}$. 
\citet{ReadTrentham2005}  find that $\Omega_{\rm H2}=2.68\times10^{-4}$. The above ranges of values are shown with grey box in
Figure \ref{fig:Omegas}. As can be seen, our results are consistent with the range of determinations described above,
especially with the results from \cite{Obreschkow+2009}.

$\bullet$ \textit{Cold gas cosmic density:} Most of the cold gas
is located in late-type galaxies, $\sim96\%$. \cite{Keres+2003} found that $\Omega_{\rm gas}=(6.34\pm1.62)\times 10^{-4}$,
which includes the resulting abundance of \HI\ from \cite{Zwaan+1997}.
\cite{Obreschkow+2009} used their best phenomenological model to the \H2-to-\HI\ ratio with the HIPASS 
results from the  \cite{Zwaan+2005} sample to 
derive $\Omega_{\rm gas}=(6.49\pm 1.18)\times 10^{-4}$, while using the values for
\HI\ and \H2\ masses from \citet{ReadTrentham2005}, we calculate that 
$\Omega_{\rm gas}=7.95\times 10^{-4}$ after correcting from helium and heavier metals. 
Our value of $\Omega_{\rm gas}=(6.85\pm 0.92)\times 10^{-4}$ is consistent with the above results. 

$\bullet$ \textit{Stellar cosmic density:} The stellar cosmic density in late-type galaxies we derive from the SDSS is  
approximately $\sim1.8$ larger than in early-type galaxies. Thus, $\sim64\%$ of the mass in stars at $z\sim0.1$ is in late-type galaxies. 
From the compilation by \citet{Madau_Dickinson2014}, the stellar cosmic density lies within 
$\Omega_{\ast}=(28.06-17.71)\times 10^{-4}$, while the derivations from \citet{Wright+2017} and \citet{Baldry+2008} are
respectively  $\Omega_{\ast}=17.14\times 10^{-4}$ and $\Omega_{\ast}=29.73\times 10^{-4}$. Our result for the cosmic 
density for all galaxies, $\Omega_{\ast}=(20.20\pm0.08)\times 10^{-4}$, is consistent with the above values. 

$\bullet$ \textit{Baryonic cosmic density:} Finally, we find that there is a factor of $\sim2.4$ more baryons in late-type
galaxies than in early-types, and thus,  $\sim71\%$ of the baryons are in late-type galaxies. \cite{ReadTrentham2005} 
found that $\Omega_{\rm bar} = 35\times10^{-4}$ which is a factor of $\sim1.3$ larger than our results. 
We find a cosmic density parameters ratio of $\Omega_{\ast} / \Omega_{\rm bar} \approx 1.3$.
Finally, our baryon density parameter is $\approx 5.4\%$ of the universal baryon fraction, $f_{\rm bar,U}=0.156$, or 
equivalently $\sim 18$ times lower than $f_{\rm bar,U}$. Most of the baryons are definitively not locked inside galaxies.

\subsubsection{Cosmic timescales}

We are now in a position to derive some relevant cosmic timescales, such as the mean galaxy depletion times. We focus  
on late-type galaxies because most 
of the star formation occurs in those galaxies. To do so, we use the observed cosmic star formation
rate (CSFR) at $z\sim0.1$. According to \citet{Madau_Dickinson2014}, who derived the CSFR from far-UV and IR rest-frame luminosities,
the CSFR is $\dot{\rho}_\ast \sim 90\times10^{-4}$ \msun\ yr$^{-1}$ Mpc$^{-3}$ after correcting to a \citet{Chabrier2003} IMF. Unfortunately,
the authors report the CSFR for all galaxies but not divided into different groups. 
The recent study by \citet{Sanchez+2019}, based on the fossil record 
analysis of a sample of more than $4\times10^{4}$ galaxies from the SDSS MaNGA survey, report similar values to the
the total CSFR of $\dot{\rho}_\ast = 114.82\pm67.61\times10^{-4}$ \msun\ yr$^{-1}$ Mpc$^{-3}$ or 
$\dot{\Omega}_\ast = \dot{\rho}_\ast/\rho_{\rm crit} = (9\pm 5)\times10^{-14}$ yr$^{-1}$ corrected to a \citet{Chabrier2003} IMF. 
The authors also derived the CSFRs for star-forming galaxies; 
$\dot{\Omega}_{\ast, \rm SFG} = (6.5\pm 3.8)\times10^{-14}$ yr$^{-1}$. 
In the following we use their value
for star-forming galaxies as a representative determination for late-type morphologies,  
that is, $\dot{\Omega}_{\ast, \rm L}\approx \dot{\Omega}_{\ast, \rm SFG}$.

We begin our discussion by estimating the mean molecular hydrogen depletion time of late-type galaxies, 
$\bar{t}_{\rm dep,L}(\H2)=\Omega_{\rm H_2, L}/\dot{\Omega}_{\rm  \ast,L}$. 
The \H2\ depletion timescale is defined as the time at which a galaxy (or a molecular cloud) would consume its \H2\ 
gas reservoir by forming stars at the current SFR. From our cosmic density 
parameters we find that  $\bar{t}_{\rm dep,L}(\H2)\approx 1.3$ Gyrs. This is consistent 
with the mean depletion time $\bar{t}_{\rm dep}(\H2)=0.96$ Gyr reported in  \citet{Saintonge+2017} 
for star-forming galaxies in a volume complete sample. Note, however, that for local individual galaxies 
the molecular depletion time could vary from $\sim 0.9$ to 3 Gyrs 
\citep[e.g.,][]{Kennicutt1998,Bigiel+2008,Leroy+2008,Leroy+2013}. Also, we estimate  
the mean total cold gas depletion time of late-type galaxies, 
$\bar{t}_{\rm dep,L}({\rm gas})=\Omega_{\rm gas,L} / \dot{\Omega}_{\rm \ast,L}$,
and find $\bar{t}_{\rm dep,L}({\rm gas})\approx 10.14$ Gyrs, that is, $\sim8$ times larger than for
the molecular gas component. The values we find for these two timescales are consistent with the proposal that, on 
average, for local late-type galaxies,  {\it i)} the global conversion of molecular gas into stars is inefficient
(recall that the \H2\ depletion times of observed local star-forming regions are actually 40-500 Myr, 
e.g., \citealp{Lada+2010,Lada+2012}); and  {\it ii)} the global conversion of atomic to molecular 
hydrogen gas is also inefficient, or equivalently, the molecular cloud formation efficiency is low. 
Thus, the mean star formation efficiency, SFE, of local late-type/star-forming galaxies is low despite the fact that 
they contain a considerable amount of interstellar gas; according to Table \ref{tab:densities}, on average approximately
 $36\%$ of the baryons in these galaxies are in form of cold gas.

According to \citet{Leroy+2008},  the SFE of a galaxy is the inverse of the neutral H gas depletion time, that is, the time required
for current star formation to consume the neutral H reservoir.  The SFE can be estimated as the product of the net efficiency of converting 
\H2\ gas into stars, ${\rm SFE}(\H2) =1 / {t}_{\rm dep}(\H2)$, 
and the net efficiency of molecular cloud formation given by the mass fraction of \H2\ with respect to the total neutral H mass in the galaxy, i.e., 
\mhm/(\mha + \mhm).  Thus, using our estimations of the cosmic parameters for late-type galaxies, we calculate the 
cosmic (mean) SFE of late-type galaxies as:
\begin{eqnarray} 
{\rm SFE}_L({\rm H}) =  {\rm SFE}_L(\H2) \times  \frac{\Omega_{\rm H_2,L}}{\Omega_{\rm HI,L}+\Omega_{\rm H_2,L}}= \\ \nonumber
 \frac{1}{\bar{t}_{\rm dep,L}(\H2)} \times \frac{\Omega_{\rm H_2,L}}{\Omega_{\rm gas,L}/1.4} = 1.4 \times \frac{\dot{\Omega}_{\ast,  \rm SF}}{\Omega_{\rm gas,L}} = \\ 
 \frac{1.4}{\bar{t}_{\rm dep,L}({\rm gas})} = 1.38\times 10^{-10} {\rm yr^{-1}}. \nonumber
\label{sfe}
\end{eqnarray}
The inverse of this efficiency is the cosmic neutral H depletion time, $\bar{t}_{\rm dep,L}({\rm H})\approx7.25$ Gyrs. Note that the 
relationship between the SFE based on the neutral H gas reservoir and the SFE based on the total cold gas reservoir is ${\rm SFE}_L({\rm H}) = 1.4\times {\rm SFE}_L({\rm gas})$,
or equivalently, $\bar{t}_{\rm dep,L}({\rm H})= \bar{t}_{\rm dep,L}({\rm gas})/1.4$. The factor 1.4 takes into account He and metals. 

We calculate also the cosmic SF timescale of late-type galaxies, which is given by the inverse of the cosmic specific SFR,
 $\bar{t}_{\rm SF,L}= \Omega_{\rm \ast,L}/\dot{\Omega}_{\rm \ast,L}\approx 20.3$ Gyrs; this is a factor of  $\sim1.5$ larger than 
the present age of the Universe. The cosmic SF timescale can be understood as the time required for the current cosmic SFR density to double 
the current cosmic stellar mass content.  Interestingly enough, the ratio $\bar{t}_{\rm dep,L}({\rm H})/\bar{t}_{\rm SF,L} = (\Omega_{\rm H2} +  \Omega_{\rm HI}) / \Omega_{\ast} = 0.36$, 
that is, the gas reservoir of late-type galaxies has not yet been dramatically consumed by star formation.
Including He and metals in the gas reservoir, this ratio increases to $\sim 0.5$.

\begin{figure*} 
	\vspace{-125pt}
	\hspace{20pt}
	\includegraphics[height=7.2in,width=6.in]{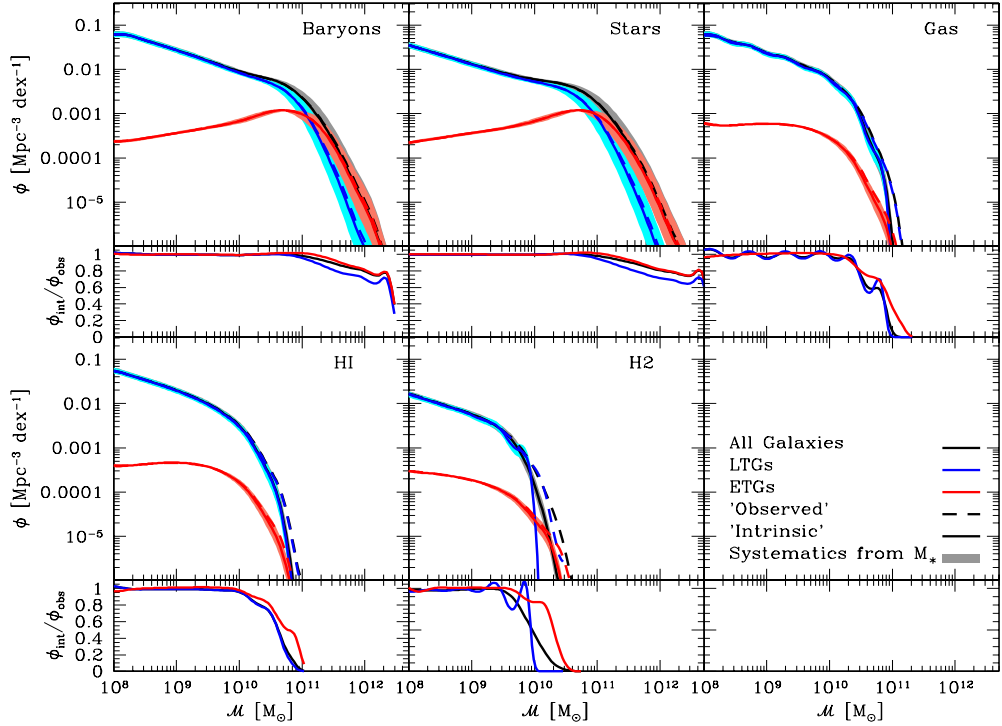}
	\vspace{-120pt}
	\caption{Impact of random and systematics errors in the baryonic, stellar, cold gas, atomic and molecular
		gas MFs for all and separately for early- and late-type galaxies. 
		The dashed lines show the ``observational'' MF from Section \ref{MFs_results} while the solid lines
		show the MF after deconvolving from random errors, i.e., the intrinsic MFs. Systematic errors are shown with the shaded areas. 
		While the impact of random errors affects notably the total cold, atomic, and molecular gas MFs,
		the impact of systematic uncertainty on \ms\ is apparently marginal on them. 
		However, the systematic uncertainties on \ms\ are noticeable in the stellar and baryonic MFs.}
	\label{fig:mfs_deconvolved} 
\end{figure*}

\section{Discussion}
\label{Discussion}

In this paper we employed a statistical approach that allows to project the observed \HI- and \H2-CPDFs into their
corresponding MFs, when using the GSMF as an interface or pivotal function, Section \ref{the_method}. 
Additionally, the cold gas and baryon MFs are obtained from the above.
 Our empirical approach makes use of the following observational data as input: 
\begin{enumerate}
	\item The local \gsmf\ over a large dynamical range and separated into early- and late-type galaxies.
	\item The observed CPDFs of \HI\ and
	\H2\ as a function of \ms, both for early- and late-type galaxies. 
\end{enumerate}
As a result, our approach provides a fully self-consistent and complete empirical description of the demographics of the local 
population of early- and late-type galaxies for a broad mass range.
Furthermore, by construction, our MFs are derived separately for early- and late-type galaxies.
As discussed in Section \ref{MFs_results}, our \HI\ and \H2\ MFs are actually
consistent with several previous determinations from radio {\it blind} or optically/infrared (selected) 
galaxy samples.
Actually, the above level of agreement is not trivial due to the chain 
of assumptions and corrections for the data sets we used here and in \citetalias{Calette+2018}, and it reinforces the robustness of 
the observational information employed. Note, however, that the above agreement is only valid
above our completeness 
limit for the GSMF of $\ms=10^{7}\msun$, which corresponds to a completeness limit of 
$\mha\sim 10^{8}\msun$ and $\mhm\sim 10^{7}\msun$ respectively for the  \HI\ and \H2\ MFs. In that regards, 
we are unable to constraint the very low mass end of the  \HI\ and \H2\ MFs.

Below, we highlight aspects that  we consider are relevant for the success of our empirical 
approach:

\begin{itemize}
	\item The \HI\ and \H2\ CPDFs for early- and late-type galaxies. We used the CPDFs   
	from \citetalias{Calette+2018}, where we derived the CPDFs from a compilation of many incomplete and 
	inhomogeneous samples, carefully homogenised to a common IMF, cosmology, 
	CO-to-luminosity conversion factor, and accounting for selection biases.
	
	\item The effect from upper limits in radio surveys. In addition to the above mentioned homogenisation and corrections, it was 
	important  to take into account
	the upper limits reported in the original sources, when radio detections were not achieved.  The fraction
	of non-detections in the compilation from \citetalias{Calette+2018} was non negligible, especially for 
	early-type galaxies. Non-detections were corrected by distance/sensitivity effects. Instead of ignoring radio non-detection or using the upper limits as the true values, as is commonly  done in the literature, 
	we derived the CPDFs by including them in our statistical analysis
	based on the non-parametric Kaplan-Meier estimator for censored data. 
\end{itemize}
		
Next, we briefly discuss below some potential caveats on our approach. Over  the next subsections we will discuss them
in more detail and show that they do not affect our main conclusions.
	
\begin{itemize}
	\item The assumption that the \HI\ and \H2\ masses are two independent random variables. 
	In reality this is not true; for example, \citet{Obreschkow+2009} showed that the \H2-to-\HI\ mass ratio 
	depends on the morphological type. Note, however, that this was partially taken into account in our approach 
	in a statistical sense. Recall that we use the observed mass CPDFs
	separately for early- and late-type galaxies, that is, the dependence with morphology
	is roughly included, as shown in \citetalias{Calette+2018}, see Figure 14 from that paper and Fig. \ref{fig:moments_from_pdfs}. 
	
	\item Differences on the mass-to-light ratios. Figure \ref{fig:GSMF_comparison} 
	in Appendix \ref{SDSS_DR7_GSMF} shows that the different mass-to-light ratios used to estimate \ms\ 
	lead to different $\gsmf$s, with differences up to 0.5--1 dex in number densities at the high-mass end
	\citep[see also][]{Bernardi+2017}. While we choose to use the geometric mean 
	over the five mass-to-light ratios described in Appendix \ref{SDSS_DR7_GSMF}, one could
	naturally question that the agreement of our \HI\ and \H2\ MFs with the observed ones is relative because using a different \gsmf\
	could result in different MFs. In Section \ref{random_and_systematic_errors} we explore and 
	quantify the impact of systematics from varying mass-to-light ratios, and show that its effect is marginal in the obtained  \HI\ and \H2\ MFs.

	\item Random errors from stellar mass estimates. Inevitably, random errors propagate 
	to our MFs resulting in a \citet{Eddington1940} bias effect. Thus, the comparison with the results based on 
	radio surveys is not trivial as they do not suffer of an  \citet{Eddington1940} bias effect due to \ms\ errors. Nonetheless, measurements
	of the \HI\ and \H2\ masses are also subject to random errors. In subsection \ref{random-errors}
	we deconvolve our MFs with the random errors, not only as a method to compare with results from radio surveys but 
	also for obtaining the intrinsic MFs to be used to constrain the predictions from galaxy formation models.   
	
	\item The morphological classification from the SDSS DR7. 
	To derive our \gsmf\ separated into early- and late-type galaxies we used the morphological classification
	based on the \citet{Huertas-Company+2011} vector machine analysis of the SDSS DR7. 
	As shown in Figure \ref{fig:fetgs}, we find that the obtained early/late-type \gsmf s using this classification are consistent
	with other determinations for the SDSS but disagree with those based on the visual 
	classification from the GAMA survey. While we explore in detail this effect in Appendix \ref{impact_of_color},
	we do not include it as one of the main source of uncertainty.  
\end{itemize}

We conclude this section by emphasising the robustness of the MFs derived when combining observational gas-to-stellar mass correlations 
from small data sets with the \gsmf, \citep[see also,][]{Lemonias+2013,Butcher+2018}. While this is an indirect method 
to study the demographics of the galaxy distribution, it is a valid and valuable approach that gives 
results that are comparable to direct observations and generalise them into a full bivariate distribution.

\subsection{The Impact of Random and Systematics Errors}
\label{random_and_systematic_errors}

When deriving stellar, \HI\ and \H2\ masses, there are two sources of errors  that will inevitably propagate over the
MFs: the random and systematic errors. In this Section, we discuss the impact
of both sources of errors on our results. 

\subsubsection{Random Errors}
\label{random-errors}

The estimation of masses from both photometric and radio observations, are subject to random 
errors. Here, we determine their impact on our resulting MFs. For simplicity, we assume that random errors follow
lognormal distributions with a constant dispersion and independent of galaxy morphology. For the stellar masses, we
assume a dispersion of $\sigma = 0.1$ dex following \citet{Behroozi+2010}, \citet{Mendel+2014}, and \citet{Rodriguez-Puebla+2017}.
For $\HI$ masses, $\sigma = 0.14$ dex,  and for \H2\ masses, $\sigma = 0.22$ dex, following \citet[][and more references therein]{Calette+2018}. 
As for the gas and baryonic masses, we assume errors of respectively  $\sigma = 0.14$ dex and $\sigma = 0.1$ dex as they are dominated
by $\HI$ and \ms\ components, respectively, especially at high masses, where random errors have a larger impact.  
Thus, our ``observational'' MFs\footnote{In the preceding sections we omit to use the term  ``observational'' MFs to avoid
confusion about our methodology. Here we use this term to refer that our determinations, similar to direct measurements of the MFs from 
galaxy surveys, suffer from random errors.} are the result of the convolution of the distribution of 
random errors and the respective intrinsic MFs. That is, our ``observational'' MFs are given by 
$\phi_{\rm obs} =  \mathcal{G}\ast \phi_{\rm int}$, where the symbol $\ast$ denotes the convolution operation,
$\mathcal{G}$ is the distribution of random errors, and $\phi_{\rm int}$ is the intrinsic MF.
For more details, the reader is referred to Appendix \ref{deconvolution}. There we 
describe our numerical algorithm for deconvolving the intrinsic MF, $\phi_{\rm int}$.

In Figure \ref{fig:mfs_deconvolved} we reproduce with dashed lines the  ``observational'' MFs derived in
Section \ref{MFs_results}. Their corresponding intrinsic MFs are shown with solid lines.
In the same figure, we present the ratios $\phi_{\rm obs}/\phi_{\rm int}$ to show the effect of the deconvolution. 
The effect of deconvolving from random errors is small at low-intermediate masses but it increases at the 
massive-end since the MFs are steeper 
\citep{Eddington1940}. This is simply because the convolution depends on the logarithmic slope of 
the intrinsic MF \citep[e.g.,][]{Cattaneo+2008}; the steeper the slope the larger the effect on
the MFs. This is also the reason why we observe a lower impact in the baryonic 
and stellar MFs compared to the \HI, \H2, and cold gas MFs; the latter fall steeply at the high-mass end. For example, 
the intrinsic \HI\ MF is a factor of $\sim4$ lower than the ``observational'' one at $\mha\sim6\times10^{10}$ \msun,
while the intrinsic  \H2\ MF is an order of magnitude lower than the  ``observational'' MF at $M_{\rm H2}\sim2\times10^{10}$ \msun. 
The  intrinsic gas MF is an order of magnitude lower than the ``observational'' MF at $\mg\sim10^{11}$ \msun.
Note that for the \HI, \H2, and cold gas MFs the impact of random errors is more noticeable in late-type galaxies than in early-type ones.

\subsubsection{Systematic Errors}
\label{systematic-errors}

In addition to random errors, systematic errors have an impact when determining the MFs. 
The IMF is one of the most important sources of systematic errors for the 
GSMF. In this paper we assumed an universal IMF given by the \citet{Chabrier2003} function. 
While there is much debate on the IMF \citep[see e.g.,][]{Bastian+2010,Conroy+2013,Bernardi+2018},
exploring the different alternatives is beyond the scope of this paper. 

The stellar masses are calculated typically using colour-dependent mass-to-light ratios based on results from
stellar population synthesis (SPS) models \citep[for a recent review see][]{Conroy2013}. 
Thus, the calculated stellar masses depend on the used SPS model. This introduces a systematic uncertainty in \ms.  
Indeed, systematics in \ms\ from SPS can be as large 
as $\sim0.25$ dex, see e.g, \citet{Perez-Gonzalez+2008,Muzzin+2009,Moustakas+2013,Rodriguez-Puebla+2017}
and references therein. Recently, \citet{Bernardi+2017} showed that systematics from SPS introduces errors that
are as large as $\sim0.5$ dex in the normalisation of the GSMF at the high mass-end. In Appendix \ref{SDSS_DR7_GSMF}
we found similar differences by using various recipes of colour dependent mass-to-light ratios. While in this paper
we calculate five different stellar masses for every galaxy, and decided to use the geometric mean of the five
as our fiducial definition of \ms, the above inevitable introduces the question of {\it which stellar mass definition shall we use
when deriving our MFs}.
Additionally, \citet{Bernardi+2017} determined that systematics in photometry are of $\sim0.1$ dex. In order to quantify the impact 
of stellar populations in our MFs, in Appendix \ref{SDSS_DR7_GSMF} we noted that 
a constant shift of $\sim\pm0.15$ dex in the stellar mass axis reproduces systematic errors in the GSMF.  In addition,
Figure \ref{GSMF} shows that the same shift in the stellar mass axis could also explain differences from photometry. 
Thus, hereafter, we will use a shift of $\pm0.15$ dex in the stellar mass axis as our fiducial model for systematic errors in the \gsmf. 
Note that we are assuming that this shift will be independent of morphology and we are ignoring systematic errors in the atomic and
molecular gas components. 

Figure \ref{fig:mfs_deconvolved} shows the impact of  systematic errors from SPS models and photometry
as the shaded areas around their corresponding $\phi_{\rm int}$ (solid lines). The effects of systematics is non-negligible 
at the massive-end of the stellar and baryonic MFs; we observe differences up to $\sim 0.6 $ dex
in their normalisations. This is approximately the same both for early- and late-type galaxies. The impact of
systematic errors in the gas, \HI\ and  \H2\ MFs is marginal; we notice a shift in their
normalisations of $\sim0.07$ at their low-mass ends but they increases
respectively to $\sim0.4$, $\sim0.4$ and $\sim0.3$ dex at their massive ends. 
The above is due to the steeper slopes observed at the high-mass end from these MFs.
In conclusion, the impact of systematic uncertainties in \ms\ is only marginal for the 
derived \HI, \H2, and cold gas MFs, making our results robust against this source of uncertainties.

As mentioned in the Introduction, the value of deriving robust MFs is that they can be used as key tools 
for constraining the processes that govern the evolution of the galaxies. However, using 
direct measurements from observations to constrain galaxy formation models is not trivial due to
random and systematic errors, as discussed here. We end this subsection by emphasising the importance of deconvolving from random 
errors and understanding the impact of systematic errors when reporting
results on galaxy demographics. 

\section{Summary and Conclusions}
\label{conclusions}

We present a self-consistent empirical approach that unifies local galaxy gas-to-stellar mass correlations
and the MFs of galaxies traced by their different baryonic components. 
We make available a \textsc{Python} code that displays tables and figures with all 
the relevant statistical distributions and correlations discussed in this paper.\footnote{{\small \url{https://github.com/arcalette/Python-code-to-generate-Rodriguez-Puebla-2020-results}}} 
Next, we summarise our main results which can be used for comparing with theoretical predictions or as input
for modeling galaxy mock catalogs:

\begin{itemize} 
	\item {\it Conditional probability distribution functions (CPDFs):} Section \ref{Models_for_HI_H2} presents the 
	functional forms for the \HI\ and \H2\ mass conditional distributions given \ms\ (the CPDFs), which are described 
	by Equations  (\ref{Sij})-(\ref{Poj}). 
	Our best-fit parameters to the empirical information presented in \citetalias{Calette+2018} are listed in Table  \ref{T0}, while
	Figures \ref{CMF_L} and \ref{CMF_E} show the data with their corresponding best fits in various stellar mass bins. 
	Theoretical predictions for the \HI, \H2\, and cold gas CPDFs 
	can be confronted with our empirically constrained distributions, for all galaxies as well as for early and late types
	in case the morphological classifications are available. If these predictions are limited in stellar and/or gas masses, 
	then our (analytical) \HI-CPDFs and \H2-CPDFs and their moments can be easily calculated over the same mass
	ranges as the theoretical predictions for a comparison. The \HI- and \H2-CPDFs combined with the GSMF allowed 
	us to calculate the respective bivariate mass distributions for all the galaxy population as plotted in Figures \ref{IsoHI} 
	and \ref{IsoH2}. 

	\item {\it Moments of the CPDFs:} 
	The (analytical) \HI- and \H2-CPDFs contain the information about any moment of the distributions. 
	Figure \ref{fig:moments_from_pdfs}  (see also Figs. \ref{IsoHI} and \ref{IsoH2}) shows the $\langle\log M_j\rangle$-$\log\ms$
	relationships, with $j=\HI, \H2$, for early- and late-type galaxies as well as for all galaxies.
	In addition we present these relationships using the arithmetic mean, $\langle M_j\rangle$. As expected,
	these relationships lie above from those calculated with the logarithmic mean, $\langle\log M_j\rangle$. 
	Moreover standard deviations can vary significantly if they are computed with respect to the arithmetic or logarithmic mean,
	which also depends on the shape of the distributions. Other statistical measures that
	can be used to characterise the population distributions are medians and percentiles, for example. 
	As mentioned above, any statistical quantity can be computed with our CPDFs and confronted with
	both theoretical and/or observational results. 

	\item {\it The Galaxy Stellar Mass Functions:} In Section \ref{Sec:GSMF}, we determined the \gsmf\ from 
	the SDSS DR7  based on the photometric 
	catalog from \citet{Meert+2015} and \citet{Meert+2016} for masses above 
	$\ms=10^9$ \msun. For masses down to $\sim3\times10^{7}\msun$, 
	we used the \texttt{low-z} SDSS DR4  \citep{Blanton+2005,Blanton+2005a}, and corrected 
	it from surface brightness incompleteness and fluctuations due to large scale structures. We determined also 
	the fractions of early- and late-type galaxies by using the SDSS DR7 morphological classification of \citet{Huertas-Company+2011}. 
	Stellar masses
	were derived from five colour-dependent mass-to-light ratios. We used as our fiducial definition 
	the geometric mean of these five stellar masses derived for each galaxy. We also determined the impact
	of systematic errors in \ms\ due to mass-to-light ratio uncertainties in our MFs. 

	\item {\it Calculated Mass Functions:} Section \ref{MFs_results} presents the results of calculating
	with our approach the MFs for atomic, molecular, cold gas
	and baryons for early- and late-type galaxies, as well as for all galaxies. 
	As discussed in Section \ref{random_and_systematic_errors}, random errors in mass determinations artificially decrease  
	 the slope of the ``observational'' MFs, an effect that 
	 affects especially the high-mass end, and that would lead to incorrect conclusions when comparing to 
	theoretical predictions. Figure \ref{fig:mfs_deconvolved} presents our MFs 
	deconvolved from random 
	errors, that is, the {\it intrinsic} MFs, for different baryon matter components, and separately for early- and late-type galaxies.
	In the same section, we studied the effects on the MFs from systematic errors in \ms, 
	also shown in Figure  \ref{fig:mfs_deconvolved}. 
	In Appendix \ref{SDSS_DR7_GSMF} and Figure \ref{GSMF} we
	showed explicitly that systematic errors in the GSMF due to mass-to-luminosity ratios and photometric uncertainties 
	are well represented by a shift in the \ms-axis of $\pm 0.15$ dex. 
	The effect of random errors in the baryonic MF is of the same order while for the gas MFs the propagated systematic errors in \ms\ have a negligible effect. 
	Note that our MFs  are complete only above a given mass limit,  $\sim 3\times 10^7$ \msun\ for the 
	GSMF, $\sim 10^7$ \msun\ for the \H2\ MF, and $\sim 10^8$ \msun\ for the \HI, cold gas, and baryonic MFs. 
\end{itemize}

From the results summarised above we highlight the following conclusions:

\begin{itemize}
	\item The low-mass slope of our \gsmf, corrected for surface brightness incompleteness, is $\alpha \approx -1.4$, 
	consistent with recent determinations based on the deeper surveys such as GAMA \citep{Wright+2017}, and 
	estimations based on the search of low surface brightness galaxies from core-collapse supernovae \citep{Sedgwick+2019}. 
	The slope for the high mass-end is shallower than previous determination most likely as the result of the 
	new photometric catalog employed in this paper \citep{Meert+2015}. 
	Similar results have been reported in \citet{Bernardi+2017}. 
	
	\item The total GSMF is well fitted by a function composed of a sub-exponential Schechter function and a 
	double power-law function. This fitting model has an error of 
	less than $\sim 2\%$ in the mass range $2\times 10^9-5\times 10^{11}$ \msun. 
	At the smallest and largest masses, the deviations increase to values above $\sim20\%$.
	In contrast, the commonly employed double Schechter function model performs considerably worse.
	
	\item Systematic errors due to stellar population synthesis models, that affect results on 
	mass-to-light ratios, introduce a systematic effect on the normalisation of the GSMF, especially at the 
	massive-end. We find differences between $\sim 0.5- 1$ dex, consistent with the result discussed in 
	\citet{Bernardi+2017}. 
	
	\item The \HI, \H2, and cold gas MFs are mostly 
	dominated by late-type galaxies. In general, we notice that our \HI\ MF is in
	good agreement with previous determinations from blind surveys. Similarly
	the \H2\ MF is consistent with previous determinations based on CO follow-up 
	optically-selected samples. When we compare to the HIPASS and ATLAS 3D surveys for  
	early-type galaxies, our \HI\ MF is consistent with those observations. However, our \H2\ MF
	for early-type galaxies is in tension at the low-mass side with the MF derived from the ATLAS 3D survey. 
	
	\item Our ``observational'' MFs were deconvolved from random errors to obtain the intrinsic MFs. 
	The effect of random errors is small at the low-mass end  
	but larger at the 
	high-mass end of our MFs. This is because 
	the convolution depends on the logarithmic slope of the intrinsic MFs. Because
	the baryonic and stellar MFs are shallower at the massive-end the effects are relatively 
	small, but the atomic, molecular and cold gas MFs have steeper slopes resulting in a larger effect.
	
	\item 
	While for the stellar (and hence baryonic) MF systematic errors due to mass-to-light ratio uncertainties introduce 
	a non-negligible effect, especially at the high-mass end, 
	for the atomic, molecular and gas MFs the effects of systematics are small. We thus conclude
	that our determinations for the gas MFs are robust against systematic errors in the the \ms\ determination.
	
	\item We determined the $z\sim 0$ cosmic densities of \HI, \H2, cold gas, stars and baryons locked in galaxies 
	calculated from the respective MFs. Our results are in good agreement 
	with previous determinations from different local censuses. 
	Most of the atomic and molecular H gas is in late-type galaxies, $\sim96\%$
	of the mass density, while this fraction decreases to $\sim70\%$ and $\sim 65\%$ for baryons and stars. We find
	that the fraction of \HI\ and \H2\ in galaxies with respect to the universal baryon fraction 
	is respectively $\sim 1 \%$ and $\sim 0.2\%$ while the respective fractions for mass in stars
	is $\sim 4\%$. Baryons in galaxies (the ionised and hot gas were not included) are $\sim 5.4\%$ of the universal 
	baryon fraction.
	
	\item Based on the values reported in the literature for the local CSFR of star-forming (late-type) galaxies,
	we estimated the cosmic \H2\ and total gas depletion times of late-type galaxies. These timescales,
	$t_{\rm dep}(\H2)\approx 1.3$  Gyr and $\bar{t}_{\rm dep,L}({\rm gas})\approx 10.14$ Gyr, respectively,
	imply that galaxies, on average, are inefficient to convert their molecular gas into stars, and are
	inefficient to transform their atomic gas into molecular gas. The depletion time for the total neutral
	hydrogen is $\bar{t}_{\rm dep,L}({\rm H}) = \bar{t}_{\rm dep,L}({\rm gas}) / 1.4\sim 7.25$ Gyrs.
	On the other hand, the average cosmic SF timescale (the inverse of the cosmic sSFR) is
	$\bar{t}_{\rm SF,L}\approx 20.3$ Gyrs, which implies that the ratio 
	$\bar{t}_{\rm dep,L}({\rm H})/\bar{t}_{\rm SF,L} = 0.38$. This shows that the gas reservoir of late-type 
	galaxies has not yet been dramatically consumed by star formation. 
\end{itemize}

Here, we provided a statistical  description for calculating any moment to characterise the gas-to-stellar mass correlations, 
the \HI- and \H2-stellar mass bivariate distributions as well as all the respective MFs. 
One of our motivations for this paper is 
to provide the community with a full self-consistent phenomenological description of the local galaxy population for various 
properties and divided into the two main morphological types
in order to be confronted with theoretical results, both from semi-analytical models and cosmological hydrodynamical simulations. 
The next generation of sensitive radio telescopes will be able to survey large samples of extragalactic sources 
in \HI\ and \H2\ gas, something that is a common practice with current optical surveys.
Thus, robust and unbiased bivariate distributions and MFs of \HI\ and \H2\ gas over large mass ranges will be routinely 
derived in the future along with the relationships of the gas contents with their optical/IR properties. 
Preparatory to that, and to pave the road to these surveys, studies based on radio follow-up observations
of (relative small) optically-selected galaxy samples provide valuable information that can be used for the gas demographics of galaxies. 
In this work, we have exploited the results from many of these studies, and by means of the conditional
 (or bivariate) approach we were able to derive the abundances of local galaxies as traced 
by different baryonic components and separated into the two main groups of galaxies, early and late types.

The present work is the second paper of a series. 
In \citetalias{Calette+2018}, we derived the CPDFs of 
\HI\ and \H2\ as a function of \ms, separately for early- and late-type galaxies, for an extensive compilation and homogenisation of 
radio data from the literature. In the present work, we made extensive use of these data. In the future,
we will use the MFs derived here to extend the galaxy-halo connection for different baryonic components, and we will
show that not only the \HI\ and \H2\ MFs derived here are in good agreement 
with radio blind or optically-selected surveys but also with the observed galaxy spatial clustering as a function of \HI\ gas mass.

\begin{acknowledgements}
ARP and VAR acknowledges support from UNAM PAPIIT grant IA104118 and from the CONACyT  `Ciencia Basica' 
grant 285721. ARC acknowledges support from CONACyT graduate fellowship. We thank to the referee
Tsutomu T. Takeuchi for a constructive report that helped to improve this paper.
\end{acknowledgements}


\begin{appendix}

 \section{Derivation of the SDSS DR7 \gsmf}
 \label{SDSS_DR7_GSMF}
 
 In this paper we derive the \gsmf\ from a spectroscopic sample of 670,722 galaxies from the SDSS DR7 based on
 the photometric estimates of the apparent brightnesses in the $g$, $r$ and $i$ band from \citet{Meert+2015} and 
 \citet{Meert+2016}. In those papers, the authors selected galaxies 
 with extinction-corrected $r-$band Petrosian magnitude between magnitude 14 and 17.77 to derive de Vacouleurs,
 S\'ersic, de Vacoulers+Exponential, and S\'ersic+Exponential fits to the observed surface brightness profiles of each
 galaxies in their SDSS DR7 catalogue. Surface brightness profiles were obtained via the \textsc{PyMorph}
 pipeline  \citep{Vikram+2010,Meert+2013}. \textsc{PyMorph} is a \textsc{python} software that uses \textsc{Sextractor} 
 \citep{Bertin_Arnouts1996} and  \textsc{Galfit} \citep{Peng+2002} to fit both one- an two-components to the seeing convolved 
 surface brightness profiles from the spectroscopic sample of SDSS DR7 galaxies. 
 \textsc{PyMorph} has been extensively tested in \citet[][see also, \citealp{Meert+2015}]{Meert+2013} showning that the algorithm 
 does not suffer from the sky subtraction problems that has been detected in previous studies based on the SDSS, in particular in 
 crowded fields\footnote{Recently, various others groups have also improved the determinations of galaxies' surface brightness profiles based on the SDSS 
 	by the improving the survey photometry, especially due to sky subtraction problems in crowded fields, 
 	\citep[see e.g.,][and more reference therein]{Simard+2011,DSouza+2015}. While in this paper we opt to use the photometric catalog
 	from \citet{Meert+2015} and \citet{Meert+2016}, \citet{Bernardi+2017} showed that, after a careful comparison, most of those studies 
 	agree up to 0.1 dex. Thus, using the photometry derived by other groups will not change significantly our results.}. 
 
 \begin{figure}
 	\vspace*{-120pt}
 	\includegraphics[height=5in,width=4in]{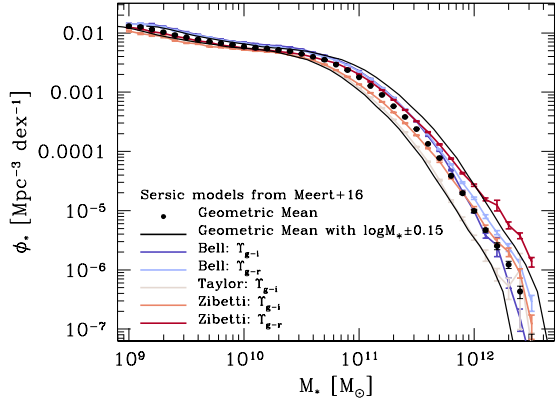}
 	\vspace*{-80pt}
 	\caption{The GSMF from our six stellar mass definitions, Equation \ref{mass_to_light_ratios}. Using 
 		different stellar masses yield to differences of $\sim0.1$ dex at low masses and as high as $\sim1$ dex 
 		at the high mass-end. In this paper we opted to use as our fiducial GSMF as the one derived from the
 		geometric mean of five different stellar masses. The solid lines show a shift 
 		of $\pm0.15$ dex in the stellar mass axis of the our fiducial GSMF, note that it recovers systematics from mass-to-light ratios. 
 	}
 	\label{fig:GSMF_comparison}        
 \end{figure}
 
 We estimate the \gsmf\ at the redshift interval between $z = 0.005$ and $z = 0.2$ by using the standard $1/V_{\rm max}$ method \citep{Schmidt_1968}:
 \begin{equation}
 \phi_{\ast}(\ms) = \frac{1}{\Delta\log\ms}\sum_{i=1}^N\frac{\omega_i(\log\ms\pm\Delta\log\ms)}{V_{{\rm max},i}},
 \end{equation}
 where $\omega_i$ is a weight factor correction that depends on the position in the sky for galaxies within the interval $\log\ms\pm\Delta\log\ms/2$, following 
 \citet[see also][]{Bernardi+2010} we assume that $\omega_i = 1.1$; and 
 \begin{equation}
 V_{{\rm max},i} = \int_{\Omega}\int^{z_{u,i}}_{z_{l,i}}\frac{d^2V_c}{dzd\Omega} dzd\Omega. 
 \end{equation}
 We denote the solid angle of the SDSS DR7 with $\Omega$ while $V_c$ refers to the comoving 
 volume enclosed within the redshift interval $[z_{l,i}, z_{u,i}]$. The redshift limits are defined as 
 $z_{l,i} = {\rm max}(0.005, z_{{\rm min},i})$ and $z_{u,i} = {\rm min}(z_{{\rm max},i}, 0.2)$; where $z_{{\rm min},i}$ 
 and $z_{{\rm max},i}$ are, respectively, the minimum and maximum at which each galaxy can be 
 observed in the SDSS DR7 sample. We estimate $z_{{\rm max},i}$ for every galaxy in the sample
 by solving iteratively the distance modulus equation:
 \begin{eqnarray}
 m_{\rm lim,r} - M_{r,i}^{0.0} = 5 \log D_{L,i}(z_{{\rm max},i}) + 25\\
  +K_{gr,i}(z_{{\rm max},i}) - E_{r,i}(z_{{\rm max},i}),
 \end{eqnarray}
 where $M_{r,i}^{0.0}$ is the Petrosian magnitude K+E-corrected at a rest-frame $z=0$, 
 $K_{gr,i}(z)$ is the $K$-correction (see Appendix \ref{Apendix_Kcorr}) and $E_{r,i} = 1.1 z$ \citep[following][]{Dragomir+2018} for
 the $i$th galaxy in the sample. 
 For the completeness limits, we use the limiting
 apparent magnitude in the $r$-band of $m_{\rm lim,r}=17.77$. Similarly, we estimate $z_{{\rm min},i}$ 
 by solving iteratively the distance modulus equation but this time using the limiting
 apparent magnitude $m_{\rm lim,r}=14$. 
 
 Errors are estimated using the jackknife technique by diving the galaxy sample into $n=300$ subsamples of
 approximately equal size and estimating a $\phi_{*,i}(\ms)$  each time. Thus errors are then given by:
 \begin{equation}
 \sigma^2 = \frac{n-1}{n} \sum_{i=1}^n\left(\phi_{*,i} - \langle\phi_{*}\rangle\right)^2,
 \end{equation}
 with $\langle\phi_{*}\rangle$ as the average of the ensemble. 
 
 Stellar masses were derived from several colour-dependent mass-to-light ratios as listed below:
 \begin{equation}
 \ms = \left\{ 
 \begin{array}{c l}
 \Upsilon^{\rm B03}_r(g-r)\cdot L_r & \mbox{\citet{Bell+2003}}\\
 \Upsilon^{\rm B03}_i(g-i)\cdot L_i & \mbox{\citet{Bell+2003}}\\
 \Upsilon^{\rm Z09}_r(g-r)\cdot L_r & \mbox{\citet{Zibetti+2009}}\\
 \Upsilon^{\rm Z09}_i(g-i)\cdot L_i & \mbox{\citet{Zibetti+2009}}\\
 \Upsilon^{\rm T11}_i(g-i)\cdot L_i & \mbox{\citet{Taylor+2011}}\\
 \end{array},\right.
 \label{mass_to_light_ratios}
 \end{equation}
 and we define our fiducial \ms\ as the geometrical mean of all the determinations in \ref{mass_to_light_ratios}:
 \begin{eqnarray} \label{Eq:fiducial-Ms}
 M_*= [M_*(\Upsilon^{\rm B03}_r) \times M_*(\Upsilon^{\rm B03}_i) \times M_*(\Upsilon^{\rm Z09}_r) \times   &\\  \nonumber
 M_*(\Upsilon^{\rm Z09}_i) \times M_*(\Upsilon^{\rm T11}_i)]^{1/5} 
 \end{eqnarray}

 \begin{figure*}
 	\vspace*{-90pt}
 	\hspace*{20pt}
 	\includegraphics[height=7.8in,width=6.5in]{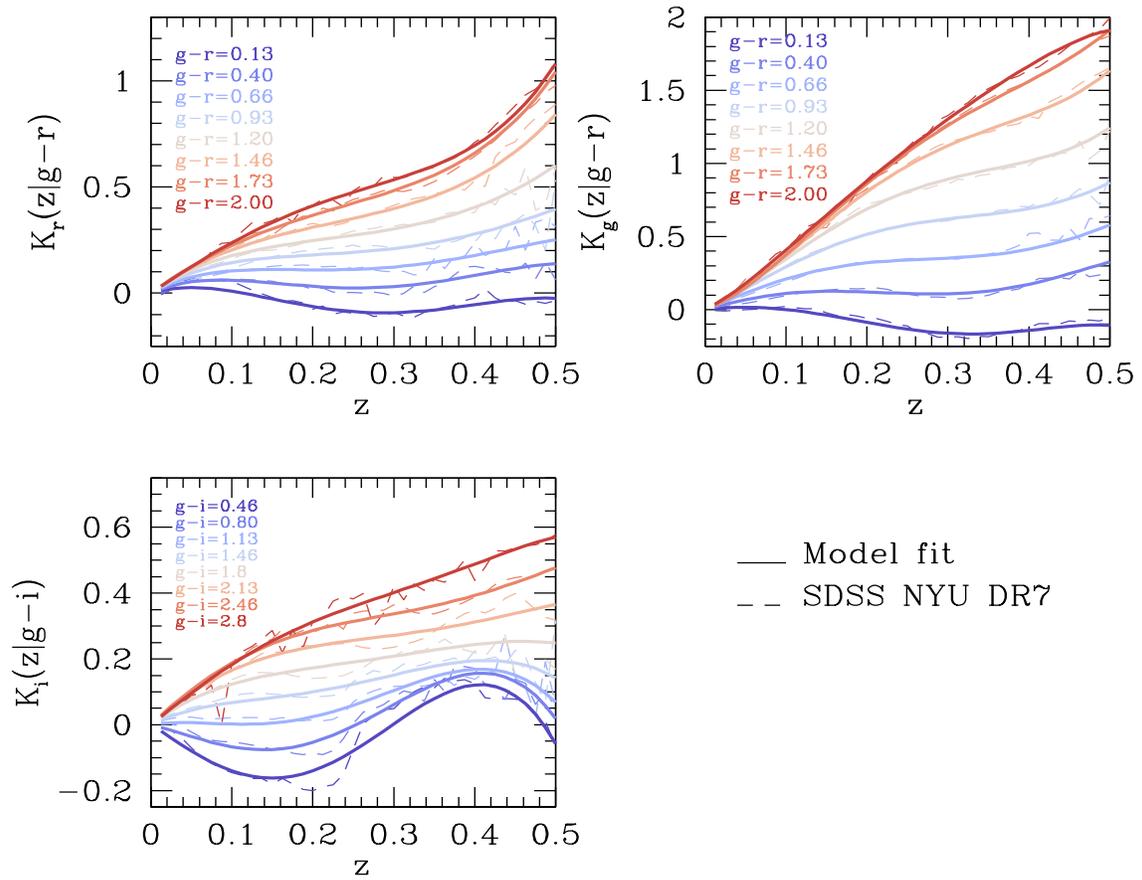}
 	\vspace*{-100pt}
 	\caption{Colour and redshift dependence of the K-corrections at a rest-frame $z=0$ for 
 		the $r$ (upper left), $g$ (upper right), and $i$ (bottom left) bands from the \textsc{k-correct} algorithm
 		\citep{BlantonRoweis2007}, dashed lines. Solid lines show our best fit models as described in the text. 
 	}
 	\label{fig:kcorr}        
 \end{figure*}
 
 We use S\'ersic apparent magnitudes to derive colours and magnitudes. We apply K+E-corrections at a rest-frame $z=0$. We use
 the values reported in \citet{Dragomir+2018} for $g$ and $i$ bands given, respectively, by $E_{g} = 1.3\times z$ and $E_{i} = 1.09\times z$. K-corrections are discuss in
 Appendix \ref{Apendix_Kcorr}.
 We applied a shift of $-0.1$ dex to the resulting masses from the colour dependent mass-to-light ratios of \citet{Bell+2003} to be consistent with the
 \citet{Chabrier2003} IMF adopted in this paper. 
 
 Figure \ref{fig:GSMF_comparison} shows the resulting $\gsmf$s as described above. The figure shows that using different recipes of deriving stellar masses
 yield to differences of $\sim0.1$ dex at low masses and between $\sim0.5-1$ dex at the high mass-end. This is consistent with the recent study by 
 \citep{Bernardi+2017} which showed that differences in mass-to-light ratios introduce discrepancies in the \gsmf\ around $\sim0.5$ dex. As a fiducial 
 estimation of the \gsmf\ in this paper we opt to utilise the geometric mean of all the masses derived based on the colour dependent mass-to-light ratios listed
 in Equation (\ref{mass_to_light_ratios}), filled circles with error bars. The black solid line shows a shift of $\pm 0.15$ dex in the stellar mass axis of our fiducial 
 \gsmf. Note that these shifts recover most of the differences observed due to 
 systematics in mass-to-light ratios. 
 
 \section{K-corrections}
 \label{Apendix_Kcorr}
 
 Figure \ref{fig:kcorr} shows the colour and redshift dependence of the K-corrections at a rest-frame $z=0$ for the 
 $r$ (upper left), $g$ (upper right), and $i$ (bottom left) bands, shown as the dashed lines, from the \nyu\ SDSS DR7 and calculated
 from the \textsc{k-correct} algorithm \citep[\textsc{v4\_1\_4}][]{BlantonRoweis2007}. In the same figure the solid lines show
 the best fit according to the following relations:
 \begin{equation}
 K_j(z|\mathcal{C}) = \mathbf{Z} \mathbf{K}_j^\mathcal{C} \mathbf{C}
 \label{k_corr}
 \end{equation}
 where $j$ denotes the $r$, $g$ and $i$ bands while $\mathcal{C}$ denotes the uncorrected $g-r$ and $g-i$ galaxy colours. A similar approach has been
 done in \citet{Chilingarian+2010}. 
 The $\mathbf{C}$ and $\mathbf{Z}$ matrices are
 respectively given by
 \begin{equation}
 \mathbf{C} = \left( 
 \begin{array}{c}
 1\\
 C\\
 C^2\\
 C^3\\
 C^4\\
 C^5\\
 \end{array}\right),
 \end{equation}
 \begin{equation}
 \mathbf{Z} = \left( 
 \begin{array}{c c c c c}
 z & z^2 & z^3 & z^4 & z^5\\
 \end{array}\right),
 \end{equation}
 while the $\mathbf{K}_j^\mathcal{C}$ matrices for the $r$, $g$ and $i$ bands are respectively given by equations (\ref{eq:Kr}), (\ref{eq:Kg}) and (\ref{eq:Ki}).
 Note that our K-corrections are polynomials of degree 5 in both colour and redshift and that in the above set of Equation  $K_j(z=0|\mathcal{C})=0$.

 \addtocounter{equation}{1}
 \setcounter{storeeqcounter}{\value{equation}}

\begin{figure*}
	\normalsize
	\setcounter{tempeqcounter}{\value{equation}} 
	\begin{IEEEeqnarray}{rCl}\setcounter{equation}{\value{storeeqcounter}} 
		K_r^{g-r} & = & \left( 
		\begin{array}{c c c c c c}
			0.894302 & 2.32866 & -0.787673 & 0.324352 & -0.239774 & 0.0444971\\
			-15.5648 & 1.544 & -2.70992 & 3.42484 & -0.280197 & -0.0221534\\
			49.7443 & -4.64543 & -8.72852 & 1.14138 & -1.76882 & 0.0702624\\
			-48.9173 & -4.95549 & 2.06966 & 14.5241 & -2.48092 & -0.322153\\
			3.65716 & 21.3194 & -0.593275 & -6.04982 & -0.157727 & 0.731093\\
		\end{array}\right)
		\label{eq:Kr}
	\end{IEEEeqnarray}
	\setcounter{equation}{\value{tempeqcounter}} 
	\vspace*{4pt}
\end{figure*}

 \addtocounter{equation}{1}
 \setcounter{storeeqcounter}{\value{equation}}

\begin{figure*}
	\normalsize
	\setcounter{tempeqcounter}{\value{equation}} 
	\begin{IEEEeqnarray}{rCl}\setcounter{equation}{\value{storeeqcounter}} 
		K_g^{g-r} & = & \left( 
		\begin{array}{c c c c c c}
			0.0786144 & 4.01535 & -0.883155 & 0.707471 & -2.05303 & 0.793141\\
			-6.81272 & 12.0599 & -10.7157 & 22.086 & -5.46384 & -1.34602\\
			-7.17353 & -52.5682 & -13.5845 & 11.2634 & -6.25812 & 2.61254\\
			86.1835 & 96.7938 & -72.2792 & -1.44621 & -5.6531 & 9.09575\\
			-106.868 & -23.5461 & 101.815 & -43.5146 & 40.8195 & -21.677\\
		\end{array}\right)
		\label{eq:Kg}
	\end{IEEEeqnarray}
	\setcounter{equation}{\value{tempeqcounter}} 
	\vspace*{4pt}
\end{figure*}

 \addtocounter{equation}{1}
 \setcounter{storeeqcounter}{\value{equation}}
 
\begin{figure*}
	\normalsize
	\setcounter{tempeqcounter}{\value{equation}} 
	\begin{IEEEeqnarray}{rCl}\setcounter{equation}{\value{storeeqcounter}} 
		K_i^{g-i} & = & \left( 
		\begin{array}{c c c c c c}
			-3.01597 & 3.287 & -0.455067 & 0.426496 & -0.242669 & 0.0283777\\
			-1.11123 & -3.04641 & -5.2804 & 2.60911 & 0.134077 & -0.0813698\\
			68.4078 & -14.6203 & -5.06879 & 0.904234 & -1.82776 & 0.47701\\
			-145.044 & 45.4714 & 8.75605 & 5.9425 & -1.32215 & -0.211679\\
			59.2903 & -12.387 & -10.8653 & -1.84054 & 0.843326 & -0.0248045\\
		\end{array}\right)
		\label{eq:Ki}
	\end{IEEEeqnarray}
	\setcounter{equation}{\value{tempeqcounter}} 
	\vspace*{4pt}
\end{figure*}

 \section{Galaxy Stellar Mass Function for Low Mass Galaxies}
 \label{GSMF-correction}
 
 \subsection{Surface Brightness Correction Completeness}
 
 In this paper we are interested in deriving the \gsmf\ over a wide dynamical mass range, i.e., from dwarf galaxies to massive
 elliptical galaxies. In Appendix \ref{SDSS_DR7_GSMF}, we describe that based on the SDSS DR7 galaxy
 sample we determined the \gsmf\ for galaxies above $\ms\sim10^{9}\msun$. In this Section, we determine the \gsmf\ for
 galaxies above $\ms\sim10^{7}\msun$. Deriving the \gsmf\ could be very challenging since the fraction of galaxies of missing
 galaxies due to surface brightness limits becomes very relevant at the faint-end of the \gsmf. Here, we follow a very simple 
 statistical approach in order to quantify the number of galaxies missed due to surface brightness incompleteness limits as described
 in \citet{Blanton+2005a}. Our galaxy sample consist of a small volume ($0.0033 < z < 0.05$) 
 carefully constructed to study very low mass/luminosity galaxies from the SDSS \nyu\ with a total of 49968 galaxies 
 \citep{Blanton+2005,Blanton+2005a}\footnote{\url{http://sdss.physics.nyu.edu/vagc/lowz.html}}. Here after we will refer to this galaxy sample as
 the \texttt{low-z} SDSS
 
 \citet{Blanton+2005a} estimated that the \texttt{low-z} SDSS galaxy sample has a completeness $>70$ percent for  
 galaxies in the effective surface brightness range of $18<\mu_{50,r}< 24$ mag arcsec$^{-2}$ and we consider galaxies only within this range.
 We assign to each galaxy a weight, $w_{\mu,j}$, which is a function of their central surface brightness
 and it takes into account the spectroscopic incompleteness ($1 / w_{s,j}$), 
 photometric incompleteness ($1 / w_{p,j}$), and tiling catalog incompleteness ($1 / w_{t,j}$) in the sample. 
 Thus, $w_{\mu,j}=w_{s,j}\times w_{p,j} \times w_{t,j}$. 
 These weights were studied in detail in \citet{Blanton+2005a} and provide the correlation between $w_{\mu,j}$ and 
 effective surface brightness, $\mu_{50,r}$, in a tabulated form, see their Table 1. We 
 use cubic spline interpolations of this Table in order to assign weight $w_{\mu,j}$ to each galaxy in the sample.

 \begin{figure*}
 	\vspace*{-120pt}
 	\includegraphics[height=4in,width=3.5in]{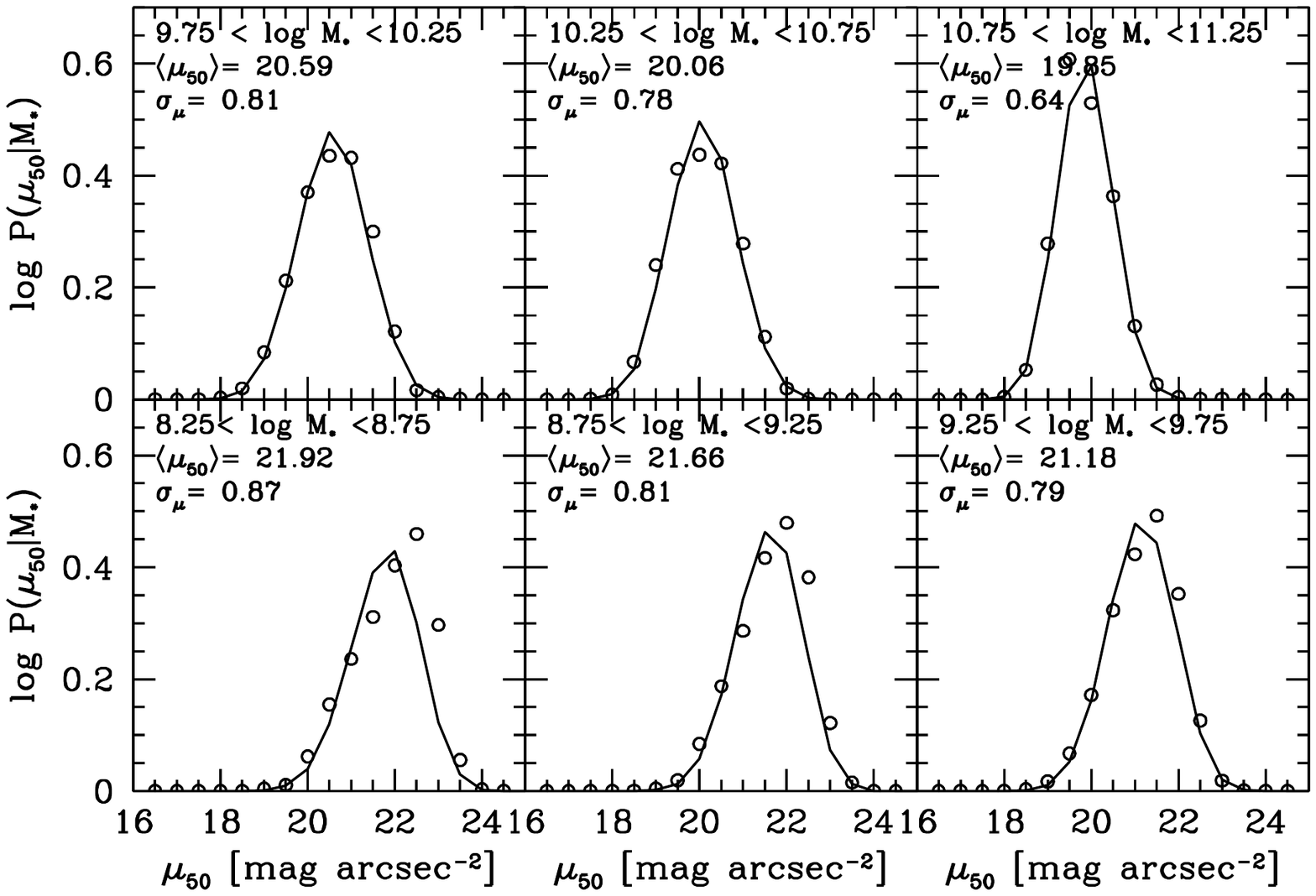}
 	\hspace*{10pt}
 	\vspace*{-50pt}
 	\hspace*{-30pt}
 	\includegraphics[height=4.5in,width=3.5in]{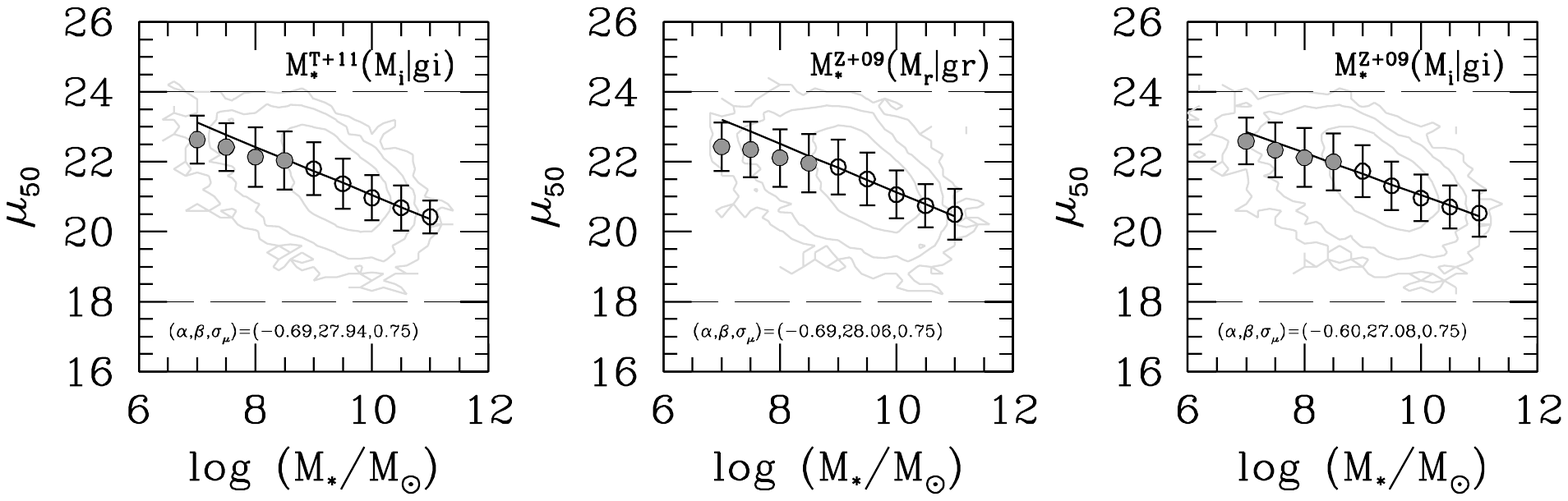}
 	\caption{ {\it Left panels:} Conditional probability distribution $P(\mu_{50,r}|\ms)$ when
 		using stellar mass estimations based on the fits to $g-i$ 
 		colors and absolute magnitudes $M_i$ from \citet{Zibetti+2009}. Empty circles
 		show the resulting distribution from observations of galaxies 
 		with Sersic index $n_s\leq2$ only. Solid lines show the fit to observations when
 		assuming a lognormal model distribution as described in the text. 
 		{\it Right panels:} SB-to-galaxy stellar mass relation for
 		three different stellar mass estimators used in this paper. Solid 
 		circles show the mean values of $\mu_{50}$ that are affected
 		by SB incompleteness while empty circles
 		show that are complete according \citet{Blanton+2005a}. 
 	}
 	\label{fig:SB-distributions}
 \end{figure*}

 \begin{figure}
 	\hspace*{20pt}
 	\includegraphics[height=3.8in,width=2.6in]{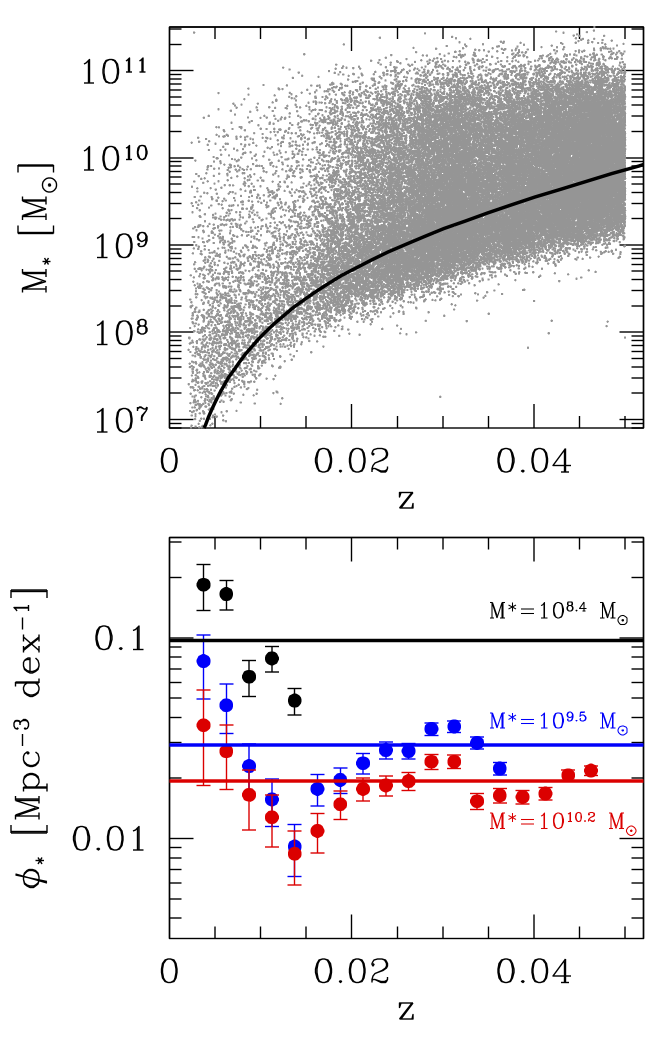}
 	\caption{{\it Upper Panel:} Distribution of galaxies in the \ms\ and redshift plane for the lowz-SDSS galaxy sample, grey
 		dots. The solid lines shows the dependence of the stellar mass completeness limit as a function of redshift for our
 		galaxy sample. {\it Bottom Panel:} The dependence of the \gsmf\ with redshift for three different stellar masses. 
 		Note that the increase and decreases in the amplitudes is due the large scale structure fluctuations. The solid lines
 		show our corrections due to large scales structures as described in the text. 
 	}
 	\label{fig:pdf_z_gsmf}        
 \end{figure}

 \begin{figure*}
 	\vspace*{-230pt}
 	\center{
 		\includegraphics[height=8.5in,width=7.in]{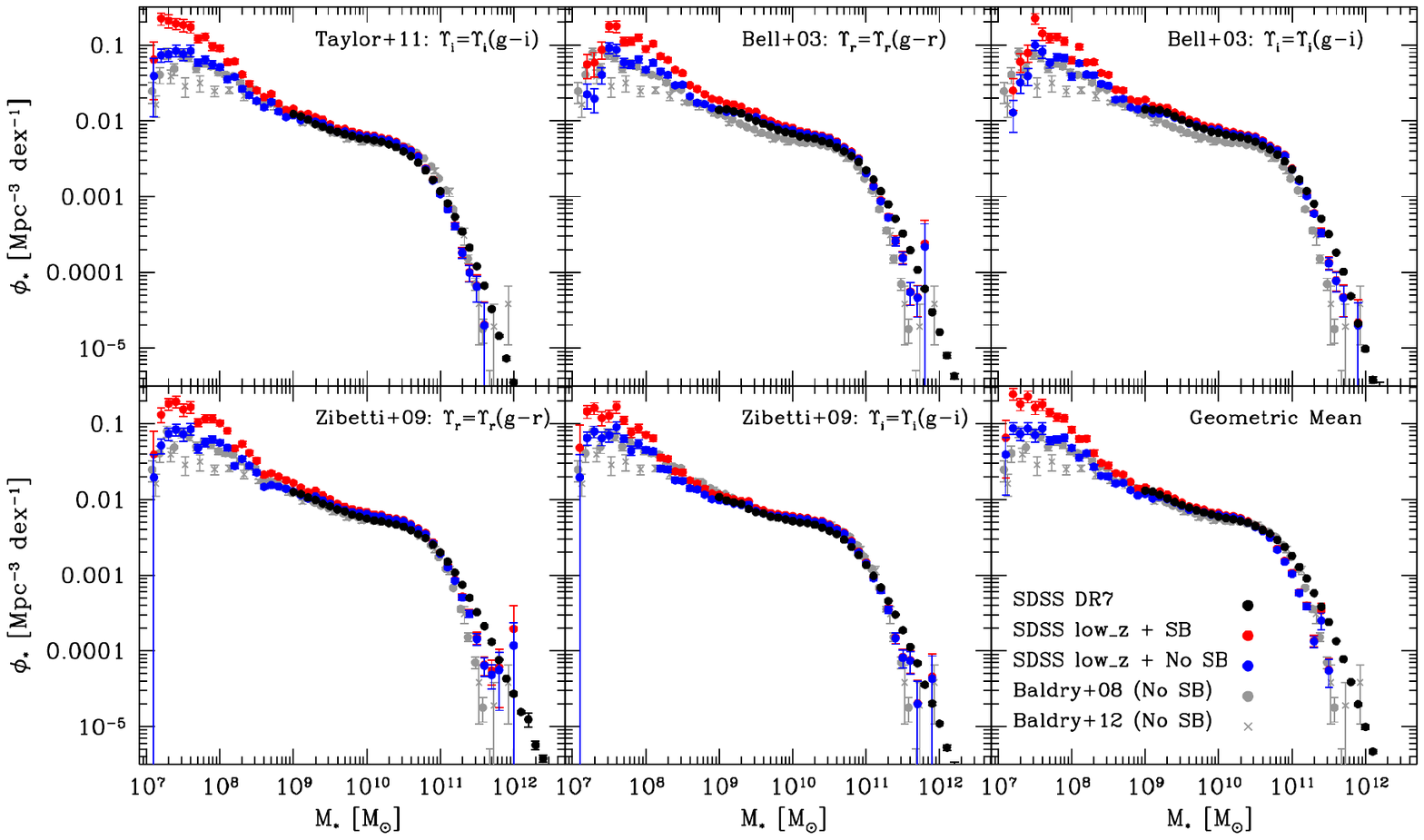}
 		\vspace*{-150pt}
 		\caption{Corrected \gsmf\ for surface brightness incompleteness, red filled circles with
 			error bars. Blue filled circles with error bars show the uncorrected \gsmf. 
 			Grey skeletal symbols and filled circles show the \gsmf\ from \citet{Baldry+2012}
 			and \citet{Baldry+2008}, respectively. The black filled circles show the \gsmf\ from the SDSS
 			DR7 derived from Appendix \ref{SDSS_DR7_GSMF}. The different panels show the various stellar mass estimators
 			used for this paper.  Note that the at $\ms\sim 10^9$ \msun\ there is a smooth transition between 
 			the  low-$z$ NYU-VAGC sample and the SDSS DR7 GSMFs.
 		} \label{fig:gsmf_SB}
 	}
 \end{figure*}

 The next step in our program is to estimate the number of missed galaxies brighter than $\mu_{50,r} =  24$ mag arcsec$^{-2}$.
 To that end, we introduce a model for the distribution of $\mu_{50,r}$ as a function of \ms.
 We define the fraction of missing galaxies brighter than $\mu_{50,r} =  24$ mag arcsec$^{-2}$ 
 as a function of stellar mass as:
 \begin{equation}
 f_{\rm loss} (\ms) = \frac {\sum\limits_{j} N_{\rm obs} (\mu_{50,j}|\ms) }{\sum\limits_{j} N_{\rm real} (\mu_{50,j}|\ms) }, 
 \end{equation}
 where  $N_{\rm real} (\mu_{50,j}|\ms)$ and $N_{\rm obs} (\mu_{50,j}|\ms)$ are the real and observed 
 number galaxies with SB between $\mu_{50,j} \pm d\mu_{50,j} $ and stellar masses 
 between $\log\ms\pm d\log\ms/2$, respectively. Thus, our problem reduces to estimate $N_{\rm real}$. 
 Let us now define $P(\mu_{50,r}|\ms)$ as the conditional probability distribution 
 of galaxies with SB $\mu_{50,r}\pm d\mu_{50,r}/2$ at a stellar mass bin $\log\ms\pm d\log\ms/2$. 
 We calculate $P(\mu_{50,r}|\ms)$ directly from our galaxy sample by dividing it into stellar mass bins of 0.5 dex.
 This is done only for galaxies with S\'ersic index $n_s\leq2$ (galaxies with $n_s>2$ are mostly of high
 SB, $\mu_{50,r} <  24$ mag arcsec$^{-2}$, so that it is not necessary to correct them for missing galaxies). 
 For each stellar mass bin, we perform an extra binning of 0.05 dex in SB. We carry out the mentioned binning 
 in \ms\ and $\mu_{50,r}$ for each of the six different stellar mass estimators described above. 
 As an example, the upper panel in Figure \ref{fig:SB-distributions} show the distributions $P(\mu_{50,r}|\ms)$ for one of our stellar 
 mass estimators (empty circles). For each \ms\ bin, we fit $P(\mu_{50,r}|\ms)$ assuming that it is 
 described by a lognormal distribution,
 \begin{equation}
 P(\mu_{50,r}|\ms) = \frac{1}{\sqrt{2\pi\sigma_\mu^2}}
 \exp\left[-\frac{\left(\mu_{50,r}-\langle\mu_{50,r}(\ms)\rangle\right)^2}{2\sigma_\mu^2}\right],
 \label{pmums}
 \end{equation}
 where $\langle\mu_{50,r}(\ms)\rangle$ and $\sigma_\mu$ are the mean SB at a given 
 stellar mass and the dispersion around it. We fit these two free parameters for each stellar mass bin.
 The best fits are plotted with solid lines. This operation is carried out for each of the stellar mass
 estimators used here. In the bottom panels of Figure \ref{fig:SB-distributions}, we show an example of the resulting best fits 
 to observations as circles with error bars in the bivariate ($\mu_{50,r}, \ms$) distribution plane (gray iso-contours) for three of our stellar mass 
 estimators. The dashed lines show our SB magnitudes limits. 
 Similarly to \citet{Baldry+2008}, we find that the relation between $\langle\mu_{50,r}\rangle$ and 
 $\log\ms$ is linear for galaxies above $\ms\sim10^{9}\msun$\footnote{In fact,
 	\citet{Baldry+2008} found that the linearity holds above masses $\ms\sim10^{8.5}\msun$. Here
 	we apply the conservative value of $\ms\sim10^{9}\msun$. Nevertheless, we have found
 	that using either \citet{Baldry+2008} or our limit, the correction for SB is
 	practically the same.} 
 (filled circles) in the right panels of Figure \ref{fig:SB-distributions}. Departures from this linearity for 
 galaxies below $\ms\sim10^{9}\msun$ (filled circles) is an indication that the 
 relation between $\langle\mu_{50,r}\rangle$ and $\log\ms$ is affected by SB incompleteness. 
 We fit the relationship between $\langle\mu_{50,r}\rangle$ and $\log\ms$ for
 galaxies above $\ms = 10^{9}\msun$ (where the missing number of low SB galaxies is negligible) as
 \begin{equation}
 \langle\mu_{50,r}\rangle = \alpha \log\ms +\beta.
 \label{mums}
 \end{equation}
 For simplicity, we assume that the dispersion around this relation, $\sigma_\mu$,
 is independent of mass and has the same value for all the mass estimators; we assign a value 
 of 0.75 dex, which is close to most of the values determined by fitting Eq. (\ref{pmums}) to the
 data from our galaxy sample for the three methods of assigning stellar masses. 
 Note that the values of  $\alpha$ and $\beta$ depend on each stellar mass estimator implying that 
 SB corrections are susceptive to systematics due
 to stellar masses estimators. The next step is to assume that the distribution of
 real galaxies, $N_{\rm real}$, can be generated from the probability
 distribution $P(\mu_{50}|\ms)$ by simply extrapolating equations \ref{mums} and \ref{pmums}
 up to $\ms\sim10^{7}$ \msun. Using the definition of $w_{\mu,j}$ (the SB completeness factor), 
 the observed distribution of galaxies, $N_{\rm obs}$, is thus
 generated from the probability distribution  $P_{\rm obs}(\mu_{50,r}|\ms) = (1/ w_{\mu}) \times P(\mu_{50,r}|\ms)$. 
 The factor of missing galaxies below the SB $\mu_{50} =  24$ mag arcsec$^{-2}$ at a
 given stellar mass is then
 \begin{equation}
 f_{\rm loss} (\ms) = 
 \int  \left(1/ w_{\mu}\right) P(\mu_{50,r}|\ms) d\mu_{50,r} \Big/  \int P(\mu_{50,r}|\ms) d\mu_{50,r}.
 \end{equation}
 
 Thus, we weight every galaxy in the sample with:
 \begin{equation}
 w_{{\rm SB},j} = w_{\mu,j} \times w_{\rm loss},
 \label{w_sb}
 \end{equation}
 where
 \begin{equation}
 w_{\rm loss} = \left\{ \begin{array}{rcl}
 1 / f_{\rm loss} & \mbox{for}
 & n_s\leq2 \\ 
 1 & \mbox{for} & \mbox{else}.
 \end{array}\right.
 \end{equation}
 
 We are now in position to estimate the \gsmf\ corrected by SB incompleteness.
 
 \subsection{The Dependence of a Stellar Mass Limit Sample with Redshift}
 \label{secc:stelim_z}
 
 In order to calculate the \gsmf\ we start by determining how the apparent magnitude 
 limit of the SDSS transforms into a stellar mass limit. In other words, given that the apparent 
 magnitude limit of the SDSS is $m_{r, {\rm lim}} = 17.77$ we compute the equivalent in terms of stellar
 mass, $M_{*, \rm lim}$. Following \citet{vandenBosch+2008}, we determined the redshift-dependent 
 absolute magnitude limit $M_{r, {\rm lim}}^{0.0}$ given the apparent magnitude limit from the SDSS $m_{r, {\rm lim}} = 17.77$
 \begin{equation}
 M_{r, {\rm lim}}^{0.0} = m_{r, {\rm lim}} - 5 \log D_{L}(z) - 25 - K_{\langle gr \rangle }(z) + E_{r}(z),
 \end{equation}
 where $D_{L}$, $K_{\langle gr \rangle}$ and $E_{r}$ are functions described in Appendix \ref{SDSS_DR7_GSMF}. Note that we have emphasised  
 the use of average colours for the $K-$correction because we are interested in the stellar mass limit for all the galaxies. Thus, the above absolute magnitude limit depends both
 on redshift and colour \citep{vandenBosch+2008}. Using the colour-dependent mass-to-light ratio $\Upsilon^{\rm Z09}_r(g-r)$ from \citet{Zibetti+2009}, 
 we transform $M_{r, {\rm lim}}^{0.0}$ into a stellar mass limit
 \begin{equation}
 M_{*, \rm lim} = -0.84 + 1.654\times \langle g-r \rangle^{0.0} - 0.4 \times \left( M_{r, {\rm lim}}^{0.0} - 4.64 \right). 
 \label{Mste_lim}
 \end{equation}
 Finally, we use the mean relationship between colour and stellar mass for blue and red galaxies as well as the fraction of red, $f_{\rm R}$, and blue galaxies, $f_{\rm B}$, 
 to compute the average colour-stellar mass relationship as
 \begin{equation}
 \langle g-r \rangle^{0.0} = f_{\rm B} (g-r)_{\rm B}^{0.0} +  f_{\rm Q} (g-r)_{\rm Q}^{0.0},
 \label{mean_color}
 \end{equation}
 where $(g-r)_{\rm B}^{0.0}$ and $(g-r)_{\rm Q}^{0.0}$ are the best fit models to the mean colour-stellar mass relationships
 of blue and red galaxies. 
 
 We paused here for a moment and described our method to derive $(g-r)_{\rm B}^{0.0}$ and $(g-r)_{\rm Q}^{0.0}$.
 To do so, we use the SDSS DR7 based on the photometric catalogue from \citet{Meert+2016}. We choose to use this catalog as
 contains many more galaxies than the SDSS DR4 and one could derive robust colour distributions. We derived the observed distribution 
 function of galaxy colours as a function of stellar mass, $P_{gr}(g-r|\ms)$, that is the observed distribution of galaxy colours at
 the range between $(g-r)^{0.0}\pm\Delta (g-r)^{0.0}/2$ and $\log\ms\pm\Delta\log\ms /2$. We divide our space into 20 bins equally spaced 
 for $(g-r)^{0.0}$ between $(g-r)^{0.0} = 0$ and $(g-r)^{0.0} = 1.4$ and into 25 bins 
 equally spaced between $\log \ms = 8.5 - 12$. 
 For galaxy stellar masses we use our fiducial definition from Appendix \ref{SDSS_DR7_GSMF}. 
 
 We assume that the distribution $P_{gr}(g-r|\ms)$ is bimodal and composed of two Gaussian distributions, this is a good approximation
 as shown by previous studies \citep[e.g.,][]{Baldry+2004,Baldry+2006}. We associate one of the modes of $P_{gr}(g-r|\ms)$
 with the distribution of blue galaxies, denoted by $P_{gr,{\rm B}}(g-r|\ms)$, while the remaining one with the distribution of red
 galaxies, denoted by $P_{gr,{\rm R}}(g-r|\ms)$. The relation between these distributions is given by:
 \begin{equation}
 \begin{split}
 P_{gr,{\rm B}}(g-r|\ms) = & f_{\rm B} (\ms) P_{gr,{\rm B}}(g-r|\ms) + \\
 & f_{\rm R} (\ms) P_{gr,{\rm R}}(g-r|\ms). 
 \end{split}
 \end{equation}
 We assume that $P_{gr,j}(g-r|\ms)$, with $ j = $ B or R, is a Gaussian distribution given by:
 \begin{equation}
 \begin{split}
 P_{gr,j}(g-r|\ms) = \frac{1}{\sqrt{2\pi\sigma_j^2(\ms)}} \times \\
 \exp\left[-\frac{\left((g-r)^{0.0} - (g-r)_{j}^{0.0}(\ms)\right)^2}{2\sigma_j^2(\ms)}\right],
 \end{split}
 \end{equation}
 where $(g-r)_{j}^{0.0}(\ms)$, with $ j = $ B or R, is the mean colour-stellar mass relationship 
 used in Equation (\ref{mean_color}) and $\sigma_j(\ms)$ is the standard deviation that depends on \ms. 
 The functional forms for $(g-r)_{j}^{0.0}(\ms)$ and $\sigma_j(\ms)$ are given by
 \begin{equation}
 (g-r)_{j}^{0.0}(\ms) = \alpha_j + \beta_j\times\log\left(\frac{\ms}{10^{11}\msun} \right),
 \end{equation}
 and
 \begin{equation}
 \sigma_j(\ms) = \lambda_j + \kappa_j\times\log(\ms),
 \end{equation}
 for $ j = $ B, R. Finally, for the fraction of red galaxies we assume that 
 \begin{equation}
 f_{\rm R} (\ms) = \frac{1}{1+ \left[ a + b\left( M_{*} / M_{\rm C} \right) \right]^{\gamma}}.
 \end{equation}
 We performed a $\chi^2$ minimisation procedure to the observed galaxy colour bimodality in order to 
 find the best fitting parameters to the functional forms described above. Our best fitting parameters are:
 $(\alpha_{\rm B}, \beta_{\rm B}, \lambda_{\rm B}, \kappa_{\rm B},\alpha_{\rm R}, \beta_{\rm R}, \lambda_{\rm R}, \kappa_{\rm R},
 a,b,\log M_{\rm C},\gamma) =$  (0.514, 0.086, 0.240, -0.015, 0.720, 0.064, -0.068, 0.014, 0.001, 1.390, 10.586, -1.001).
 
 The upper panel of Figure \ref{fig:pdf_z_gsmf} shows the dependence of $M_{*, \rm lim}$ with redshift. The region above $M_{*, \rm lim}$
 is the area above which the NYU-VAGC galaxy sample is a volume-limited sample that is complete in stellar mass. The small grey dots show individual 
 galaxies from the NYU-VAGC sample in the case that stellar masses were determine using the geometric mean of all our stellar mass estimators. 
 
 \subsection{Volume and large scale structure corrections}
 
 \begin{figure}
 	\vspace*{-80pt}
 	\center{
 		\includegraphics[height=4.4in,width=3.5in]{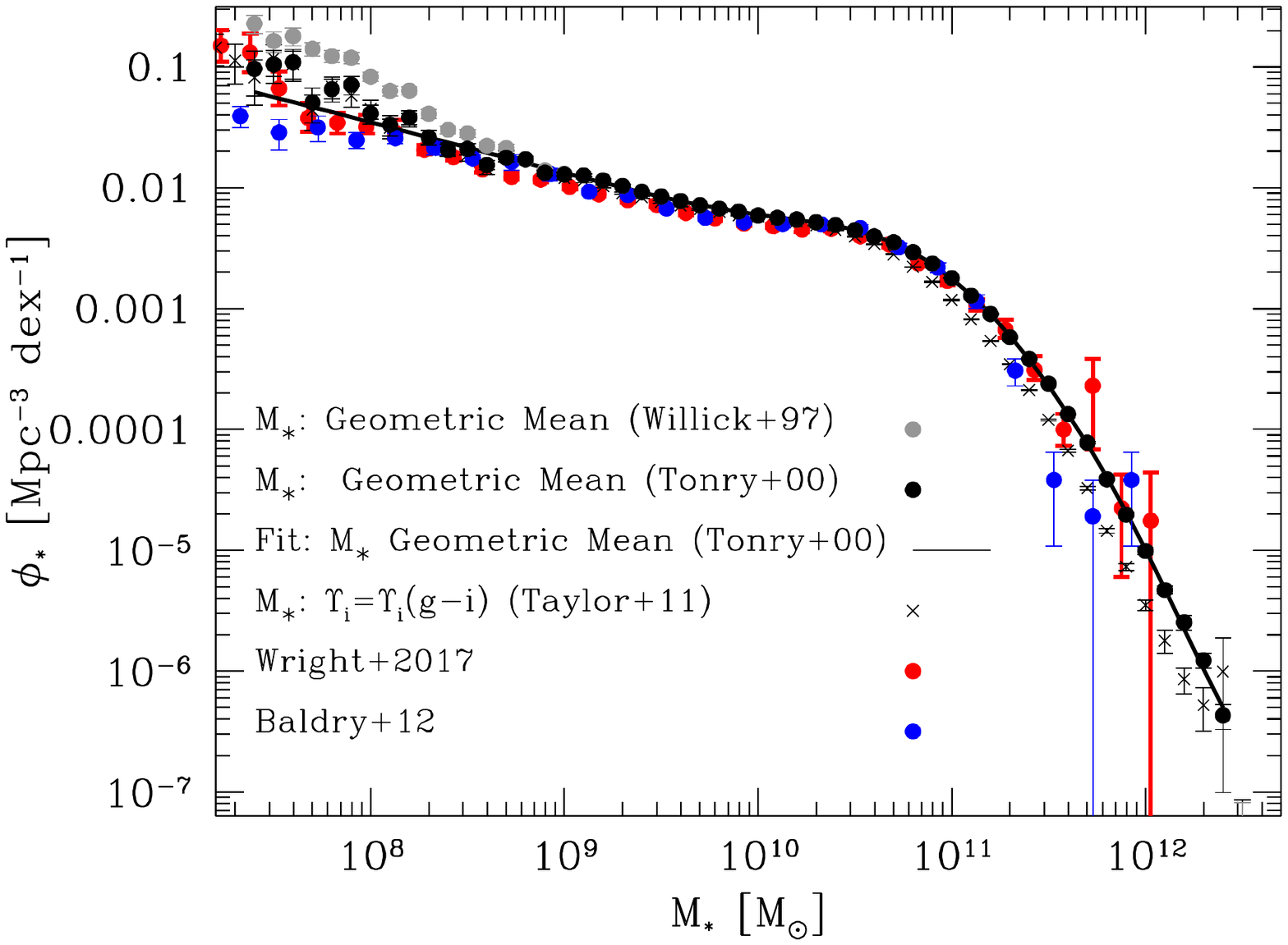}
 		\vspace*{-70pt}
 		\caption{Corrected \gsmf\ for SB and for the flow model, filled circles with
 			error bars. This is our fiducial \gsmf. The corresponding best fit model is shown with the solid line, see
 			Section \ref{best-fitGSMF}. Grey filled circles shows when using the flow model from \citet{Willick+1997}. We also
 			show our results when using the mass-to-light ration from \citet{Taylor+2011} and
 			compared to \citet{Baldry+2012} who used the same mass estimator. Notice that both mass functions are consistent 
 			between each other. For completeness we compared to \citet{Wright+2017}.  
 		} \label{fig:gsmf_final}
 	}
 \end{figure}

 In a volume-limited sample that is complete in stellar mass, we can derive the \gsmf\ as the number of
 observed galaxies, $N_{\rm gals}$, per unit of comoving volume $V$ with stellar masses between 
 $\log \ms\pm \Delta \log\ms/2$, that is:  $\phi_{\ast}(\ms) \Delta \log\ms = N_{\rm gals} / V$. 
 Once we determined the dependence of the stellar mass limit with redshift, $M_{*, \rm lim}$, from the SDSS NYU VAGC sample, 
 we can use the above idea by defining various volume-limited subsamples that are complete in stellar mass. These subsamples were defined by 
 dividing the galaxy redshift range
 covered by the NYU VAGC, $0.0033<z<0.005$, into 20 bins. Therefore, the GSMF for the $j$th volume-limited subsample at the redshift range 
 $z\pm\Delta z/2$ that is complete in stellar mass can be estimated for the mass bin $\log \ms \pm \Delta \log \ms/2$ as
 \begin{equation}
 \phi_j(M_{\ast},z)  \Delta \log \ms= \frac{N_{{\rm gals},j}(M_{\ast},z)}{V(z-\Delta z/2) - V(z+\Delta z/2)}.
 \end{equation}
 
 We tested the above methodology with realistic mock galaxy catalogues. To do so, 
 we use the $N$-body Bolshoi-Planck simulation \citep{Klypin+2016}, and halo catalogues
 described in \citep{Rodriguez-Puebla+2016}. We use the semi-emperical modelling from 
  \citep{Rodriguez-Puebla+2017} in order to assign galaxies to dark matter halos/subhalos. 
 The galaxies in the catalogue were projected into the redshift space through a lightcone. 
  We use the dependence of the stellar mass limit with redshift described in Appendix \ref{secc:stelim_z}
  and include galaxies only within the same redshift range as the NYU VAGC in order
  to reproduce the observed distribution of galaxies in the \ms\ and redshift plane for the lowz-SDSS
  sample. Our results show that the above methodology recovered the original \gsmf\ with differences
  not larger than $\sim5\%$. In addition, we have calculated the \gsmf\ using the 
  Stepwise Maximum Likelihood method \citep{Efstathiou+1988} and found very similar similar results (not shown) 
  as those reported based on our methodology.
 
 Finally, we calculate the \gsmf\ corrected by surface brightness incompleteness by 
 \begin{equation}
 \phi_{{\rm SB},j}(\ms,z)  \Delta \log \ms= \frac{N_{{\rm SB,gals},j}(M_{\ast},z)}{V(z-\Delta z/2) - V(z+\Delta z/2)},
 \end{equation}
 where 
 \begin{equation}
 N_{{\rm SB,gals},j} = \sum _{i=1}^{N_{{\rm gals},j}} w_{{\rm SB},i},
 \end{equation}
 and $w_{{\rm SB},i}$ is our SB incompleteness correction given by Equation (\ref{w_sb}).
 
 The bottom panel of Figure \ref{fig:pdf_z_gsmf} shows $\phi_{{\rm SB},j}(M_{\ast},z_j)$ for three different stellar masses, $\ms=10^{8.4}\msun$, $10^{9.5}\msun$ 
 and $10^{10.2}\msun$. Fluctuations in the amplitud of the GSMF shows that the distributions of galaxies is not uniform across the redshift distribution because of
 environmental effects arising from large scale structures. In order to minimise the above effect, we compute the weighted mean of the GSMF. 
 In other words, we derive the total GSMF as 
 \begin{equation}
 \langle \phi_{\rm SB}(M_{\ast}) \rangle = \sum_{j=1}^{N=20} \phi_{{\rm SB},j}(M_{\ast},z_j) \times w_j
 \end{equation}
 where $w_j = N_{{\rm gals},j} / \sum_j N_{{\rm gals},j} $ the fraction of galaxies at the $j$th volume-limited subsample centred at the
 redshift bin $z\pm\Delta z/2$
 for the mass bin $\log \ms \pm \Delta \log \ms/2$. The solid line in Figure \ref{fig:pdf_z_gsmf} shows 
 the resulting value of $\langle \phi_{\rm SB} \rangle$ for the masses discussed above. 
 
 Figure \ref{fig:gsmf_SB} compares the resulting GSMFs when SB corrections are applied $\phi_{{\rm SB},*}$ (red filled circles)
 and when we ignore SB corrections $\phi_{*}$ (blue filled circles) for each of the six stellar mass 
 definitions used here. As expected, the SB correction increases the number density of low-mass galaxies.
 For higher masses than $\sim 3\times 10^9$ \msun, this correction is negligible. For comparison, we reproduce with grey filled circles
 the GSMF reported in \citet{Baldry+2008}, who used also the  \texttt{low-z} NYU-VAGC sample 
 but for the DR4 as well as the \citet{Baldry+2012} from the GAMA survey with the skeletal symbols. In non of them SB corrections were applied. 
 Finally, in all the panels of Figure \ref{fig:gsmf_SB} we reproduce the GSMF from the main 
 SDSS DR7 derived in Appendix \ref{SDSS_DR7_GSMF}. Observe how the GSMFs constructed from the \texttt{low-z} NYU-VAGC sample
 and the ones constructed from the main  SDSS DR7 samples match extremely well at $\ms\sim 10^9$ \msun, but the latter overcomes the
 former at high stellar masses due to the larger volume covered by the SDSS DR7. 
 
 Finally, we briefly describe our final GSMF. For galaxies below $\ms=10^{9}\msun$, we use the GSMF derived from the  
 \texttt{low-z} NYU-VAGC sample, while for galaxies above $\ms=10^{9}\msun$, we use the GSMF from the SDSS DR7 
 based on the photometric catalog from \citet{Meert+2015}.
 We apply a simply correction in our GSMF for passing from the \citet{Willick+1997} distance flow model to the \citet{Tonry+2000} one. 
 Figure 12 from \citet{Baldry+2012} shows that after adjusting the \citet{Baldry+2008} GSMF to the \citet{Tonry+2000} distances both MFs are 
 in excellent agreement. With that information, we first note that our fiducial (uncorrected) GSMF (bottom right panel from Figure \ref{fig:pdf_z_gsmf}) is very similar to the \citet{Baldry+2008} GSMF, and thus we assume that the impact of correcting by 
 \citet{Tonry+2000} distances is equivalent to 
 rescale it to the \citet{Baldry+2012} GSMF. Based on the above, we rescale our SB-corrected GSMF as 
 $ \phi_{\rm SB, T00} = \langle \phi_{\rm SB} \rangle\times \phi_{\rm B08} /  \phi_{\rm B12} $. Recall that
 our fiducial \gsmf\ uses stellar masses from the geometric mean of all stellar masses described by Equation (\ref{mass_to_light_ratios}).  
 
 Figure \ref{fig:gsmf_final} shows our final GSMF, $\phi_{\rm SB, T00}$, as the black filled circles with error bars. The filled
 grey symbols show the GSMF, $\langle \phi_{\rm SB} \rangle$, in which the \citet{Willick+1997} model flow is utilised. We also compare
 to \citet{Baldry+2012} and \citet{Wright+2017} determinations. Note that after distance and SB corrections our
 fiducial GSMF is in good agreement with the observed low-mass end 
 slope of the GAMA survey. For comparison we present our corrected
 GSMF but when using the \citet{Taylor+2011} mass-to-light ratios. Note that in this case our  GSMF is consistent with 
 the \citet{Baldry+2012} GSMF.

\subsection{The impact of galaxy classification: the criteria for separating the galaxy population into two groups}
\label{impact_of_color}

\begin{figure*} 
	\vspace{-190pt}
	\hspace{-10pt}
	\includegraphics[height=7.8in,width=6.2in]{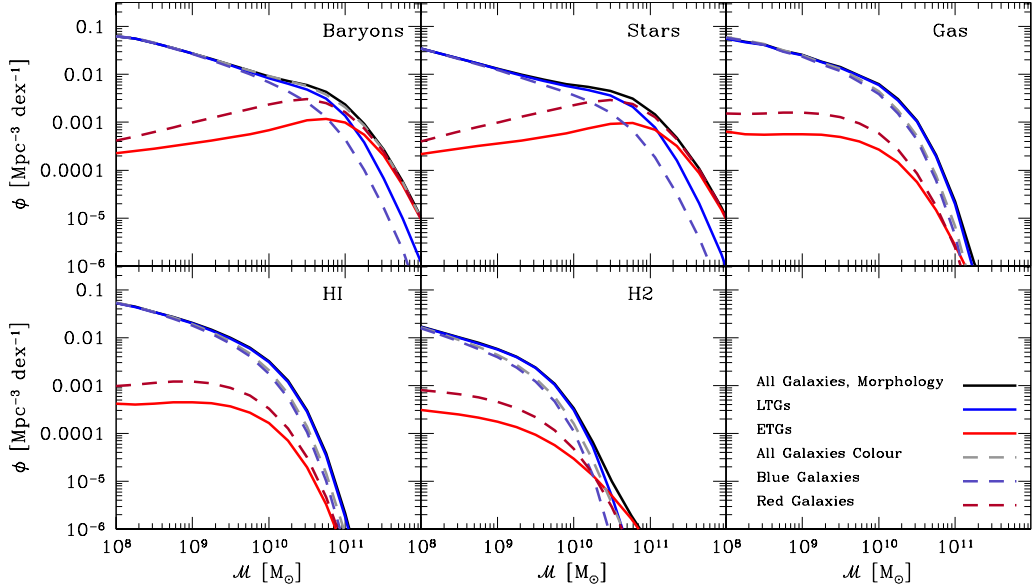}
	\vspace{-120pt}
	\caption{Impact on the MFs due to the use of two different criteria for the division of
		the galaxy population.
		The solid lines show the original MF from Section \ref{MFs_results}, based on
		galaxy morphology, while the dashed lines show the results when using galaxy colour. The 
		classification according to galaxy colours results in a overabundance of red galaxies
		compared to early-types, especially at intermediate and low masses.}
	\label{fig:mfs_color_cirteria} 
\end{figure*}

For our goal of projecting gas scaling correlations (more precisely, the gas CPDFs) into gas MFs separately for 
early- and late-type galaxies, the derivation of the fraction of early-type galaxies as a function of \ms, $f_E(\ms)$,
was an important step. 
As discussed in Section \ref{fraction_ETG}, based on
the morphological classification from \citet{Huertas-Company+2011}
we found the GSMFs of early- and late-type galaxies that are in good agreement with the results based on the 
visual classification from \citet{Nair_Abraham2010}, and with the classification based on concentration
utilised in \citet{Bernardi+2010}. In contrast, we found that our GSMFs of early- and late-type galaxies are in 
tension when comparing to those from the GAMA survey with their visual morphological classification, but interestingly
enough,  they agree with the GAMA GSMFs when we use a $g-r$ color criterion to separate our galaxies into
the two populations.  
Recall that for the GAMA classification, Sa galaxies are included into their early-type group since their visual classifications combines 
S0 and Sa galaxies \citep{Kelvin+2014,Moffett+2016}, contrary to our definition; see Section \ref{fraction_ETG} for 
more details. Thus, understanding the impact of using different criteria to separate
the galaxy population into two main groups is of great importance in our study. Following, we study the impact 
of using galaxy color instead of morphology in order to give a rough idea of 
what would it be the result of using a very different proxy to galaxy morphology (a one close to the GAMA survey, for instance). 

The lower panel  of Figure \ref{fig:fetgs} presented the fractions of early-type galaxies as well as of 
red galaxies as a function of \ms. The fraction of red galaxies is clearly larger than the one of early-type galaxies
at all masses. Based on the SDSS DR7 sample described in Section \ref{local-GSMF+correlations}, 
we found that the great majority of the galaxies that are classified as early-type are actually red; the fraction 
of early-type galaxies with blue colours has a maximum at $\ms\sim 8\times10^{10}\msun$ representing only $\sim5\%$
of the population. In contrast, the fraction of red galaxies classified as late-types is larger than $\sim10\%$
at practically all masses, and it peaks at $\ms\sim 2\times10^{10}\msun$ with a contribution of $\sim 50\%$
(we also observe a second peak at the massive-end $\ms\sim 4\times10^{11}\msun$).  
Similar results have been reported in previous studies 
\citep[see e.g.,][]{Masters+2010a}. Additionally, note
that we ignored the effects of reddening due to extinction from the galaxy inclination, which would misclassify 
galaxies based on their colours  \citep[see e.g.,][]{Masters+2010b}. Therefore, from the physical point of view, 
the separation of the galaxy population by colour is, perhaps, not as ``clean" or reliable as morphology. 

Figure \ref{fig:mfs_color_cirteria} presents the resulting MFs when using the fraction
of red galaxies, $f_r(\ms)$, as a proxy for early-type galaxies, dashed lines. The solid lines reproduce the results from 
Fig. \ref{mfs_comp_to_obs}, where our morphology-based fraction, $f_E(\ms)$, was used. 
Notice that for \HI, \H2, and cold gas mass, not only the MFs of blue and red
galaxies are different to their morphological counterparts but also the total MFs. The above
can be understood in the following terms. 
Using the fraction of red galaxies as a proxy of early-type galaxies
results in a large fraction of red galaxies misclassified  as
late-types as discussed above. However, the above has a
lager impact for early-type galaxies with low to intermediate masses than at high masses, while for late-type galaxies, 
the major impact is from intermediate to high masses. As a consequence, on one hand, the resulting \HI\ and \H2\ MFs 
see an increase of early-type galaxies at their low-mass ends. 
Interesting enough, the use of $f_r$ instead of $f_E$ would produce \HI\ and \H2\ MF of early-type galaxies
in better agreement with the inferences of the ATLAS 3D sample.  On the other hand, lowering the fraction of late-type 
galaxies at intermediate-high masses, which have significantly larger gas fractions than early-type galaxies, 
affects the projected total \HI\ and \H2\ MFs, and they 
would be in tension with direct observations, especially with the \HI\ MF from the ALFALFA and HIPASS surveys. 

Finally, we emphasise that the above does not imply that using galaxy colours will lead to 
incorrect inferences of the gas MFs but that combining two different criteria 
for dividing the galaxy population will lead to a very different results that, perhaps,
will be in tension with the observations. Thus, the success of our determinations is 
in part that we are using data sets that are consistent between each other in that
regards the morphological separation into two galaxy subpopulations.

 \section{Deconvolution Algorithm}
 \label{deconvolution}
 
 Individual mass estimates are subject to 
 random errors. Thus, every MF that is inferred from observations 
 through indirect estimations of any type of mass (we will
 denote this as $\phi_{\rm obs}$) is the result of the random errors over the
 intrinsic mass, (it will be denoted by $\phi_{\rm int}$). Formally, we can represent the observed $\phi_{\rm obs}$ as 
 the convolution of $\phi_{\rm int}$:
 \begin{equation}
 \phi_{\rm obs}(M)=\int \mathcal{G}(\log M-\log x)\phi_{{\rm int}}(x)d\log x.
 \label{conv_gsmf}
 \end{equation}
 We will assume that random errors have a lognormal distribution, denoted by $\mathcal{G}(\log M-\log x )$:
 \begin{equation}
 \mathcal{G}(\log M -\log x)=\frac{1}{\sqrt{2\pi\sigma^2}}\exp\left[-\frac{1}{2\sigma^2}\log^2\left(\frac{M}{x}\right)\right],		
 \end{equation}
 where $\sigma$ are the 1-$\sigma$ statistical fluctuations, in either directions, in the inferred galaxy masses. 
 Note that in Equation (\ref{conv_gsmf}) the units for $\phi_{\rm obs}$ and  $\phi_{\rm int}$ are in Mpc$^{-3}$dex$^{-1}$.
 
 The basic idea of our algorithm is simple. We start by defining the following relation:
 \begin{equation}
 \phi_{ {\rm int}}^j(M) = \phi_{ {\rm int}}^{j-1}(M) \int \mathcal{G}(\log M -\log x)\frac{\phi_{\rm obs}}{\phi_{ {\rm conv}}^{j-1}} (x)d\log x,
 \end{equation}
 where 
 \begin{equation}
 \phi_{ {\rm conv}}^{j-1}(x) = \int \mathcal{G}(\log x-\log y)\phi_{ {\rm int}}^{j-1}(y)d\log y,
 \end{equation}
 with $\phi_{ {\rm int}}^{j-1}$ denoting the $j$th iterated intrinsic MF. Note that as
 $\phi_{ {\rm conv}}^{j-1}$ approaches to $\phi_{{\rm obs}}$ the above equation converges to the maximum
 likelihood solution for $\phi_{ {\rm int}}^{j-1}$, in other words, we have found the numerical
 solution to the intrinsic MF, $\phi_{{\rm int}}$.
 The zero-th iteration is defined as convolution of the observed MF with the lognormal distribution $\mathcal{G}$ described above:
 \begin{equation}
 \phi_{ {\rm int}}^0(\ms)=\int \mathcal{G}(\log\ms-\log x)\phi_{\rm obs}(x)d\log x.
 \end{equation}
 We declare that the $\phi_{{\rm int}}^{j}$ has converged when the parameter $\Delta \leq 7\%$, defined
 as the relative error between the observed MF and the  $j$-th iterated intrinsic MF convolved with the random
 error distribution:
 \begin{equation}
 \Delta = \frac{100\%}{N}\sum_i\left| 1 - \frac{\int \mathcal{G}(\log\ms-\log x)\phi_{ {\rm int},i}^{j}d\log x}{\phi_{ {\rm obs},i}}\right|.
 \end{equation}
 The summation in the above definition goes over all the tabulated values of individual reports of the observed MF $\phi_{{\rm obs}}$.
 By trial and error we found that the value of $\Delta = 7\%$ is a compromise between accuracy and 
 efficiency. Typically, $\Delta = 7\%$ was reached in less than 10 iterations.

\end{appendix}


\bibliographystyle{pasa-mnras}
\bibliography{Bibliography} 

\label{lastpage}
\end{document}